\documentclass[a4paper]{article}

\usepackage[margin=1in]{geometry} 

\usepackage[utf8]{inputenc}
\usepackage[T1]{fontenc}

\usepackage{mathtools}
\usepackage{amsmath}
\usepackage{amsthm}
\usepackage{amssymb}
\usepackage{physics}
\usepackage{float}
\usepackage[colorlinks=true]{hyperref}
\usepackage{booktabs}
\usepackage{array}
\usepackage{bm}
\hypersetup{%
    unicode,
    pdfauthor={Author One, Author Two, Author Three},
    pdftitle={QAOA},
    pdfkeywords={qaoa, qosf, quantum optimization},
}
\usepackage{authblk}
\usepackage{multirow}
\usepackage[sort&compress,numbers,square]{natbib}
\usepackage{graphicx, color}
\graphicspath{{img/}}

\definecolor{OurRed}{RGB}{184,80,75}
\definecolor{OurBlue}{RGB}{88, 118, 185}
\definecolor{OurPurple}{RGB}{103, 82, 158}
\hypersetup{%
    colorlinks=true,
    linkcolor=OurBlue,
    citecolor=OurRed,
    urlcolor=OurPurple,
    pdftitle={A Review on Quantum Approximate Optimization Algorithm and its Variants},
    pdfnewwindow=true,
}

\title{A Review on Quantum Approximate Optimization Algorithm and its Variants}

\author{Kostas Blekos\thanks{mplekos@upatras.gr}}
\affil{Department of Physics, School of Natural Sciences, University of Patras, Patras 26504, Greece; email: mplekos@upatras.gr}
\author{Dean Brand}
\affil{Department of Physics, Stellenbosch University, Stellenbosch 7600, South Africa; email: deanbrand@protonmail.com}
\author{Andrea Ceschini}
\affil{Department of Information Engineering, Electronics and Telecommunications, Sapienza University of Rome, Rome 00184, Italy; email: andrea.ceschini@uniroma1.it}
\author{Chiao-Hui Chou}
\affil{Department of Engineering Science and Ocean Engineering, National Taiwan University, Taipei 10617, Taiwan; email: 
s110112020@gmail.com}
\author{Rui-Hao Li}
\affil{Department of Physics, Case Western Reserve University, Cleveland, Ohio 44106, USA; email: rxl527@case.edu}
\author{Komal Pandya}
\affil{Department of Computer Science and Engineering, Indian Institute of Technology Patna, Patna 801106, India; email: komal.lame@gmail.com}
\author{Alessandro Summer}
\affil{School of Physics, Trinity College Dublin, Dublin 2, Ireland; email: summera@tcd.ie}

\newcommand{\G}{\mathcal{G}}    
\newcommand{\V}{\mathcal{V}}    
\newcommand{\E}{\mathcal{E}}    
\newcommand{\h}{\hat{H}}        
\newcommand{\U}{\hat{U}}        
\newcommand{\vc}[1]{\vb{#1}}     
\newcommand{\bs}[1]{\boldsymbol{#1}} 

\newcommand{\maqaoa}{ma-QAOA}	

\begin{document}
    \maketitle
   
    \begin{abstract}
	The Quantum Approximate Optimization Algorithm (QAOA) is a highly promising variational quantum algorithm that aims to solve combinatorial optimization problems that are classically intractable.
    This comprehensive review offers an overview of the current state of QAOA, encompassing its performance analysis in diverse scenarios, its applicability across various problem instances, and considerations of hardware-specific challenges such as error susceptibility and noise resilience.
    Additionally, we conduct a comparative study of selected QAOA extensions and variants, while exploring future prospects and directions for the algorithm.
    We aim to provide insights into key questions about the algorithm, such as whether it can outperform classical algorithms and under what circumstances it should be used.
    Towards this goal, we offer specific practical points in a form of a short guide.
    \end{abstract}
    
    \textit{All authors contributed equally to this work.}
    
    \tableofcontents
    \listoffigures{}
    \listoftables{}

    \section{Introduction}%
    \label{sec:intro}
    
Although fault-tolerant quantum computers are still years away, significant progress has been made in the development of the Noisy Intermediate-Scale Quantum (NISQ) machines, and there is a growing interest in finding useful algorithms meant to be run on these near-term quantum devices.
As such, Variational Quantum Algorithms (VQAs)~\cite{mccleanTheoryVariationalHybrid2016, bhartiNoisyIntermediatescaleQuantum2022, cerezoVariationalQuantumAlgorithms2021,bravo-prietoVariationalQuantumLinear2020,jiangNearoptimalQuantumCircuit2017,baaquieQuantumClassicalHybridAlgorithms2023} have been proposed to take advantage of current quantum systems through a hybrid quantum-classical optimization routine.
The hybrid loop of a VQA involves a parameterized quantum circuit to be run on a quantum computer and an optimizer that can update the parameters on a classical machine by minimizing the cost function constructed based on the outputs of the quantum circuit.
In this way, VQAs often have the advantage of having shallow quantum circuits, making them less susceptible to noise in NISQ devices.
To date, VQAs have found use cases in various areas, including quantum chemistry simulations, machine learning, and optimization~\cite{kokailSelfVerifyingVariationalQuantum2019, bauerQuantumAlgorithmsQuantum2020, lloydQuantumEmbeddingsMachine2020, delgadoVariationalQuantumAlgorithm2021, amaroFilteringVariationalQuantum2022}.

In particular, the Quantum Approximate Optimization Algorithm (QAOA)~\cite{farhiQuantumApproximateOptimization2014, farhiQuantumApproximateOptimization2015} is one of the most promising VQAs that has attracted great interest in recent years.
QAOA is designed to find approximate solutions to hard combinatorial optimization problems on quantum computers: it encodes the Hamiltonian related to the problem into a quantum circuit and leverages adiabatic time evolution and layering to optimize the variational parameters of the circuit, such that the approximate solution to the problem can be constructed by measuring the QAOA circuit with the optimal set of parameters.
The fundamental building block, a single layer of the QAOA circuit, consists of a cost layer associated with the problem Hamiltonian and a mixer layer whose corresponding Hamiltonian does not commute with the problem Hamiltonian.
The performance of QAOA is typically measured by the approximation ratio $C_\text{QAOA}/C_{\max}$, i.e., the ratio of the cost associated with the solution output by QAOA to that of the true optimal solution.
Theoretically, such an approximation ratio increases with increasing layers $p$, as QAOA recovers the adiabatic evolution in the $p\to \infty$ limit.

QAOA is suitable for finding good approximated solutions to several optimization problems, such as Maximum Cut (MaxCut)~\cite{farhiQuantumApproximateOptimization2014}, Maximum Independent Set (MIS)~\cite{choiTutorialQuantumApproximate2019,zhouQuantumApproximateOptimization2020}, Binary Paint Shop Problem (BPSP)~\cite{streifBeatingClassicalHeuristics2021}, Binary Linear Least Squares (BLLS)~\cite{borleQuantumApproximateOptimization2021}, Max E3LIN2~\cite{farhiQuantumApproximateOptimization2015}, Multi-Knapsack~\cite{awasthiQuantumComputingTechniques2023}, and, more generally, Quadratic Unconstrained Binary Optimization (QUBO) problems~\cite{moussaUnsupervisedStrategiesIdentifying2022}. 
Consequently, applications of QAOA in the real world are many and far-reaching. 
Some recent examples include portfolio optimization~\cite{hodsonPortfolioRebalancingExperiments2019, bakerWassersteinSolutionQuality2022}, tail assignment~\cite{vikstalApplyingQuantumApproximate2020}, object detection~\cite{liHierarchicalImprovementQuantum2020}, maximum likelihood detection of binary symbols over a multiple-input and multiple-output channel~\cite{cuiQuantumApproximateOptimization2021}, text summarization~\cite{niroulaConstrainedQuantumOptimization2022}, maximum independent set~\cite{ebadiQuantumOptimizationMaximum2022}, factorization (Variational Quantum Factoring
algorithm)~\cite{anschuetzVariationalQuantumFactoring2018, karamlouAnalyzingPerformanceVariational2021}, protein folding~\cite{mustafaVariationalQuantumAlgorithms2022}, and wireless scheduling~\cite{choiQuantumApproximationWireless2020}.

However, at the moment, literature still contains many conflicting opinions on various aspects of the algorithm, such as for which problems, if any, QAOA can outperform classical algorithms and whether it can provide any practical quantum advantage under the noise and errors of current quantum devices.
Here we extensively study the available literature in order to provide a comprehensive review of the current status of QAOA and summarize existing results in different aspects of the algorithm.
This review aims to be a guide for using QAOA, providing insights into key questions about the algorithm, that is, whether QAOA can outperform classical algorithms and under what circumstances it should be used.
Additionally, we provide meaningful insights on QAOA's potential for achieving quantum advantage, and discuss promising research directions for the future.
In particular, we focus on the following aspects: a survey of various extensions and variants of the QAOA ansatz, strategies to improve parameter optimization, efficiency, and performance analysis of the algorithm in various problem instances, and hardware-specific issues including the effects of noise and hardware-tailored implementations.
Moreover, we also implement and assess the efficiency and performance of some promising QAOA variants on the MaxCut problem, which is a paradigmatic combinatorial optimization problem commonly used to benchmark the potential of the algorithm~\cite{villalba-diezImprovementQuantumApproximate2021}.

The remainder of this paper is organized as follows. 
The Background section (Section~\ref{sec:background}) provides an overview of relevant hard combinatorial problems such as the MaxCut problem, General QUBO Problems, and related algorithms, including the Variational Quantum Eigensolver (VQE) and Quantum Adiabatic Algorithm (QAA). 
The Analysis section (Section~\ref{sec:analysis}) provides a detailed examination of various aspects of QAOA, including the ansatz, computational efficiency, quality of solution, effects of noise and errors, and hardware-specific implementations (see Figure~\ref{fig:general_scheme}).
Our Experimental Results (Section~\ref{sec:experimental}) provide quantitative evaluations and performance comparisons between different QAOA variants.
In Section~\ref{sec:discussion} we summarize our findings, highlight possible applications for QAOA, discuss its potential quantum advantage and examine future directions for the research.
Finally, in Section~\ref{sec:guide}, we provide insights for a practical guide to QAOA by answering key questions about the algorithm, such as which QAOA variant or ansatz to use for a specific problem and how to effectively optimize it.

    \section{Background}%
    \label{sec:background}
    
Generally speaking, combinatorial optimization problems concern finding the optimal solution among a set of feasible solutions, given some constraints on a discrete set of variables.
The objective function can either be minimized or maximized, and it can be seen as a (possibly weighted) sum of the clauses satisfied by a feasible solution.

Some typical combinatorial optimization problems include the Knapsack, Traveling Salesman, and MaxCut problem~\cite{korteCombinatorialOptimization2012}.
However, due to the combinatorial nature of such problems, the solution space explodes with respect to the number of inputs, and the optimization process quickly becomes intractable.
Generally, finding the exact solution to many combinatorial optimization problems belongs to the NP complexity class~\cite{hammerBooleanMethodsOperations1968}.
This means that classical algorithms cannot efficiently retrieve the optimal solution since the time required scales exponentially with the number of inputs.

In this context, approximate optimization algorithms are employed to find a good approximate solution in polynomial time~\cite{ausielloComplexityApproximation1999}, which can be formulated as follows.
Given a combinatorial optimization problem defined on $n$-bit binary strings of the form $\textbf{x} = x_1 \cdots x_n$, where the goal is to maximize a given classical objective function $C(\textbf{x}):{\{0,1\}}^n \rightarrow \mathbb{R}_{\geq 0}$, an approximate optimization algorithm aims to find a solution $\mathbf{x^*}$ such that the approximation ratio $\alpha$, defined as
\begin{equation}
    \alpha = \frac{C(\mathbf{x^*})}{C_\text{max}},
\end{equation}
with $C_\text{max} = \max_\mathbf{x} C(\mathbf{x})$, reaches some desired value. 
Ideally, the value should be as close to 1.
Despite the fact that the solution found by approximate algorithms may not be optimal, it generally comes with some optimality guarantees, which are typically lower bounds on the quality of the solution.
For example, an algorithm is said to be $\alpha$-approximated for a problem if and only if it can find a solution within a factor $\alpha$ ($\leq 1$) of the optimal solution for every instance of the problem~\cite{vaziraniApproximationAlgorithms2003}.
Thus, should such an algorithm exist, the above criterion certifies that the approximate solution is at least $\alpha$ times the optimum.
However, for some optimization problems, the gap between the approximate solution and the optimal one cannot be reduced in polynomial time, suggesting the difficulty of finding tight lower bounds with respect to the optimal solution.
This is known as the ``hardness of approximation'', and it implies that finding a polynomial time approximation for the underlying problem is impossible unless P = NP~\cite{khotInapproximabilityNPcompleteProblems2011}.
A comprehensive list of state-of-the-art approximation algorithms for some key combinatorial optimization problems can be found in~\cite{vaziraniApproximationAlgorithms2003}.

More formally, as outlined in Section~\ref{sec:qubo}, many optimization problems can be transformed into a quadratic unconstrained binary optimization (QUBO) form.
However, QUBO problems are usually NP-complete~\cite{kochenbergerUnifiedModelingSolution2004}, meaning that finding the solution classically requires traversing a solution space that grows exponentially with the problem size.
On the other hand, quantum computing promises to enable exponentially faster computation due to the superposition nature of qubits.
A quantum system's exponentially growing Hilbert space can naturally accommodate the solution space of a combinatorial optimization problem and, therefore, may provide advantages in solving such problems over classical machines.
The Quantum Approximate Optimization Algorithm (QAOA) is designed to tackle QUBO problems by utilizing a quantum circuit to find approximate solutions. 
The objective is to address the inherent hardness of approximation present in classical computation by leveraging the capabilities of QAOA. 
It should be noted, however, that while QAOA has the potential to be applied to a wide range of optimization problems, its effectiveness is dependent on the specific problem characteristics (more details in Sections~\ref{sec:analysis_performance} and \ref{sec:efficiency}).


\subsection{MaxCut Problem Overview}
\begin{figure}[!ht]
    \centering{\includegraphics[width=1\columnwidth]{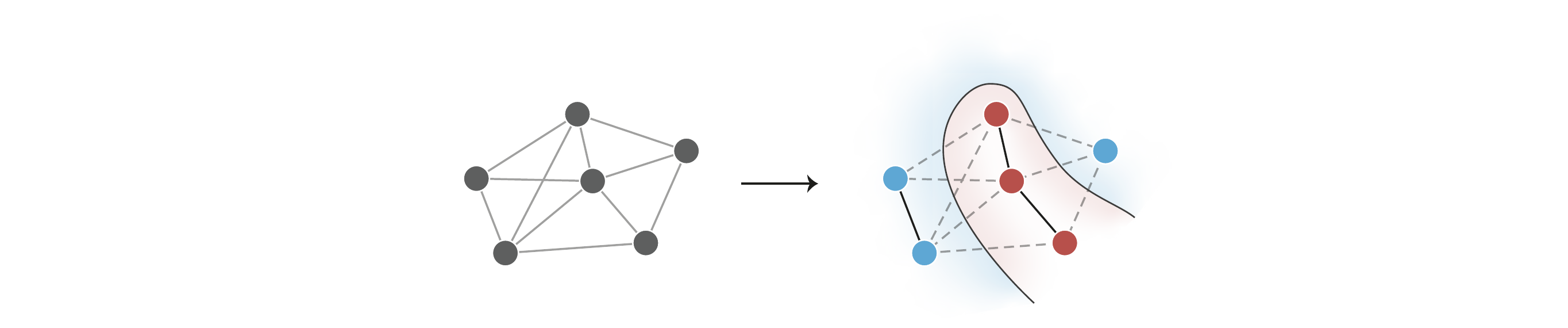}}
    \caption[Illustration of a 6-nodes graph and its maximum cut]{Left: A problem graph with 6 vertices and 11 equal-weight edges. Right: The solution to the MaxCut problem, where the vertices are partitioned into two groups (red and blue) such that the number of edges crossed by the cut (black curve) is maximized, which is 8.}%
    \label{fig:maxcut}
\end{figure}

The MaxCut problem is one of the most well-known optimization problems, and it is hereby discussed in detail.
It involves finding a cut in a graph such that the vertices of the graph are divided into two complementary subsets, and the sum of the weights of the edges crossed by the cut is maximized.
The MaxCut problem can be formulated as follows:
Given an undirected graph $\G = (\V,\E)$, where $\V$ is the set of vertices, $\E$ is the set of edges, and $w_{ij}$ is the weight corresponding to the edge $(i,j) \in \E$, which connects the vertices $i$ and $j$.
The objective of the MaxCut is to partition the graph vertices $x_i$, for $i = 1,\ldots,\lvert\V\rvert$, into two complementary subsets labeled by 0 and 1, such that the weighted sum of the edges connecting vertices in different partitions, defined as
\begin{equation} \label{eq:objective}
    \ C(\mathbf{x}) = \sum_{i,j = 1}^{\abs{\V}} w_{ij}x_i(1-x_j),
\end{equation}
is maximized, where $w_{ij} > 0$, $w_{ij} = w_{ji}$, $\forall (i,j) \in \E$, and $x_i \in \{0,1\}$.
An example of the MaxCut problem is illustrated in Figure~\ref{fig:maxcut}.
With general weights $w_{ij}$, the problem is commonly known as the weighted-MaxCut; the plain MaxCut problem is a special case of the weighted-MaxCut where $w_{ij} = 1$ for all $(i,j) \in \E$.

Since finding a cut that yields the maximum value of the objective function, $C$ is an NP-hard problem~\cite{kochenbergerUnconstrainedBinaryQuadratic2014}, our best hope for a polynomial-time algorithm lies in an approximate optimization approach.
This means finding a partition $\mathbf{x}^*$ which yields a value $C(\mathbf{x}^*)$ that is as close as possible to the maximum value $C_\text{max} = \max_\mathbf{x} C(\mathbf{x})$.
\citet{hastadOptimalInapproximabilityResults2001} conjectured that achieving an approximation ratio higher than $16/17 \simeq 0.9412$ is NP-hard, highlighting the hardness of approximation for the MaxCut problem.
Currently, the best-performing classical algorithm for the MaxCut problem is the Goemans-Williamson algorithm, which delivers a solution with $\alpha \simeq 0.878$~\cite{karloffHowGoodGoemans1999, goemansWorstcaseComparisonValid1995} (Section~\ref{sec:classical_algorithms}).
For these reasons, the MaxCut is considered the paradigmatic example of hard combinatorial optimization problems.
As a result, the aspiration is to find good and efficient approximate solutions to MaxCut by leveraging the power of quantum algorithms.
In this regard, QAOA is a quantum algorithm that has shown promise in finding solutions to hard combinatorial optimization problems such as MaxCut.
More details of the performance analysis of QAOA and the comparison to its classical counterparts will be provided in Section~\ref{sec:efficiency}.

\subsection{QUBO Problems and Applications}%
\label{sec:qubo}
A more general formulation of the MaxCut problem is represented by Quadratic Unconstrained Binary Optimization (QUBO) problems.
QUBO problems belong to the NP-complete class. This mathematically assures that any NP-complete problem can be mapped to a QUBO one in polynomial time. \citet{lucasIsingFormulationsMany2014} discussed how all the Karp's 21 NP-complete problems can be mapped to a QUBO one. 
Among them some relevant optimization problems  other than  MaxCut
are Graph Coloring~\cite{tabiQuantumOptimizationGraph2020}, Number Partitioning, and Quadratic Knapsack~\cite{gloverQuantumBridgeAnalytics}.
An extensive list of QUBO applications is presented in~\cite{kochenbergerUnifiedModelingSolution2004}.
In a QUBO problem, the vector of unknowns $\mathbf{x} = (x_1,\ldots,x_n)$ is represented by decision variables taking discrete binary values, so that $\mathbf{x} \in \{0,1\}^n$.
Moreover, a QUBO problem is defined by a square symmetric matrix $\mathbf{Q} \in \mathbb{R}^{n \times n}$.
Given the cost function
\begin{equation}
    \label{eq:qubocost}
    C(\mathbf{x}) = \mathbf{x}^T \mathbf{Q} \mathbf{x} = \sum_{i, j=1}^{n} Q_{ij}x_i x_j,
\end{equation}
the aim of a QUBO problem is to find the optimal vector $\mathbf{x}^*$ such that 
\begin{equation}
    \mathbf{x}^* = \arg\min_{\mathbf{x} \in \{0,1\}^n} C(\mathbf{x}).
\end{equation}
QUBO can also be defined as maximization problems instead of minimization ones by simply inverting the sign of the cost function $C(\mathbf{x})$, i.e.,\ by flipping the sign of the coefficients in $w$: 
\begin{equation}
    \min_{\mathbf{x} \in \{0,1\}^n} C(\mathbf{x}) = \max_{\mathbf{x} \in \{0,1\}^n} -C(\mathbf{x}) = \max_{\mathbf{x} \in \{0,1\}^n} \sum_{i, j=1}^{n} (- Q_{ij})x_i x_j.
\end{equation}
It is important to note that QUBO problems are unconstrained, namely there are no constraints on the variables $\mathbf{x}$.

QUBO instances can also be seen as a one-to-one correspondence with Ising models~\cite{lucasIsingFormulationsMany2014}.
Moreover, \citet{lodewijksMappingNPhardNPcomplete2020} discusses several mappings from NP-hard problems to QUBO problems and corrects the errors in the approach taken in~\cite{lucasIsingFormulationsMany2014}.
It also expands the range of NP-complete and NP-hard optimization problems that can be formulated as QUBO problems.
Ising problems replace the original QUBO variables $\mathbf{x} \in \{0,1\}^n$ with Ising variables $\mathbf{z} \in \{-1,1\}^n$, such that $z_i = 2x_i - 1$ for $i=1,\ldots,n$.
The final Ising Hamiltonian, which depends on $\mathbf{z}$, is equivalent to Eq.~\eqref{eq:qubocost} except for a constant irrelevant to the optimization.
A more detailed explanation of the relationship between QUBO and Ising models can be found in~\cite{gloverQuantumBridgeAnalytics}.
Inspired by the adiabatic theorem, annealing methods are used to find the ground state of a physical system. Similarly, solutions to Ising problems are often carried out with annealing techniques~\cite{mohseniIsingMachinesHardware2022}.
Due to its equivalence with Ising models, QUBO represents a family of problems suitable to be solved by adiabatic quantum computing through quantum annealing~\cite{dateEfficientlyEmbeddingQUBO2019}.

As previously mentioned, many optimization problems can be reformulated as QUBO problems. 
Although QUBO problems are limited to quadratic interactions between variables, they can be extended to higher-order terms. 
For example, let us consider a problem with a third-order term $x_ix_jx_k$. 
To convert this problem into a quadratic one, we can introduce an ancillary variable (also called ``gadget'') $x' \coloneqq x_ix_j$, and express the original term as $x_ix_jx_k = x'x_k$. 
This allows us to rewrite the entire problem in terms of quadratic interactions between variables.
Optimization problems with higher order interaction are in general refereed to as Polynomial Unconstrained Binary Optimization (PUBO) problems. \citet{babbushResourceEfficientGadgets2013} analysed how to efficiently map PUBO problems to QUBO ones.
The fundamental reason why QAOA is focused on QUBO instances is linked to the hardware constraints. 
However, in case the device can implement gates on more than two qubits it could be more advantageous to reduce the number of interaction by increasing their order. 
It was shown that arbitrary combinatorial optimization problems can be mapped to PUBO problems through dualizing constraints~\cite{herrmanLowerBoundsCircuit2021}. 
Alternatively, they can also be formulated in the Lechner-Hauke-Zoller (LHZ), or parity model, which is a lattice gauge model with nearest neighbor four-body interactions~\cite{lechnerQuantumAnnealingArchitecture2015, leibTransmonQuantumAnnealer2016, lechnerQuantumApproximateOptimization2020}.

\subsection{Classical Algorithms for MaxCut Problem}%
\label{sec:classical_algorithms}

Despite the MaxCut problem being a well-known optimization problem in computer science with a practical significance, finding the optimal solution is computationally challenging, as it is an NP-hard problem \cite{korteCombinatorialOptimization2012}. 
This means that no known algorithms can solve it in polynomial time concerning the size of the input. 
However, several approximation algorithms and heuristics can provide good solutions in a reasonable time for practical problem sizes \cite{ausielloComplexityApproximation1999}.

For example, greedy algorithms are widely used because of their simplicity and efficiency: they make locally optimal choices at each iteration step in the hope of finding a globally optimal solution, but such a myopic strategy of focusing only on the current step often leads to suboptimal solutions, especially when the problem exhibits complex interactions between different variables or features.
Local search algorithms are heuristics techniques that can overcome this limitation by systematically exploring the search space. 
They start with an initial solution and iteratively improve it by considering a neighborhood of the current solution and moving to the best neighboring solution; the quality of the solution found profoundly depends on the initial solution and the quality of the neighborhood explored.
However, local search algorithms may still get trapped in suboptimal solutions, especially if the search space is large or complex. 
Simulated annealing, a specific type of local search algorithm, can help avoid getting stuck in local optima. 
It uses a temperature parameter to control the probability of accepting a worse solution.
In the context of the MaxCut problem, simulated annealing can provide good solutions, but it can be slow for large problem instances.
Genetic algorithms are also popular metaheuristic algorithms that generate an initial population of possible solutions and apply genetic operators like selection, crossover, and mutation to generate new solutions, which are then evaluated using an objective function. 
They can provide reasonable solutions for complex optimization problems. However, they can be slow for large problem instances due to the need to maintain a population of solutions and evaluate each solution using the objective function.
Another popular heuristic algorithm for MaxCut is spectral clustering, that involves using the eigenvectors of the graph Laplacian matrix to partition the graph into two parts; the eigenvectors of the Laplacian matrix capture the global structure of the graph and are able to identify the most significant cuts in the graph.
Spectral clustering can provide good solutions for many MaxCut instances, despite being sensitive to the choice of the number of eigenvectors used for partitioning and the spectral gap between the eigenvalues. 
In addition, spectral clustering can be computationally expensive for large graphs.
Optimization techniques like linear programming and semidefinite programming (SDP) are based on formulating the problem as a mathematical program and solving it using optimization algorithms.
These techniques can provide strong theoretical guarantees on the quality of the solution, but they may require significant computational resources to solve large instances of the problem.

In this regard, a prominent classical approximation algorithm for the MaxCut problem is the Goemans-Williamson algorithm, based on Semidefinite Programming (SDP) relaxations. 
The algorithm transforms the MaxCut problem into an SDP problem by relaxing the binary constraints on the membership of each node in one of the two sets into a real-valued vector in a high-dimensional space.
The solution of the SDP relaxation provides a set of real-valued vectors that can be rounded randomly to obtain a feasible solution for the original problem. 
The randomized rounding procedure maps each vector to one of the two sets with probability proportional to its squared length.
Notably, the algorithm guarantees an approximation ratio of at least 0.87856, ensuring that the obtained cut weights at least 87.856\% of the optimal cut's weight~\cite{karloffHowGoodGoemans1999, goemansImprovedApproximationAlgorithms1995}. 
Under the Unique Games Conjecture, this approximation ratio is the best possible for any polynomial-time algorithm~\cite{khotOptimalInapproximabilityResults2007}.

The Goemans-Williamson algorithm can be summarized in the following steps~\cite{goemansImprovedApproximationAlgorithms1995}:

\begin{enumerate}
    \item Given a graph $G = (V, E)$ with $n$ vertices and edge weights $w_{ij}$, formulate the MaxCut problem as a QUBO that maximizes the objective function: $\sum_{i,j<i} w_{ij}x_i(1-x_j)$, where $x_i \in \{0, 1\}$ indicating which side of the cut vertex $i$ belongs.
    \item Relax the QP by replacing binary variables $x_i$ with unit vectors $y_i \in \mathbb{R}^n$ whose elements could be continuous variables, and $x_ix_j$ with $y_i^Ty_j$, where the superscript $T$ denotes the transpose operation. 
	This gives a semidefinite program (SDP) that maximizes the objective function: $\sum_{i,j<i} w_{ij}(1-y_i^Ty_j)$, subject to $\forall i \in \{0,\dots, n\}$, $y_i^Ty_i = 1$, and $Y = (y_i^Ty_j)$ is positive semidefinite.
    \item Solve the SD using a polynomial-time algorithm such as interior point methods to obtain an optimal solution $Y^*$.
    \item Choose a random vector $r \in \mathbb{R}^n$ from a Gaussian distribution and $\forall i$, let $h_i = \operatorname{sgn}(r^Ty_i)$, where $\operatorname{sgn}(x) = 1$ if $x \geq 0$ and $-1$ otherwise. 
	This gives a partition of $V$ into two sets: $S_+ = {i | h_i = 1}$ and $S_- = {i | h_i = -1}$.
    \item Return the cut $(S_+, S_-)$ as the output of the algorithm.
\end{enumerate}

Gaining a deep understanding of the classical approaches proposed to approximate the MaxCut problem can offer valuable insights into the problem's complexity and the limitations of classical computing resources.
Furthermore, with the advent of quantum computing, this knowledge can inspire the development of novel quantum algorithms that can leverage the power of quantum mechanics to solve this problem more efficiently.

\subsection{Variational Quantum Algorithms}

QAOA is part of a broader category of quantum algorithms known as Variational Quantum Algorithms (VQAs)~\cite{cerezoVariationalQuantumAlgorithms2021,weigoldPatternsHybridQuantum2021}.
These algorithms share the properties of being targeted primarily at optimization problems on NISQ devices, using the variational principle in quantum theory.
The variational principle is used to find the lowest expectation value which can be obtained for a particular observable, typically the ground state energy, with respect to a trial wave function.
This trial wave-function makes the principle variational, as it is parameterized by a set of values which allows for a general wave-function form to be fitted to the system and the minimum expectation value to be found.
This is expressed in terms of a Hamiltonian $\h$, and trial wave-function $\ket{\psi}$, to find the ground state energy of the system $E_0$, which is bounded as follows,
\begin{equation}
    E_0 \leq \frac{\expval{\h}{\psi}}{\braket{\psi}}.
\end{equation}
Given this form, the objective of variational algorithms is to find a parametrization of $\ket{\psi}$, which minimizes the expectation value of the Hamiltonian.
This is achieved by approximating the eigenvector, $\ket{\psi}$, of the Hermitian operator, $\h$, with the lowest eigenvalue, $E_0$, by iteratively improving upon an ansatz.
The initial trial wave function and its first set of parameters form the ansatz, with the parameters typically chosen at random within a range expected to be reasonable in the context of the quantum system.

The selection of the ansatz form is a problem of its own which has many possible approaches depending on the Hamiltonian of the system, as explored in~\cite{tillyVariationalQuantumEigensolver2022,fedorovVQEMethodShort2022,anandQuantumComputingView2022} but is inspired by the context of the problem to be solved.
For example, in quantum chemistry, finding the ground state of a helium atom involves having an ansatz wave-function of the product of two hydrogen atom wave-functions, which is then improved upon from there~\cite{cruzeiroInteractivelyApplyingVariational2016}.
Apart from the ``problem-inspired ansatze'', another architecture is the so-called hardware-efficient ansatze~\cite{kandalaHardwareefficientVariationalQuantum2017}, which build arbitrary unitaries using single- and two-qubit gates that are native to the hardware in use.
It has the advantage of reducing circuit depth and being versatile enough to solve a wide range of problems as it is ``problem-agnostic''.
Overall, this ambiguity and freedom of choice in forming an ansatz for a variational problem allows for a smooth transition to quantum computers being used, as they can encode Hamiltonians as the sum of unitary operations.
This linear combination of unitary operators also allows for creating an ansatz wave function that can be easily parameterized through Bloch sphere rotation angles.

This principle translates very smoothly to quantum computers, as qubits are elementary and versatile manifestations of wave functions, which are measured to obtain the expectation value and energy of the system at the end of the quantum circuit after going through a set of parameterized quantum gates to alter the system and its wave-functions.
This idea's simplest and most direct implementation is the Variational Quantum Eigensolver (VQE).

\subsubsection{Variational Quantum Eigensolver (VQE)}
\label{sec:VQE}

Variational Quantum Eigensolver (VQE) is a quantum algorithm that employs a hybrid system, integrating both quantum and classical computing resources, to solve the eigenvalue problem for a given Hamiltonian operator. 
This technique was initially presented by \citet{peruzzoVariationalEigenvalueSolver2014} as an alternative to the quantum phase estimation algorithm by implementing a quantum chemistry problem on a hybrid system consisting of a photonic quantum processor and a conventional computer.
This work was improved, and its theoretical framework was reinforced and extended in a subsequent work by \citet{mccleanTheoryVariationalHybrid2016}.

The VQE algorithm, as all VQAs, operates via a parameterized quantum circuit, or ansatz, characterized by a set of parameters, $\bm{\theta}$.
A systematic method is required to vary the ansatz parameters until an optimal solution is found to implement the variational principle on a quantum computer.
The entire action of this variational operation can be represented by a unitary gate $U(\bm{\theta})$.
The ansatz acts on the initial state of the quantum circuit of $N$ qubits, $\ket{\psi_0}$, typically taken to be the ground state $\ket*{\vc{0}}$ (also expressed as $\ket{0}^{\otimes N}$), and generates an output $U(\bm{\theta})\ket{\psi_0} = \ket*{\psi(\bm{\theta})}$.

From this construction, it is clear that $U(\bm{\theta})\ket{\psi_0}$ is a normalized wave-function, which allows for the expression of the optimization problem as
\begin{equation}
    \lambda = \min_{\bm{\theta}} \expval{U^\dagger(\bm{\theta})\h U(\bm{\theta})}{\psi_0}.
\end{equation}
This is the state that is iteratively optimized by varying the parameters $\bm{\theta}$ towards an optimal set of parameters, $\bm{\theta}^*$, to obtain the optimized expectation value,
\begin{equation}
    \lambda_\text{min} = E_0 \approx \expval{\h}{\psi(\bm{\theta}^*)}.
\end{equation}
This algorithm can be easily scaled up to include more complexity, parameters, variational gates, and entanglement schemes~\cite{simExpressibilityEntanglingCapability2019}. 
These more expressive variational forms allow for better fine-tuning of the ansatz, resulting in a more accurate output state estimation of the eigenvalues.

The VQE algorithm, as a VQA, relies on both a quantum and classical part. The quantum part consists of estimating a quantum circuit, and it evaluates the desired quantum states for the given set of parameters.
Since it is not optimized to calculate the variations in the parameters towards an optimized set, this part is dealt with in the classical part of the algorithm. 
Indeed, the mathematics involved in optimization calculations is very efficient on modern classical computers.

Combining quantum and conventional computers to handle different components of a larger problem is known as hybrid quantum-classical computing and is the backbone of any VQA~\cite{mccleanTheoryVariationalHybrid2016}.

Moreover, it is a powerful framework for many use cases, especially with NISQ devices, as quantum computers are less efficient and fault-tolerant than their conventional counterparts with tasks such as optimizing parameters.
So the burden can be shared between the two to maximize overall performance.
This hybrid regime allows for the outsourcing of tweaking the parameters to a classical computer which then passes the values back to the quantum computer to calculate the eigenvalues.
The classical computer calculates the new parameters through optimization methods typically based on gradient descent~\cite{swekeStochasticGradientDescent2020,wierichsGeneralParametershiftRules2022}, which calculates a hyperplane of error or deviation from the ideal solution to find a minimum point that indicates the highest accuracy of the model.
An attempt to experimentally prove the efficiency of this hybrid method was carried out by \citet{otterbachUnsupervisedMachineLearning2017} by training a weighted MaxCut problem on 19 qubits.

However, it should be noted that hybrid computing does not always provide the most efficient solution, as some algorithms are more powerful when using only quantum hardware, as demonstrated by \citet{magannFeedbackBasedQuantumOptimization2022}.
\citet{kandalaHardwareefficientVariationalQuantum2017} have used a medium-sized quantum computer to optimize Hamiltonian problems with up to six qubits and over one hundred Pauli terms, determining ground-state energy for molecules up to BeH$_2$.
The approach used a variational quantum eigensolver, efficiently prepared trial states tailored to available quantum processor interactions, and a robust stochastic optimization routine.
Their results help elucidate requirements for scaling the method to larger systems and bridging gaps between high-performance computing problems and their implementation on quantum hardware.
Recent VQE variations have also been proposed to efficiently tackle combinatorial optimization problems,  similar to those targeted by the QAOA~\cite{amaroFilteringVariationalQuantum2022,otterbachUnsupervisedMachineLearning2017}.

\subsubsection{Quantum Adiabatic Algorithm (QAA)}
\label{sec:QAA}
As mentioned in Section~\ref{sec:qubo}, the adiabatic theorem is vital in solving optimization problems.
The adiabatic theorem states that starting in the ground state of a time-dependent Hamiltonian, if the Hamiltonian evolves slowly enough, the final state will be the ground state of the final Hamiltonian. 
Moreover, the adiabatic theorem can be generalized to any other eigenstate as long as there is no overlap (degeneracy) between different eigenstates across the evolution.
The $n$-th eigenstate of the initial Hamiltonian will evolve into the $n$-th eigenstate of the final one.
The Quantum Adiabatic Algorithm (QAA)~\cite{farhiQuantumComputationAdiabatic2000, farhiQuantumAdiabaticEvolution2001} was developed based on this principle to solve optimization problems on a quantum computer.
It also falls under a more general computational paradigm called Adiabatic Quantum Computing (AQC)~\cite{albashAdiabaticQuantumComputation2018}.
Specifically, we have an initial Hamiltonian $\h_M$, whose ground state is typically easy to prepare, and a final Hamiltonian $\h_C$, whose ground state encodes the solution to the optimization problem of interest.
The adiabatic evolution path is then encapsulated in the transitional Hamiltonian, which is expressed as $\h(t) = f(t)\h_C + g(t)\h_M$ with some slowly varying control functions such as $f(t) = t/T$ and $g(t) = 1 - t/T$, where $t\in[0, T]$ and $T$ is the total evolution time.
The evolution operator will then be $\U(t)\eqqcolon e^{-i\int^t_0 \mathrm{d}\tau\h(\tau)}$.
It is worth noting that although the QAA requires a continuous evolution of the state, it can be emulated on a gate-based quantum computer by Trotterizing $\U(t)$ in sufficiently small steps, namely, decomposing $\U(t)$ into a sequence of small steps through the Trotter-Suzuki formula:
\begin{equation}
\label{eq:trotter_qaa}
    \U(t) \approx \prod_{k=0}^{r-1}\exp \left[-i\h(k \Delta\tau) \Delta\tau\right] = \prod_{k=0}^{r-1}\exp\left[-i f(k \Delta\tau) \h_C \Delta\tau\right] \exp \left[-ig(k \Delta\tau)\h_M \Delta\tau\right]
\end{equation}
where $\Delta\tau\eqqcolon t / r$. 
Here, we notice that as $k$ increases, $f(k \Delta\tau)$ increases while $g(k \Delta\tau)$ decreases. 
Therefore, the time steps of $\h_C$ ($\h_M$) will decrease (increase) linearly.

 \citet{crossonDifferentStrategiesOptimization2014} conducted numerical simulations of the QAA for over 200,000 instances of MAX 2-SAT on 20 qubits with a unique optimal solution.
They selected a subset of instances for which the success probability was less than $10^{-4}$ at $T = 100$ and proposed three strategies to increase the success probability for all of these instances.
The first strategy was to run the adiabatic algorithm more rapidly, which increased the success probability at shorter times for all instances.
The second strategy was initializing the system in a random first excited state of the problem Hamiltonian, producing an average success probability close to the upper bound for most hard instances.
The third strategy involved adding a random local Hamiltonian to the middle of the adiabatic path, which often increased the success probability.
These strategies were also tested on the QAA version of the Grover search algorithm, but they did not improve the success probability. 
The authors concluded that their strategies might be helpful only for particularly challenging instances and could be tested on a quantum computer with higher qubit numbers.


\subsubsection{Barren plateaus}\label{sec:barren_plateaus}
In training parameterized quantum circuits, such as VQE and QAOA, there is a major challenge related to the cost function landscape. 
In fact, for certain families of parameterized quantum circuits, the landscape of the cost function can be flat, meaning that the gradients concerning the trainable parameters are exponentially small in the number of qubits, causing the optimization process to stall.
This is known as the problem of barren plateaus~\cite{mccleanBarrenPlateausQuantum2018}, i.e., the gradient of the cost function with respect to any parameter vanishes exponentially in the number of qubits.
Formally, a cost function $C(\boldsymbol{\theta})$ exhibits a barren plateau if, for all trainable parameters $\theta_i \in \boldsymbol{\theta}$, the variance of the partial derivative of the cost function vanishes exponentially in the number $n$ of qubits:
\begin{equation}
\label{eq:barren_plateaus}
    \text{Var}_{\boldsymbol{\theta}}[\partial_i C(\boldsymbol{\theta})] \leq F(n),
\end{equation}
with $F(n) \in O(b^{-n})$ for some constant $b > 1$. 
Eq.~\eqref{eq:barren_plateaus} implies that the gradient of the cost function will be, on average, exponentially small: due to Chebyshev’s inequality, the probability that the partial derivative $\partial_i C(\boldsymbol{\theta})$ deviates from its average (of zero) by a value larger than a given constant $c$, with $c > 0$, is bounded by $\text{Var}_{\boldsymbol{\theta}}[\partial_i C(\boldsymbol{\theta})]$, as illustrated in Eq.~\eqref{eq:barren_plateaus2}.
\begin{equation}
\label{eq:barren_plateaus2}
    \text{Pr}[\abs{\partial_i C(\boldsymbol{\theta})} \geq c] \leq \frac{1}{c^2} \text{Var}_{\boldsymbol{\theta}}[\partial_i C(\boldsymbol{\theta})]
\end{equation}
Quantum circuit training strategies are becoming increasingly crucial for all variational algorithms.

\subsection{The Quantum Approximate Optimization Algorithm (QAOA)\label{sec:qaoa_desc}} 
The QAOA was first introduced by \citet{farhiQuantumApproximateOptimization2014} as a VQA able to find approximate solutions to the MaxCut problem, suitable to be run on NISQ devices.
Inspired by the Trotterized version of QAA (Section~\ref{sec:QAA}), QAOA was designed to be a variational algorithm with repeated cost and mixer layers, namely these steps, instead of following some $f, g$ functions, are trained variationally.
Hence, it comes with a repeated cost and mixer layers, denoted $\U_C(\gamma_k)$ and $\U_M(\beta_k)$, respectively, where $k$ denotes the $k$-th layer. These layers are analogous to the exponentiated operators on the right-hand side of Eq.~\eqref{eq:trotter_qaa}. However, instead of following predefined $f$ and $g$ functions, the parameters $\gamma_k$ and $\beta_p$ are trained variationally.
In this sense, QAOA can be regarded as a discretized version of QAA and a special case of VQE (Section~\ref{sec:VQE}). 
The key idea behind QAOA is to encode the objective function of the optimization problem into the cost Hamiltonian $\h_C$ to search for an optimal bitstring $\mathbf{x^*}$ that will give a good approximation ratio $\alpha$ with a high probability.
In fact, the cost function $C(\mathbf{x})$ can be mapped to a cost Hamiltonian $\h_C$ such that
\begin{equation}
    \h_C \ket{\mathbf{x}} = C(\mathbf{x}) \ket{\mathbf{x}},
\end{equation}
where $\mathbf{x}$ is the quantum state encoding the bitstring $\bf x$.

\begin{figure}[!ht]\label{matrix_elements}
    \centering{\includegraphics[width=1\columnwidth]{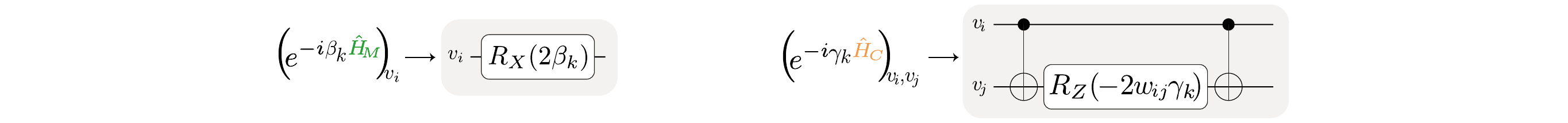}}
    \caption[Implementation of the elements of cost and mixing Hamiltonians]{Implementation of the elements of mixer (left) and cost (right) layers based on the cost and mixer Hamiltonians, $\h_C$ and $\h_M$. By $\big(e^{-i\beta_k \h_M}\big)_{v_i} \eqqcolon \big(\hat{U}_M(\beta_k)\big)_{v_i}$ we mean the element of $\hat{U}_M(\beta_k)$ generated by the vertex $v_i$, i.e., $e^{-i\beta_p (\h_M)_{v_i}}=e^{-i\beta_k X_i} = R_{X_i}(2\beta_k)$, where $\beta_k$ is the variational parameter for the layer $k$. Similarly for $\big(e^{-i\gamma_k \h_C}\big)_{v_i, v_j}$, the cost unitary is given by $e^{-i\gamma_k (\h_C)_{ij}} = R_{Z_iZ_j}(-2w_{ij}\gamma_k)$, which can be decomposed as shown on the right.}
    \label{fig:qaoa}
\end{figure}

The original QAOA consists of the following steps:
\begin{enumerate}
    \item Define a cost Hamiltonian $\h_C$ such that its highest energy state encodes the solution to the optimization problem. 
    Define also a mixer Hamiltonian $\h_M$ that does not commute with $\h_C$. 
    Typically, for the MaxCut problem of a graph $\G = (\V, \E)$, $\h_C$ and $\h_M$ are given as:
    \begin{subequations}
    \begin{align}
        \h_C &= \frac{1}{2} \sum_{(i,j) \in \E} w_{ij} (I - Z_iZ_j), \label{eq:h_c}  \\
        \h_M &= \sum_{j \in \V} X_j, \label{eq:h_m}
    \end{align}
    \end{subequations}
    where $I$ is the identity operator, $Z_j$ ($X_j$) is the Pauli-Z (-X) operator acting on the $j$-th qubit. 
    In the problem Hamiltonian $\h_C$, diagonal in the computational basis, each binary variable $x_i \in \{0,1\}$ in the MaxCut problem is mapped to a Pauli-Z operator $Z_i$ in the following way: 
    \begin{equation}
        x_i \rightarrow \frac{1}{2}(1 - Z_i).
    \end{equation}
    The Hamiltonian $\h_C$ in Eq.~\eqref{eq:h_c} corresponds precisely to the objective function in Eq.~\eqref{eq:objective}.

    \item Initialize the circuit in the state $\ket{s}$:
	\begin{equation}
	    \ket{s} = \ket{+}^{\otimes n} = \frac{1}{\sqrt{2^n}} \sum_{\vc x \in \{0,1\}^n} \ket{\vc x},
	\end{equation}
	where $n$ is the number of qubits and $n = \lvert \V \rvert$. 
    The state $\ket{s}$ corresponds to the highest energy state of the Pauli-X basis, i.e.,\ to the highest energy state of the mixer Hamiltonian $\h_M$.

    \item Construct the circuit ansatz by defining and applying the unitaries:
    \begin{subequations}
        \begin{align}
            \hat U_C(\gamma) &= e^{-i\gamma \h_C} = \prod_{i=1, j<i}^n R_{Z_iZ_j}(- 2w_{ij}\gamma), \\
            \hat U_M(\beta) &= e^{-i\beta \h_M} = \prod_{i=1}^n R_{X_i}(2\beta),
        \end{align}
    \end{subequations}
    where $\gamma$ and $\beta$ are variational parameters of the circuit. 
    We call $\U_C(\gamma)$ and $\U_M(\beta)$ the cost and mixer layers, respectively.
    A single QAOA layer comprises one cost and one mixer layer, which can be further stacked to build a deeper circuit with more layers.
    As shown in Figure~\ref{fig:qaoa}, each element in the mixer layer can be implemented with a single rotation gate $R_X$. In contrast, each of the two-qubit Pauli-Z interactions in the cost layer is implemented through two CNOT gates sandwiching a local rotation gate $R_Z$.

    \item Define the total number of QAOA layers, $p \geq 1$. 
    Initialize the $2p$ variational parameters $\bs{\gamma} = (\gamma_1,\gamma_2,\ldots,\gamma_p)$ and $\bs{\beta} = (\beta_1,\beta_2,\ldots,\beta_p)$ such that $\gamma_k \in [0,2\pi)$ and $\beta_k \in [0,\pi)$ for $k = 1, \dots, p$. 
    The final state output by the circuit is therefore given by
    \begin{equation}
        \ket{\psi_p(\bs{\gamma},\bs{\beta})} = e^{-i\beta_p \h_M} e^{-i\gamma_p \h_C} \ldots e^{-i\beta_1 \h_M} e^{-i\gamma_1 \h_C} \ket{s}.
    \end{equation}

    \item The expectation value of the Hamiltonian $\h_C$ with respect to the ansatz state $\ket{\psi_p(\bs{\gamma},\bs{\beta})}$, which corresponds to the cost obtained by the quantum algorithm for the underlying problem, is calculated through repeated measurements of the final state in the computational basis:
	\begin{equation}
	    F_p(\bs{\gamma},\bs{\beta}) = \expval{\h_C}{\psi_p(\bs{\gamma},\bs{\beta})}
	\end{equation}

    \item A classical optimization algorithm is employed to iteratively update the parameters $\bs{\gamma}$ and $\bs{\beta}$. The goal of the aforementioned routine is to find the optimal set of parameters $(\bs{\gamma}^*,\bs{\beta}^*)$ such that the expectation value $F_p(\bs{\gamma},\bs{\beta})$ is maximized:
	\begin{equation}
	    (\bs{\gamma}^*,\bs{\beta}^*) = \arg\max_{\bs{\gamma},\bs{\beta}} F_p(\bs{\gamma},\bs{\beta})
	\end{equation}
\end{enumerate}

\begin{figure}[!ht]
    \centering{\includegraphics[width=1.\columnwidth]{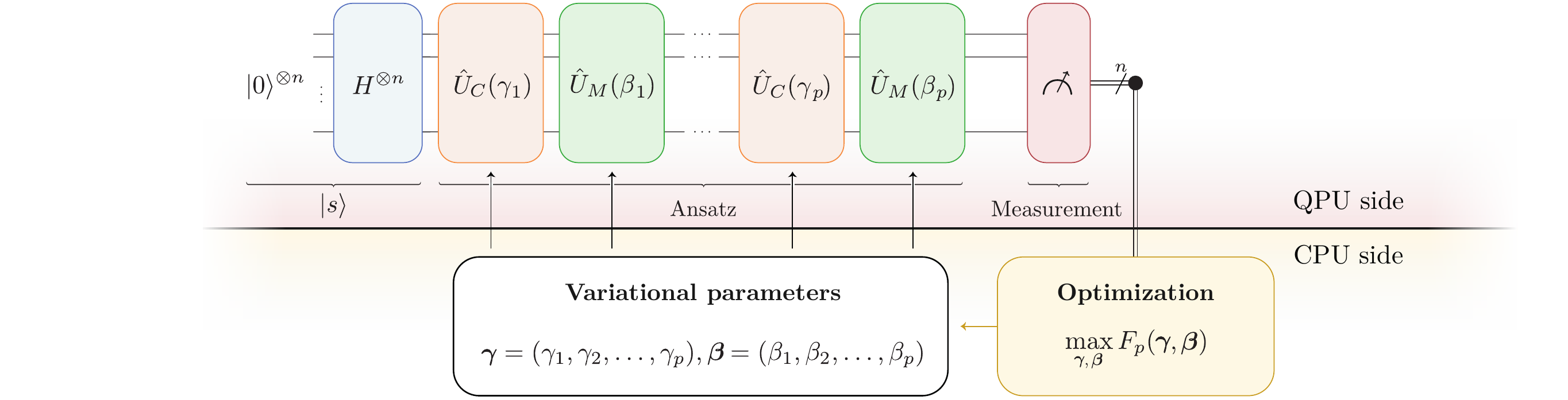}}
    \caption[Scheme of the hybrid workflow of QAOA]{Schematic of the hybrid workflow of QAOA with $p$ layers.}%
    \label{fig:qaoascheme}
\end{figure}

An example of a QAOA circuit instance is shown in Figure~\ref{fig:qaoascheme}. 
At the end of the optimization procedure, the approximation ratio $\alpha$ will be given by:
\begin{equation}
    \alpha = \frac{F_p(\bs{\gamma}^*,\bs{\beta}^*)}{C_\text{max}},
\end{equation}
and the state $\ket{\psi_p(\bs{\gamma}^*,\bs{\beta}^*)}$ will encode the solution to the optimization problem.


It is worth noting that, instead of the ground state, QAOA is typically initialized with the highest energy eigenstate of the mixer Hamiltonian, such as in the case of MaxCut.
Nevertheless, the adiabatic theorem still holds in such cases.
Apart from the digitized adiabatic quantum computing approach, which the standard QAOA implementation (Section~\ref{sec:qaoa_desc}) is based on, an analog version of QAOA that can be run on quantum annealers was recently proposed by \citet{barrazaAnalogQuantumApproximate2022}.

    \section{QAOA Analysis}%
    \label{sec:analysis}
    \begin{figure}[ht!]
    \centering{\includegraphics[width=1.\columnwidth]{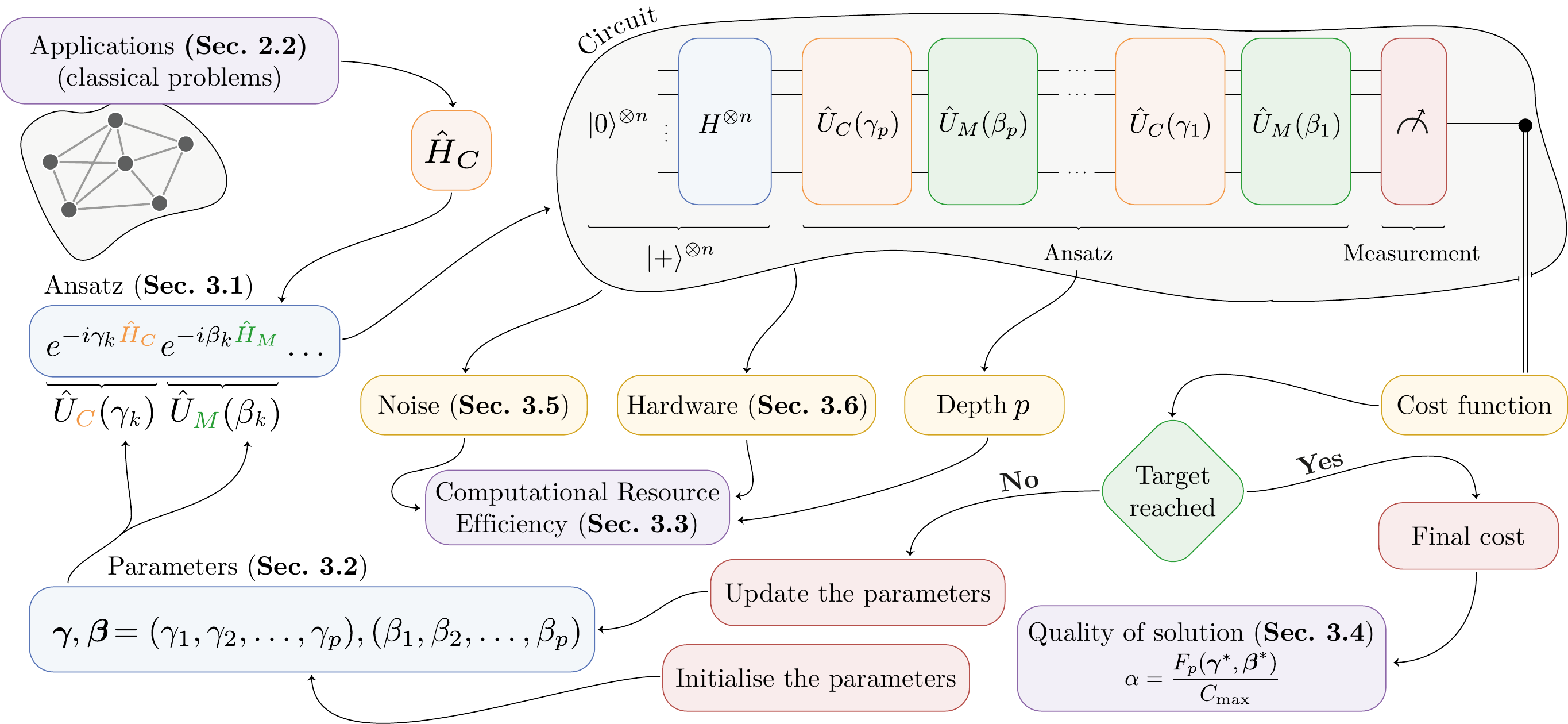}}
    \caption[General scheme of QAOA and its features]{General scheme of QAOA and its features.}%
    \label{fig:general_scheme}
\end{figure}

The QAOA has generated significant interest in the quantum computing community as a promising method for solving combinatorial optimization problems. 
This section comprehensively analyzes QAOA and its various associated aspects (Figure~\ref{fig:general_scheme}).
Our analysis covers a range of topics, including
ansatz variants (Section~\ref{sec:analysis_an}), parameter optimization strategies (Section~\ref{sec:analysis_parameter_optimization}), computational resource efficiency (Section~\ref{sec:analysis_performance}), quality of solution (Section~\ref{sec:efficiency}), noise and error considerations (Section~\ref{sec:noiseanderrors}), and hardware-specific approaches (Section~\ref{sec:hardware_specific_suggestions}).
We evaluate the strengths and limitations of QAOA in light of recent advancements and studies in the literature.
Our analysis aims to illuminate the capabilities and potential of QAOA and emphasize the challenges that must be addressed to harness its power in practical applications fully.


    \subsection{Ansatz Variants}
    \label{sec:analysis_an}
    
An ansatz is an educated guess about the form of an unknown function that is made in order to facilitate the solution of an equation or other problem.
In the context of QAOA, the ansatz that needs to be made is about the structure of the quantum circuit, which defines the operators
\begin{equation}
    \U(\bs\gamma, \bs\beta) = e^{-i\beta_p\h_M}e^{-i\gamma_p\h_C}\cdots e^{-i\beta_1\h_M}e^{-i\gamma_1\h_C},
\end{equation}
where such operators are layered $p$ times.
The choice of ansatz typically depends on the problem type, such as combinatorial problems represented as graphs~\cite{liImplementingGraphTheoreticFeature2022}, or problems strongly influenced by hardware design~\cite{proiettiNativeMeasurementbasedQuantum2022,rabinovichIonNativeVariational2022,rajakumarGeneratingTargetGraph2022}. 
However, ansatz design must balance specificity and generality to avoid overfitting and maintain applicability to a wide range of problems.
For this reason, designing optimal ansatze for QAOA is an extensively researched and widely investigated topic.
This section introduces and discusses various prominent designs and variations of the QAOA ansatz. These variations encompass several approaches to constructing the optimization methodology and address many of the shortcomings of the original algorithm. These variations are summarized in Table \ref{tab:ansatz_variations}.

{
\renewcommand{\arraystretch}{2.5}
\begin{table}[htp!]
    \centering
    \footnotesize
    \begin{tabular}{>{\raggedright}p{0.15\linewidth}p{0.4\linewidth}p{0.35\linewidth}}
	\toprule
	\textbf{Ansatz} & \textbf{Main Idea} & \textbf{Enhancement \& Applications}\\
	\midrule
	ma-QAOA~\cite{herrmanMultiangleQuantumApproximate2022} & Multi-angle ansatz with a unique parameter for each element of cost and mixer Hamiltonians & Improves approximation ratio for MaxCut while reducing circuit depth \\
	QAOA+~\cite{chalupnikAugmentingQAOAAnsatz2022} & Augments traditional QAOA with an additional multi-parameter problem-independent layer & Higher approximation ratios for MaxCut on random regular graphs \\
    DC-QAOA~\cite{chandaranaDigitizedcounterdiabaticQuantumApproximate2021,wurtzCounterdiabaticityQuantumApproximate2022} & Adds a problem-dependent counterdiabatic driving term to the QAOA ansatz & Improves the convergence rate of the approximation ratio while reducing circuit depth \\
	ab-QAOA~\cite{yuQuantumApproximateOptimization2022} & Incorporates local fields into the operators to reduce computation time & Computation time reduction for combinatorial optimization \\
	ADAPT-QAOA~\cite{zhuAdaptiveQuantumApproximate2020} & Iterative version of QAOA with systematic selection of mixers based on gradient criterion & Can be problem-specific and addresses hardware constraints \\
	Recursive QAOA~\cite{bravyiObstaclesVariationalQuantum2020} & Non-local variant of QAOA that iteratively reduces problem size by eliminating qubits & Overcomes locality constraints and achieves better performance \\
	QAOAnsatz~\cite{hadfieldQuantumApproximateOptimization2019} & Extends the original formulation with broader families of operators and allows for encoding of constraints & Adaptable to a wider range of optimization problems with hard and soft constraints  \\
	GM-QAOA~\cite{bartschiGroverMixersQAOA2020} & Uses Grover-like selective phase shift mixing operators & Solves $k$-Vertex Cover, Traveling Salesperson Problem, Discrete Portfolio Rebalancing \\
	Th-QAOA~\cite{goldenThresholdBasedQuantumOptimization2021} & Replaces standard phase separator with a threshold function & Solves MaxCut, Max $k$-Vertex Cover, Max Bisection \\
	Constraint Preserving Mixers~\cite{fuchsConstraintPreservingMixers2022} & Constructs mixers that enforce hard constraints & Solves optimization problems with hard constraints \\

    WS-QAOA~\cite{eggerWarmstartingQuantumOptimization2021} & Modifies the initial state and mixer Hamiltonian based on the optimal solution to the relaxed QUBO problem & Solutions guaranteed to retain the GW bound for the MaxCut problem \\
	FALQON~\cite{magannFeedbackBasedQuantumOptimization2022} & Uses qubit measurements for feedback-based quantum optimization, avoiding classical optimizers & Produces monotonically improving approximate solutions as circuit depth grows while bypassing classical optimization loops \\
	FALQON+~\cite{magannLyapunovcontrolinspiredStrategiesQuantum2022} & Combines FALQON's initialization with QAOA for better parameter initialization & Improves initialization of standard QAOA for non-isomorphic graphs with 8 to 14 vertices \\
	FQAOA~\cite{yoshiokaFermionicQuantumApproximate2023} & Utilizes fermion particle number preservation to intrinsically impose constraints in QAOA process & Improves performance in portfolio optimization, applicable to Grover adaptive search and quantum phase estimation \\
	Quantum Dropout~\cite{wangQuantumDropoutEfficient2022}	& Selectively drops out clauses defining the quantum circuit while keeping the cost function intact  & Improves QAOA performance on hard cases of combinatorial optimization problems \\
	ST-QAOA~\cite{wurtzClassicallyOptimalVariational2021} & Uses an approximate classical solution to construct a problem instance-specific circuit & Achieves same performance guarantee as the classical algorithm, outperforms QAOA at low depths for MaxCut problem \\
    Modified QAOA~\cite{villalba-diezImprovementQuantumApproximate2021} & Modifies cost Hamiltonian with conditional rotations & Improves approximation ratio for MaxCut at $p=1$\\
	\bottomrule
    \end{tabular}
    \caption{Summary of ansatz strategies for improving QAOA.}
    \label{tab:ansatz_variations}
\end{table}
}

\subsubsection{Multi-Angle QAOA} \label{subsec:ma-qaoa}
A straightforward approach to enhanced ansatz design was introduced by \citet{herrmanMultiangleQuantumApproximate2022}.
The authors proposed a multi-angle ansatz for QAOA (ma-QAOA), which improves the approximation ratio by increasing the number of variational parameters.
In ma-QAOA, new parameters are introduced into the circuit so that each element of the cost and mixer layers has its angle instead of one angle for the cost operator and one for the mixer operator, as follows: 
%
\begin{subequations}
\begin{align}
    \U_C(\bs\gamma_l) &= e^{-i\sum_{a=1}^m\gamma_{l,a}\h_{C,a}} = \prod_{a=1}^m e^{-i\gamma_{l,a}\h_{C,a}} \label{eq:ma-qaoa1}  \\
    \U_M(\bs\beta_l) &= e^{-i\sum_{v=1}^n\beta_{l,v}\h_{M,v}} = \prod_{v=1}^n e^{-i\beta_{l,v}\h_{M,v}}, \label{eq:ma-qaoa2}
\end{align}
\end{subequations}
where $\bs\gamma_l = (\gamma_{l,1}, \gamma_{l,2}, ..., \gamma_{l,m})$, $\bs\beta_l = (\beta_{l,1}, \beta_{l,2}, ..., \beta_{l,n})$, $l$ is the QAOA layer, $n$ is the number of qubits (nodes) and $m$ is the number of clauses (edges) of the problem (graph).
The authors referred to the matrices $\h_C$ and $\h_M$ as $C$ and $B$, respectively.
The total number of circuit parameters becomes $(n + m)p$, where $p$ is the number of QAOA layers.
The vanilla QAOA can be treated as a special case of ma-QAOA, where all the parameters of a given cost or mixer layer have the same value.
As such, ma-QAOA was revealed to be more potent than the standard algorithm: the value of the approximation ratio $\alpha$ achieved by the ma-QAOA is better or equal to that of the vanilla QAOA\@.
Despite a more complex parameter optimization, empirical results suggest that ma-QAOA may require shallower circuits.
In this regard, a follow-up study by \citet{shiMultiAngleQAOADoes2022} proposed to reduce the number of ma-QAOA parameters by exploiting the natural symmetries of the input graphs.
This approach reduced approximately 33\% of the parameters while having little to no impact on the objective function.
Moreover, a connection between ma-QAOA and Continuous-Time Quantum Walks (CTQW) on dynamic graphs was investigated by  \citet{herrmanRelatingMultiangleQuantum2022}, who showed that ma-QAOA is equivalent to a restriction of CTQW on dynamic graphs. 
A possible advantage of relating ma-QAOA to CTQW on dynamic graphs is that well-studied CTQW phenomena, such as hitting times, might be investigated to improve our understanding of ma-QAOA and help find optimal parameters.


\subsubsection{QAOA+} \label{subsec:qaoap}
Another example of an improvement in the ansatz is the work of \citet{chalupnikAugmentingQAOAAnsatz2022}.
To address the problem of the originally proposed form of the QAOA ansatz not providing sufficient performance advantage over classical counterparts in problems such as MaxCut~\cite{hastingsClassicalQuantumBounded2019}, the authors proposed an alternative ansatz, which they call QAOA+.
This variant augments the traditional $p = 1$ QAOA ansatz with an additional multi-parameter problem-independent layer of parameterized $ZZ$ gates and a layer of mixer $X$ gates.
The QAOA+ ansatz allows one to obtain higher approximation ratios than $p = 1$ QAOA while keeping the circuit depth below that of $p = 2$ QAOA with comparable performance, as benchmarked on the MaxCut problem for random regular graphs.
Moreover, it showed a similar level of performance to $p = 2$ QAOA.
The added circuit depth beyond the vanilla QAOA grows only in the number of qubits used as a set of $2N-1$ parameters for $N$ qubits. 
They additionally showed that the proposed QAOA+ ansatz, while using a more significant number of trainable classical parameters than the standard QAOA, in most cases outperforms the alternative multi-angle QAOA ansatz in~\cite{herrmanMultiangleQuantumApproximate2022}.

\subsubsection{Digitized counterdiabatic QAOA}\label{subsec:dc-qaoa}

In pursuit of reducing computational complexity, and thereby circuit depth, of QAOA, \citet{chandaranaDigitizedcounterdiabaticQuantumApproximate2021} propose a new variant of the algorithm coined Digitized Counterdiabatic QAOA (DC-QAOA). This QAOA variant is built on the adiabatic evolution of the Hamiltonians in vanilla QAOA resulting in unnecessary computational cost and circuit depth, which is difficult to implement on near-term devices. This method utilizes counterdiabatic (CD) driving to speed up the optimization process of the variational algorithm. This is achieved through the extension of the time evolution operator to include an additional variational parameter,
\begin{equation}
U(\gamma, \beta) \to U(\gamma, \beta, \alpha),
\end{equation}
that represents a CD operator. This operator is represented as
\begin{equation}
U_\mathrm{CD}(\alpha) = \prod_{J=1}^L \exp(-i\alpha A_t^q),
\end{equation}
where $A_t^q$ is the respective $q$-local CD operator chosen from the CD pool $A$. This pool of operators is defined through the nested commutator approach of the adiabatic gauge potential \cite{claeysFloquetEngineeringCounterdiabaticProtocols2019} as
\begin{equation}
A_\lambda^{(l)} = i\sum_{k=1}^l \alpha_k(t)\underbrace{{[H_a, [H_a, \ldots [H_a,}}_{2k-1} \partial_\lambda H_a]]].
\end{equation}
The authors applied this QAOA variant to problems such as Ising models, classical optimization problems, and the $k$-spin model, demonstrating that it outperforms the standard QAOA in all cases.

\citet{wurtzCounterdiabaticityQuantumApproximate2022} propose a similar algorithm, CD-QAOA, also inspired by the use of counterdiabaticity to accelerate the convergence of QAOA to minimize circuit depth and improve solution quality.

\subsubsection{Adaptive bias QAOA} \label{subsec:ab-qaoa}
Inspired by the previous work that introduced bias fields in quantum annealing~\cite{grassQuantumAnnealingLongitudinal2019}, \citet{yuQuantumApproximateOptimization2022} proposed a modified version of QAOA called the adaptive bias QAOA (ab-QAOA), which incorporates the adaptive bias fields into the mixer operators of QAOA to accelerate the convergence of the algorithm.
Essentially, in this approach, $n$ additional parameters $\{h_j\}$ that comprise the bias fields are introduced in the $n$-qubit QAOA circuit, which enter both the modified mixer Hamiltonian
\begin{equation}
    \hat H_M^\text{ab}(\{h_j\}) = \sum_{j\in \V}\qty(X_j - h_j Z_j),
\end{equation}
and the initial state that is the product ground state of $\hat H_M^\text{ab}(\{h_j\})$.
These local fields are not optimized but rather updated according to the following prescription,
\begin{equation}
    h_j \to h_j - \ell \left(h_j - \ev{Z_j}{\psi_p^\text{ab}}\right),
\end{equation}
where $\ell$ is the learning rate and $\ket{\psi_p^\text{ab}}$ is the state output by the level-$p$ ab-QAOA circuit, that is,
\begin{equation}
    \ket{\psi_p^\text{ab}} = \prod_{k=1}^j e^{-i\beta_k\hat H_M^\text{ab}(\{h_j\})}e^{-i\gamma_k \hat H_C}\ket{\psi_0^\text{ab}(\{h_j\})}.
\end{equation}
The method was shown to substantially reduce the computation time of QAOA for a fixed level of accuracy and the same number of gates.
The computation time of the ab-QAOA converging to a desired accuracy was polynomially shorter than that of the vanilla QAOA. 
Moreover, such improvement further increases with the problem size, paving the way for the quantum advantage of QAOA in combinatorial optimization problems.

\subsubsection{ADAPT-QAOA} \label{subsec:adapt-qaoa}
\citet{zhuAdaptiveQuantumApproximate2020} addressed the ansatz selection problem by proposing an iterative version of QAOA called the Adaptive Derivative Assembled Problem Tailored-QAOA (ADAPT-QAOA).
Instead of the standard mixer Hamiltonian, the ADAPT-QAOA
systematically selects the QAOA mixer from a pre-defined pool of operators $\hat{A}_k$ that changes from one layer to the next:
\begin{equation}
\label{eq:adapt}
\ket{\psi_p(\bs{\gamma},\bs{\beta})} = \left( \prod_{k=1}^{p} e^{-i\beta_k \hat{A}_k} e^{-i\gamma_k \h_C} \right) \ket{s}.
\end{equation}
In each step, the operator $\hat{A}_k$ is selected by maximizing the gradient of the commutator of the pool operator and the cost Hamiltonian over the ansatz of the previous step, namely maximizing the gradient of
\begin{equation*}
    -i\bra{\psi_{k-1}(\bs{\gamma},\bs{\beta})} e^{i\h_C\gamma_k}[\h_C, \hat{A}_k]e^{-i\h_C\gamma_k}\ket{\psi_{k-1}(\bs{\gamma},\bs{\beta})},
\end{equation*}
where $\gamma_k$ is initialized to a certain value $\gamma_0$. 
Once $\hat{A}_k$ is selected, all parameters are optimized again, and if the cost function has not reached a target value, a new layer can be added similarly.
In simulations on the MaxCut problems, ADAPT-QAOA converged faster than the standard QAOA while reducing the number of CNOT gates and optimization parameters by about 50\% each, particularly when entangling gates were included in the operator pool.
Such a speedup is attributed to the concept of shortcuts to adiabaticity~\cite{guery-odelinShortcutsAdiabaticityConcepts2019,chaiShortcutsQuantumApproximate2022}.
This concept has been crucial in the enhancement of many ansatz designs which provide improved variations of QAOA, such as digitized and counterdiabatic frameworks~\cite{headleyApproximatingQuantumApproximate2022,chandaranaDigitizedcounterdiabaticQuantumApproximate2021,wurtzCounterdiabaticityQuantumApproximate2022}. 
However, the drawback of the method is that the selection of the mixing operators requires an additional number of measurements that depends on how big the operator pool is.

\subsubsection{Recursive QAOA} \label{subsec:rqaoa}
Performance of QAOA can be limited by the $Z_2$ symmetry of the QAOA states and the geometric locality of the ansatz; that is, the cost operators include interactions only between nearest neighbor qubits concerning the underlying graph.
To address this, \citet{bravyiObstaclesVariationalQuantum2020} proposed the Recursive QAOA (RQAOA) as a non-local variant of QAOA that iteratively reduces the size of the problem.
At each step, RQAOA uses the output distribution of QAOA to compute the $ZZ$-correlations of all pairs of edges in the graph, i.e., $M_{ij} = \expval{Z_iZ_j}{\psi(\bs{\gamma}, \bs{\beta})}, \ \forall (i,j) \in \E$. 
Then, it selects the pair(s) with the largest magnitude of the correlation and imposes a parity constraint, 
\begin{equation}
        Z_j = \text{sgn}(M_{ij})Z_i.
\end{equation}
This effectively eliminates one or more qubits from the Hamiltonian by imposing a constraint on them. 
RQAOA then reruns the QAOA circuit on the reduced Hamiltonian and repeats the process until the problem reaches a predefined cutoff size. 
At this point, RQAOA solves the remaining problem exactly using classical methods and reconstructs the final solution by reinserting the eliminated qubits.
While RQAOA is less studied compared with the vanilla version, research interest is increasing as it emerges as a promising QAOA variant on NISQ devices~\cite{patelReinforcementLearningAssisted2022,bravyiHybridQuantumclassicalAlgorithms2022,otterbachUnsupervisedMachineLearning2017}. 
For example, \citet{baeRecursiveQAOAOutperforms2023} compared the performance of the level-1 QAOA with that of the RQAOA applied to the MaxCut problem on complete graphs with $2n$ vertices.
They analytically demonstrated that in this particular scenario, the level-1 RQAOA achieves the approximation ratio 1, while the approximation ratio of the original QAOA at $p=1$ is strictly upper bounded as
\begin{equation*}
\alpha\leq 1-\frac{1}{8n^2}.
\end{equation*}

\subsubsection{Quantum alternating operator ansatzes}\label{subsec:qaltop}
The process of ansatz design in QAOA is very versatile and can be extended to much more far-reaching constructions.
One such extension is the remodeling of QAOA to the Quantum Alternating Operator Ansatz (QAOAnsatz) by \citet{hadfieldQuantumApproximateOptimization2019}.
As introduced in Section~\ref{sec:qaoa_desc}, in QAOA, the ansatz structure alternates between applying unitaries based on the cost and mixer Hamiltonians.
The extended framework allows for alternating between a more general set of operators.
This extension is based on considering general parameterized families of unitary operators instead of only those corresponding to time evolution under a fixed local Hamiltonian.

This altered ansatz structure allows for a broader set of problems to be solved by this family of algorithms, especially in optimization problems that have hard constraints that always need to be satisfied, defining feasible subspaces, and soft constraints which need to be minimized in their violations~\cite{hadfieldQuantumApproximateOptimization2019}.
This ansatz supports representing a larger, potentially more useful, set of states than the original formulation, with potential long-term impact on a broad array of application areas~\cite{hadfieldAnalyticalFrameworkQuantum2022}.

\begin{figure}[!ht]
    \centering{\includegraphics[width=1\columnwidth]{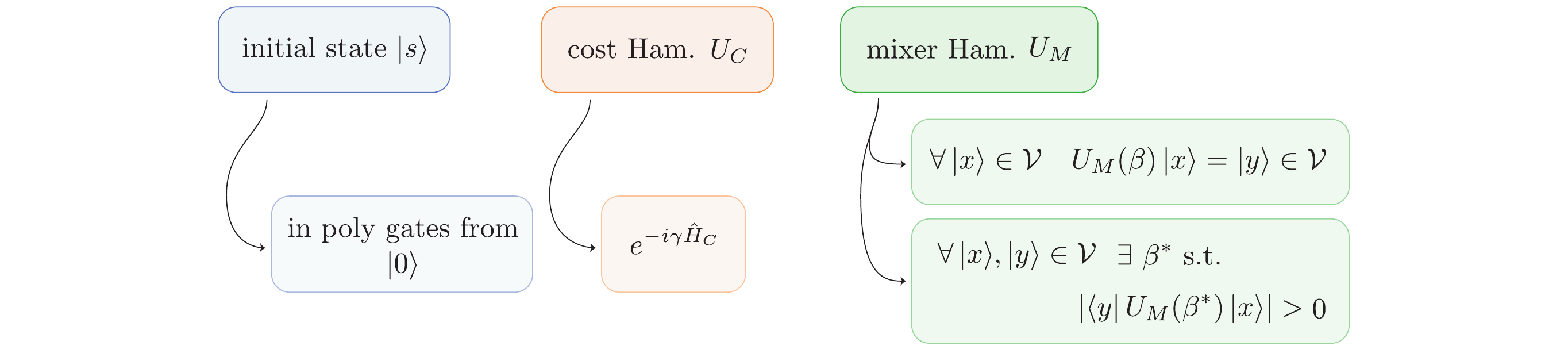}}
    \caption[Representation of the QAOAnsatz]{Representation of the QAOAnsatz.}%
    \label{fig:QAOAnsatz}
\end{figure}
A novelty of the QAOAnsatzes is their different approach in encoding the constraints of a graph into the ansatz.
While the standard approach is to add ``penalties'' to the cost Hamiltonian, \citet{hadfieldAnalyticalFrameworkQuantum2022} proposed to modify the mixer Hamiltonian into XY-mixer, partial mixer, and others to encode different constraints.
It is worth noting that a recent effort to unify these two different approaches into a single method was carried out by \citet{ruanQuantumApproximateOptimization2023} in the so-called Unified Quantum Alternating Operator Ansatz (UQAOA).

As depicted in Figure~\ref{fig:QAOAnsatz}, QAOAnsatz proposes an alternative and more generic way of defining the different parts of the ansatz.
A QAOAnsatz circuit is characterized by two families of parameterized operators in a Hilbert space $\mathcal{H}$: one of phase-separation operators, $\hat{U}_P(\gamma)$, that depends on the objective function $f$; and one of mixing operators, $\hat{U}_M(\beta)$, which depends on the domain and its structure, where $\beta$ and $\gamma$ are real parameters.
A depth-$p$ circuit consists of $p$ alternating applications of operators from these two families, i.e.,
\begin{equation}
    Q_p(\bs\gamma, \bs\beta) = \hat{U}_M(\beta_p)\hat{U}_P(\gamma_p)\cdots \hat{U}_M(\beta_1)\hat{U}_P(\gamma_1).
\end{equation}
QAOAnsatz consists of states representable by the application of this type of circuit to a suitable initial state $\ket{s}$:
\begin{equation}
    \ket{\psi(\bs\gamma, \bs\beta)} = Q_p(\bs\gamma, \bs\beta) \ket{s}.
\end{equation}
It provides a framework that allows almost any combinatorial optimization problem to be modeled as a QAOA problem, with the objective functions encoded as the sum of Pauli operator-based Hamiltonians, forming the phase separator and mixing unitary operators.
Many combinations and variations of these mixer unitaries have been analyzed in recent literature.
For example, the original formulation of QAOA~\cite{farhiQuantumApproximateOptimization2014,farhiQuantumApproximateOptimization2015} made use of transverse field based $X$-mixers for unconstrained problems, while the XY-model Ring and Clique Mixers are effective for Hamming weight constrained problems~\cite{cookQuantumAlternatingOperator2020,wangMixersAnalyticalNumerical2020}. 
\citet{cookQuantumAlternatingOperator2020} compared the efficacy of classical states, and Dicke states as initial states, assessed the impact of two distinct XY-Hamiltonian mixing operators, and conducted an analysis of solution distributions via Monte Carlo sampling.
Their findings indicate that Dicke states enhance performance compared to easily-prepared classical states. 
They suggest that the complete graph mixer outperforms the ring mixer, though with a trade-off between improved performance and more extended circuit depth. 
An intriguing aspect of their study was that the standard deviation of solution distributions decreases exponentially with the number of rounds, which has implications for the feasibility of finding better solutions in deeper algorithm rounds. 
Despite this, they found that high-quality solutions share patterns with a discretized version of the quantum adiabatic algorithm, suggesting potential avenues for efficient angle selection strategies.

Below we introduce a few notable QAOAnsatz variants with different mixer/phase separator designs.

\paragraph{Grover Mixer QAOA:}
For both constrained and unconstrained optimization problems, the use of Grover-like selective phase-shifting operators has shown to be effective.
One such example is the work of \citet{bartschiGroverMixersQAOA2020}, in which the Grover Mixer QAOA (GM-QAOA) variant is proposed.
This variation to QAOAnsatz makes use of the Grover-like selective phase shift mixing operators, inspired by the Grover Search quantum algorithm~\cite{groverFastQuantumMechanical1996,groverFixedPointQuantumSearch2005}.
This variant works for any NP optimization application, which can be efficiently prepared with an equal superposition of all feasible solutions.
This design works incredibly well for constraint optimization problems where not all possible variables are feasible solutions, such as $k$-Vertex-Cover.

The GM-QAOA variant has significant benefits over the original algorithm, such as not being susceptible to Trotterization errors or any other Hamiltonian simulation errors, due to its operators being implemented exactly using only standard gate sets.
The design of the variant also allows solutions sharing an objective value to be sampled with the same amplitude, which significantly increases efficiency and stability.

The authors demonstrate the prowess of this framework on a set of significant optimization problems.
One such problem is the Traveling Salesman Problem, in which an efficient algorithm is presented to prepare a superposition of all possible permutations of $n$ numbers over $O(n^2)$ qubits.
Another problem to which GM-QAOA is applied in this demonstration is the hard constraint $k$-Vertex-Cover problem, as a standard benchmark against other combinatorial optimization problems.
Finally, in the problem of discrete portfolio rebalancing, the application of GM-QAOA is demonstrated to outperform other existing QAOA approaches, as it can restrict mixing to the feasible subspace and provide transitions between all feasible states.

\paragraph{Threshold QAOA:}
Another variation within the alternating operator ansatz framework is provided by \citet{goldenThresholdBasedQuantumOptimization2021}, based on the above discussion on various QAOA mixers.
The authors presented a variation of QAOA in which the standard phase separation operator is replaced by a threshold function, returning a value of 1 for solutions with an objective value above a defined threshold and a value of 0 otherwise.
This variation is coined Threshold QAOA (Th-QAOA), in which the threshold value is varied to reach a quantum optimization algorithm.
Although this algorithm is versatile enough to be constructed using any previously studied QAOA mixers, in this work, the authors focused on the combination with the Grover mixers, which have shown to be effective for solving constrained~\cite{akshayReachabilityDeficitsQuantum2020} and unconstrained~\cite{bartschiGroverMixersQAOA2020} optimization problems.
This application of Th-QAOA combined with the Grover mixer is called GM-Th-QAOA\@.
It was demonstrated that the algorithm could be classically simulated up to 100 qubits with relative ease because of memory optimization techniques applicable to implementing the algorithm.
The authors found that this variation of QAOA outperforms other variants in terms of the approximation ratio for a range of optimization problems, including MaxCut, Max-$k$-Vertex-Cover, and Max Bisection.

\paragraph{Constraint Preserving Mixers:}
A framework for constructing mixing operators that enforce hard constraints in quantum optimization problems was introduced by \citet{fuchsConstraintPreservingMixers2022}.
The authors generalized the XY-mixer, designed to preserve the subspace of ``one-hot'' states, to the case of subspaces given by a number of computational basis states.
The underlying mathematical structure is exposed to minimize the cost of the mixer in terms of CNOT gates, mainly when Trotterization is considered.
This work also introduces efficient decomposition algorithms for basis gates and analyzes several examples of more general cases.
\citet{goviaFreedomMixerRotation2021} proposed the free axis mixer quantum alternating operator ansatz called Free Axis Mixer-QAOA (FAM-QAOA), which adds additional variational parameters in the XY-plane of the mixer Hamiltonian.
They explore the Hilbert space expansion and the Z-phase error mitigation, showing that the ansatz outperforms the standard QAOA, especially at low depths.

\subsubsection{Warm-starting QAOA}
Inspired by the recent progress in the study of the continuous relaxations of NP-hard combinatorial optimization problems, \citet{eggerWarmstartingQuantumOptimization2021} proposed to ``warm-start'' the QAOA parameters based on the solution to the relaxed QUBO problem, i.e., one with continuous variables instead of binary ones. 
They considered two types of continuous-value relaxations, quadratic programming (QP), which corresponds to the cases of QUBO where the matrix $\vb Q$ in Eq.~\eqref{eq:qubocost} is positive semidefinite, and semidefinite programming (SDP) when $\vb Q$ is not positive semidefinite. 
In the simplest variant of Warm Starting QAOA (WS-QAOA), one replaces the initial $n$-qubit equal-superposition state $\ket{+}^{\otimes n}$ with the state
\begin{equation}
    \ket{s^*}  = \bigotimes_{i=0}^{n-1} \hat R_Y(\theta_i) \ket{0}^{\otimes n},
\end{equation}
where $\theta_i = 2\arcsin(\sqrt{c_i^*})$, with $c_i \in [0,1]$ is the $i$-th coordinate of the optimal solution to the continuous-valued relaxation QP. 
Moreover, the mixer Hamiltonian is modified accordingly to $\hat H_M^{\text{(ws)}} = \sum_{i=0}^{n-1} \hat H_{M,i}^{\text{(ws)}}$, where
\begin{equation}
    \hat H_{M,i}^{\text{(ws)}} = -\sin(\theta_i)\hat X - \cos(\theta_i)\hat Z = \mqty(2c_i^* - 1 & -2\sqrt{c_i^*(1 - c_i^*)} \\ -2\sqrt{c_i^*(1 - c_i^*)} & 1 - 2c_i^*).
\end{equation}
Therefore, the action of the mixer unitary $\hat U_M^\text{(ws)} = e^{-i\beta \hat H_M^\text{(ws)}}$ on qubit $i$ can be implemented via single-qubit rotations $\hat R_Y(\theta_i) \hat R_Z (-2\beta) \hat R_Y (-\theta_i)$.

In another variant of WS-QAOA, the optimum of the continuous-valued relaxation is randomly rounded before using it as the initial state. 
A notable example is the Goemans-Williamson random-hyperplane rounding of SDP relaxations for the MaxCut problem.
The most significant advantage of this approach is that it provides initial states that already have the best approximation guarantee available classically in polynomial time and can retain the GW bound at any number of layers $p$.
Therefore, the solution output by WS-QAOA is at least as good as that given by GW rounding.
Furthermore, WS-QAOA can be readily incorporated into the workflow of recursive QAOA, and this variant (WS-RQAOA) was shown in numerical simulations to give the best MaxCut results for both random and full-connected graphs with 20 and 30 nodes.
The WS-QAOA approach thus provides an advantage over the standard QAOA at low depth, which is particularly important for implementing on the NISQ devices.

\subsubsection{FALQON}
In the literature discussed thus far, the QAOA and its variants have primarily been hybrid algorithms, using quantum computation alongside classical optimization of variational parameters.
This proves beneficial in the NISQ era of quantum computers, as they are not fault-tolerant or stable enough to handle an entire optimization problem independently and do not need to.
However, this may hinder performance in future quantum computers, as there will be an efficiency bottleneck in passing information back and forth between quantum and classical computers.
The solution to this potential problem lies in the quantum computer's ability to encompass the full spectrum of processing and optimization.

An introduction to this framework is given in the work by \citet{magannFeedbackBasedQuantumOptimization2022}, in which the authors introduced a feedback-based strategy for quantum optimization.
This algorithm is designed around qubit measurements to assign values to quantum circuit variational parameters constructively.
This procedure results in an estimate of combinatorial optimization solutions that improves monotonically with the circuit depth.
Crucially, this measurement-based feedback loop enables approximate solutions without classical optimization, as the entire process can be done on the quantum device.
This purely quantum optimization loop is achieved through a direct connection to Quantum Lyapunov Control (QLC)~\cite{magannLyapunovcontrolinspiredStrategiesQuantum2022}, which is a control strategy that uses feedback to identify controls to manipulate the drive dynamics of a quantum system.

The authors demonstrated the capabilities of this algorithm. They coined the Feedback-based ALgorithm for Quantum OptimizatioN (FALQON) on combinatorial optimization problems such as MaxCut on 3-regular and randomly generated non-isomorphic graphs.
The FALQON is defined similarly to the original QAOA, with the ``drift'' and ``control'' Hamiltonians, $H_p$ and $H_d$, respectively, rather than cost and mixer ones, although they have similar forms.
The authors begin by considering a quantum system whose dynamics are governed by
\begin{equation}
    i\dv{t}\ket{\psi(t)} = \left( H_p + H_d\beta(t)\right)\ket{\psi(t)},
\end{equation}
and seek to minimise $\langle H_p \rangle = \langle \psi(t) | H_p | \psi(t) \rangle$, which is done by designing $\beta(t)$ such that
\begin{equation}
    \dv{t}\langle \psi(t) | H_p | \psi(t) \rangle = A(t)\beta(t) \leq 0
\end{equation}
where $A(t) \equiv \langle \psi(t) | i[H_d, H_p] | \psi(t) \rangle$ and $\beta(t) = -A(t)$. 
This is an essential step in circumventing the need for a classical optimizer since the updated parameters are taken directly from the expectation value measurements.

Although this algorithm is designed for purely quantum computation for fault-tolerant devices that do not yet exist, it still finds use in NISQ devices as a way to improve the initialization of the standard QAOA\@.
In the original QAOA, the initial parameters are chosen at random. However, with the help of FALQON, these seed parameters can be better chosen after a few iterations of the purely quantum algorithm. 
It was demonstrated that this hybridization of FALQON and QAOA, coined the FALQON+~\cite{magannLyapunovcontrolinspiredStrategiesQuantum2022}, allows QAOA to start from a parameter set that has a higher success probability and smoother solution landscape. 
Performing the algorithm on a set of non-isomorphic graphs over 8 to 14 vertices, the authors find a significant increase in approximation ratio and success probability with a minimal increase in circuit depth and noise degradation. 
This makes the FALQON+ algorithm highly suitable for NISQ devices as a warm-start technique that utilizes the best features of both the fully quantum and hybrid approaches.

\subsubsection{Fermionic QAOA}

\citet{yoshiokaFermionicQuantumApproximate2023}
proposed the Fermionic Quantum Approximate Optimization Algorithm (FQAOA) to solve combinatorial optimization problems with constraints, utilizing fermion particle number preservation to impose these constraints intrinsically.
Many such problems feature constraints that can negatively impact optimization algorithms when treated as soft constraints in the cost function. 
FQAOA tackles this issue by using fermion particle number preservation to impose constraints throughout the QAOA process intrinsically. 
The authors offer a systematic guideline for designing the ``driver'' Hamiltonian ($H_d$) for problem Hamiltonians with constraints.
In the context of the quantum adiabatic theorem, the driver Hamiltonian is used to slowly transform a system Hamiltonian from its initial state to a cost function-based problem Hamiltonian, ultimately leading to optimal solutions of the cost function from the ground state of $H_d$ (Section~\ref{sec:QAA}).
They suggest choosing the initial state to be a superposition of states satisfying the constraint and the ground state of the driver Hamiltonian.
In the FQAOA ansatz, the mixer unitary is generated by the driver Hamiltonian which is carefully designed to satisfy specific conditions, ensuring that the constraints of the combinatorial optimization problem are intrinsically imposed.
The driver Hamiltonian introduces hybridization between different basis states by representing non-local fermions with hopping terms. 
This hybridization allows the algorithm to effectively explore different solutions in the search space.

FQAOA has demonstrated substantial performance advantages over existing methods in portfolio optimization problems. 
According to the authors, the Hamiltonian design guideline is valuable for both QAOA and Grover adaptive search and quantum phase estimation in solving constrained combinatorial optimization problems. 
This compatibility enables the application of existing software tools developed for fermionic systems in quantum computational chemistry to address constrained optimization challenges.


\subsubsection{Other approaches}

\citet{wangQuantumDropoutEfficient2022} proposed a method to deal with hard combinatorial optimization problems where the energy landscape is rugged and the global minimum locates in a narrow region of the cost function landscape.
In such a problem, the global minimum must satisfy a set of clauses $C$, encoded in the cost Hamiltonian of QAOA.
The authors found that the problem mainly originates from the QAOA circuit rather than the cost function. 
So they decided to exploit the combinatorial nature of the problem by selectively dropping out the clauses defining the quantum circuit, 
\begin{equation}
    \hat H_{C'} = \sum_{c_i\in C' \subset C}\hat H_{c_i}, 
\end{equation}
while keeping the cost function intact to ensure the uniqueness of the global minimum.
The quantum dropout of clauses helps smoothen the energy landscape, making optimizing parameters easier.
The numerical results confirmed QAOA’s performance improvements with various types of quantum dropout implementation and that the dropout of clauses in the circuit does not affect the solution.

\citet{wurtzClassicallyOptimalVariational2021}
introduced the Spanning Tree QAOA (ST-QAOA) to solve MaxCut problems using an ansatz derived from an approximate classical solution. 
In the ST-QAOA ansatz, a classical solver is first used to find an approximate solution for the MaxCut problem, typically represented by a spanning tree of the graph. 
This approximate solution is then used to construct a problem instance-specific circuit with $r$ rounds of gates. 
The circuit is designed to reflect the problem structure, using classical algorithm insights to tailor the quantum circuit specifically for the given problem instance.
When $r = 1$, the ST-QAOA is guaranteed to match the performance of the classical solver. 
As the number of rounds increases, the ST-QAOA ansatz approaches the exact solution to the MaxCut problem. 
This approach achieves the same performance guarantee as the classical algorithm and can outperform the vanilla QAOA at low depths. 

An additional modification to the ansatz construction of QAOA is given by the work of \citet{liQuantumOptimizationNovel2020}, in which the authors proposed modifications to both the QAOA ansatz as well as the prescription for choosing the variational parameters of the quantum circuit.
First, the Gibbs objective function is defined and shown to be superior to the energy expectation value, $\expval{E}$, as an objective function for optimizing variational parameters.
Second, the authors describe an Ansatz Architecture Search (AAS) algorithm for searching the discrete space of quantum circuit architectures near QAOA to find a better ansatz.
The Gibbs objective function is defined as follows,
\begin{equation}
	f = -\log\langle e^{-\eta E}\rangle,
\end{equation}
where $\eta > 0$ is a hyper-parameter based on the general properties of the class of problems, $E$ is the energy of the Ising model used to encode the optimization problem, and $f$ has a form similar to the Gibbs free energy in statistical mechanics; hence the name.
The exponential profile rewards the optimization procedure for increasing the probability of low energy and de-emphasizes the shape of the probability distribution at higher energies.
The authors propose greedy search as an affordable strategy for AAS\@.
A model $\mathcal{I}$ is defined on a graph $\G^\mathcal{I}$ with $m$ vertices, while $\G^\mathcal{A}$ is the ansatz graph obtained from $\G^\mathcal{I}$ by removing some edges.
The QAOA prescription is to set $\G^\mathcal{A} = \G^\mathcal{I}$ and search architectures by searching through graphs obtained by removing edges from $\G^\mathcal{I}$.
Given an instance $\mathcal{I}$, the search starts with $\mathcal{G}^\mathcal{A} = \mathcal{G}_m$ at level 0.
Then level by level, ansatz\"{e} are expanded by removing one two-qubit gate from the best ansatz of the previous level, scored, and the best of them is selected as the output of this level.
The output architectures at level $l$ have $l$ two-qubit gates (i.e., \ edges of the graph) removed.
It was found that applying these modifications to a complete graph Ising model results in a 244.7\% median relative improvement in the probability of finding a low-energy state while using 33.3\% fewer two-qubit gates.

\citet{villalba-diezImprovementQuantumApproximate2021} also proposed a modification to the cost Hamiltonian.
Their improved ansatz, ``Modified QAOA'', performs a conditional rotation of $\gamma$ to each node connected to another if the second is in state $\ket{1}$.
This is done by concatenation of two $U_3(\gamma/2,0,0)$ and $U_3(-\gamma/2,0,0)$ gates.
They follow this with a conditional CX rotation to each pair of nodes.
They tested their approach in simulations for up to 30 network nodes and for $p = 1$ and observed significant increase of the approximation ratio achieved; compared to the vanilla QAOA\@.

    \subsection{Parameter Optimization}%
    \label{sec:analysis_parameter_optimization}

{\renewcommand{\arraystretch}{1.7}
\begin{table}[htp!]
  \centering
  \footnotesize
  \begin{tabular}{p{0.4\linewidth}p{0.5\linewidth}}
 \toprule
  \textbf{Problem Addressed} & \textbf{Approach}\\
 \midrule
\multirow{7}{*}{Initial Parameter Search} & 
Heuristic strategies for initializing optimizations \cite{zhouQuantumApproximateOptimization2020}\\ 
& Layer-by-layer optimization \cite{leeParametersFixingStrategy2021}\\
& Warm-start techniques using GNNs \cite{jainGraphNeuralNetwork2021}\\
& Transferability among different QAOA instances based on local characteristics of subgraphs \cite{galdaTransferabilityOptimalQAOA2021}\\
& Parameter reusability for similar problem instances \cite{shaydulinMultistartMethodsQuantum2019,shaydulinQAOAKitToolkitReproducible2021}\\
& Parameter initialization using TQA method \cite{sackQuantumAnnealingInitialization2021}\\
\midrule
\multirow{4}{*}{Gradient-free Parameter Optimization} & 
Comparison of gradient-based and gradient-free methods \cite{fernandez-pendasStudyPerformanceClassical2022,bonet-monroigPerformanceComparisonOptimization2023,pellow-jarmanComparisonVariousClassical2021}\\
& Genetic algorithm approach for optimization \cite{acamporaGeneticAlgorithmsClassical2023}\\
& Robust control optimization techniques \cite{dongRobustControlOptimization2020}\\
\midrule
\multirow{8}{*}{Gradient-based Parameter Optimization} & 
Gradient-based approaches with machine learning techniques \cite{crooksPerformanceQuantumApproximate2018}\\
& Gradient-based optimization with tensor networks \cite{streifTrainingQuantumApproximate2020}\\
& Stochastic Gradient Descent in quantum context \cite{swekeStochasticGradientDescent2020}\\
& Surrogate-model based optimization \cite{sungUsingModelsImprove2020}\\
& Policy gradient-based reinforcement learning algorithm for QAOA \cite{yaoPolicyGradientBased2020}\\
& BFGS optimization algorithm for QAOA \cite{lotshawEmpiricalPerformanceBounds2021}\\
\midrule
\multirow{5}{*}{Machine Learning for Parameter Optimization} & 
Parameter correlation and machine learning models \cite{alamAcceleratingQuantumApproximate2020}\\
& Meta-learning for QAOA optimization \cite{wangQuantumApproximateOptimization2021,wilsonOptimizingQuantumHeuristics2021}\\
& Clustering for setting QAOA parameters \cite{moussaUnsupervisedStrategiesIdentifying2022}\\
& GNN-based prediction of QAOA parameters \cite{deshpandeCapturingSymmetriesQuantum2022}\\
& RL and KDE techniques for parameter optimization \cite{khairyLearningOptimizeVariational2020,patelReinforcementLearningAssisted2022}\\
\midrule
\multirow{3}{*}{Remedies for Barren Plateaus} & 
Incremental growth of circuit depth during optimization \cite{skolikLayerwiseLearningQuantum2021}\\
& Parameter concentraion as an inverse polynomial with respect to the problem size \cite{akshayParameterConcentrationsQuantum2021}\\
\midrule
\multirow{4}{*}{Parameter Concentration \& Symmetry} & 
Inverse polynomial concentration of optimal parameters \cite{akshayParameterConcentrationsQuantum2021}\\
& Exploiting symmetry in objective functions, cost Hamiltonians, and QAOA parameters to improve optimization \cite{shaydulinExploitingSymmetryReduces2021,zhouQuantumApproximateOptimization2020,shiMultiAngleQAOADoes2022,shaydulinClassicalSymmetriesQuantum2021}\\
\midrule
Analytical Solutions for Optimal Parameters & 
Deriving analytical solutions for optimal parameters \cite{wangQuantumApproximateOptimization2018}\\
 \bottomrule
  \end{tabular}
  \caption{Summary of approaches in QAOA parameter optimization.}%
  \label{tab:param_optim}
\end{table}
}



In this section, we discuss key aspects of parameter optimization in QAOA.
This covers various approaches to finding good initial parameters, which becomes increasingly important as the depth and complexity of the QAOA circuit increase. The original proposal of the algorithm suggests a random selection of initial parameters within a range believed to be close to the optimal parameters. However, this method can often hinder the algorithm's performance, especially when the cost function landscape is rugged and contains numerous local minima. Therefore, developing efficient and reliable parameter optimization strategies is crucial to achieving optimal QAOA performance. This section also covers the selection of appropriate classical optimization algorithms for investigating the parameter space and strategies for overcoming barren plateaus (Table~\ref{tab:param_optim}).

\subsubsection{Finding good initial parameters}\label{sec:parameter_initialisation}

To address the issue of a random selection of initial parameters, researchers have proposed various techniques to find better initial parameters.
One such approach is provided by \citet{zhouQuantumApproximateOptimization2020}, who developed an efficient parameter optimization procedure for QAOA applied to MaxCut problems. 
They proposed heuristic strategies for initializing optimizations, allowing them to find quasi-optimal $p$-level QAOA parameters in $O[\text{poly}(p)]$ time compared to random initialization, which requires $2^{O(p)}$ optimization runs for similar performance. 
Their heuristic strategies, such as INTERP and FOURIER heuristics, showed promising results in finding optimal parameters for large p-level QAOA.

\citet{leeParametersFixingStrategy2021} proposed a parameter fixing strategy for QAOA training to address the difficulty of finding optimal parameters at large p values and achieve higher approximation ratios. 
The algorithm initializes the QAOA circuit with optimal parameters from previous layers, finds the best parameters for each layer, fixes them, and then adds another layer on top of the previous one with new parameter values to optimize.

\citet{jainGraphNeuralNetwork2021} proposed a more efficient initialization of QAOA using Graph Neural Networks (GNNs). 
This approach builds on the precedent of warm-start techniques, aiming to initiate the initialization process closer to the target parameters. 
The GNN approach generalizes across graph instances and increases graph sizes, speeding up inference time across graphs. 
After GNN initialization, the authors explore several optimizers, including quantum aware/agnostic methods and machine learning techniques such as reinforcement learning, making the training process an end-to-end differentiable pipeline.

Another technique to improve initial parameters is based on parameter transferability across graphs, as proposed by \citet{galdaTransferabilityOptimalQAOA2021}. 
They showed that optimal QAOA parameters converge around specific values, and their transferability among different QAOA instances can be predicted based on local characteristics of the subgraphs composing the original graph. 
Building on this idea, \citet{shaydulinMultistartMethodsQuantum2019} proposed reusing optimal QAOA parameters for a given problem as an initial point for similar problem instances, demonstrating that this approach not only improves solution quality but also reduces the number of evaluations required.
\citet{shaydulinQAOAKitToolkitReproducible2021} also developed QAOAKit, a Python framework that includes a set of pre-optimized parameters and circuit templates for QAOA, leveraging parameter transferability to generate high-quality initial guesses for parameter optimization.

\citet{sackQuantumAnnealingInitialization2021} suggested initializing QAOA parameters based on the Trotterized quantum annealing (TQA) method, parameterized by a single variable, the Trotter time step. 
They established a heuristic way of finding the optimal time step based on TQA protocol performance, showing that this method of initialization can avoid the issue of false minima for a wide range of time steps, allowing QAOA to find solutions comparable to the best outcomes obtained from an exponentially scaling number of random initializations.

\subsubsection{Selecting an appropriate optimizer}

Selecting an appropriate optimization algorithm is crucial for enhancing the performance of QAOA. 
This subsection reviews various optimization approaches for improving QAOA parameters, including gradient descent, policy gradient, BFGS algorithm, and machine learning techniques.
We categorize these methods into three broad categories: gradient-free, gradient-based, and ML.

\paragraph{Gradient-free methods}

In optimization algorithms, gradient-free methods have gained attention due to their computational efficiency.
For example, \citet{mccleanTheoryVariationalHybrid2016} found that significantly fewer functional evaluations are required when using gradient-free methods in VQE.
\citet{shaydulinMultistartMethodsQuantum2019} were among the first to benchmark different gradient-free optimizers on QAOA.
They compared the solution quality of QAOA produced by six different gradient-free optimizers, including BOBYQA, COBYLA, NEWUOA, Nelder-Mead, PRAXIS, and SBPLX.
It was found that for a fixed number of functional evaluations allowed, using BOBYQA within the APOSSM framework, which allows for parallel optimization, led to the best performance.
Despite this, QAOA's performance was severely challenged with increasing layers across all gradient-free optimizers tested, suggesting the hardness of parameter optimization even at modest circuit depths.
More recently, \citet{fernandez-pendasStudyPerformanceClassical2022} investigated the performance of twelve different classical optimizers for QAOA optimization and found that gradient-based methods like Adam and SPSA have computational times up to two orders of magnitude higher compared to gradient-free methods such as COBYLA, Powell and Nelder–Mead, despite achieving similarly good results.



In a broader context of hybrid quantum-classical algorithms, \citet{bonet-monroigPerformanceComparisonOptimization2023} tested four commonly used gradient-free optimization methods: SLSQP, COBYLA, CMA-ES, and SPSA, on finding ground-state energies of a range of small chemistry and material science problems. 
The study, although not explicitly focused on QAOA, demonstrated the necessity for tailoring and hyperparameter-tuning known optimization techniques for inherently-noisy variational quantum algorithms and highlighted that the variational landscape that one finds in a VQA is highly problem- and system-dependent.

Similarly, in a study focused on variational hybrid quantum-classical algorithms, specifically the variational quantum linear solver, \citet{pellow-jarmanComparisonVariousClassical2021} examined the impact of several gradient-free and gradient-based classical optimizers on the performance of these algorithms.
They analyzed both the average rate of convergence and the distribution of average termination cost values of the classical optimizers, considering the effects of noise. 
Their findings indicate that realistic noise levels on NISQ devices pose a significant challenge to the optimization process, negatively affecting all classical optimizers. 
However, they found that the gradient-free optimizers, Simultaneous Perturbation Stochastic Approximation (SPSA) and Powell’s method, and the gradient-based optimizers, AMSGrad and BFGS, performed the best in the noisy simulation and were less affected by noise than other methods. 
In particular, SPSA emerged as the best-performing method. 
Conversely, the COBYLA, Nelder–Mead, and Conjugate-Gradient methods were the most heavily affected by noise, with even slight noise levels significantly impacting their performance. 
The study suggests that if noise levels can be significantly improved, gradient-based methods, which performed better than the gradient-free methods with only shot-noise present, may be preferred in the future.

\citet{acamporaGeneticAlgorithmsClassical2023} addressed the remaining shortcomings of the above gradient-free optimization methods with a new approach. 
The authors proposed an evolutionary approach to optimization using genetic algorithms.
A genetic algorithm is a search heuristic that reflects the process of natural selection where the fittest individuals are selected for reproduction to produce offspring of the next generation. 
The authors expressed that such a population-based heuristic could more efficiently process candidate solutions and converge to an optimal parameter set for a QAOA circuit.
This evolutionary approach is demonstrated on noisy quantum devices and compared with popular gradient-free optimizers (COBYLA, Nelder–Mead, Powell’s modified method, and SPSA) when solving the MaxCut problem for graphs with 5 to 9 nodes. 
The authors found that the proposed genetic algorithm statistically outperforms other gradient-free algorithms in achieving higher approximation ratios for the same QAOA circuits.
Despite these promising results, the authors noted that genetic algorithms are known to have limitations, such as scalability in the number of parameters to learn and premature convergence, encouraging more research to build upon this proposal for a more reliable gradient-free optimization solution.

In another recent work, \citet{chengErrormitigatedQuantumApproximate2023} introduced a novel gradient-free optimizer called Double Adaptive-Region Bayesian Optimization (DARBO), which demonstrated robustness against measurement noise and quantum noise.
This optimizer explores the QAOA landscape using a Gaussian process surrogate model and iteratively suggests the optimal parameters restricted to two auto-adaptive regions.
Incorporating the two adaptive regions makes it more robust to noise and different initial parameters.
Upon benchmarking against other optimizers (Adam, COBYLA, SPSA) for the MaxCut problem on weighted 3-regular graphs, DARBO showed superior performance in simulations with measurement shot noise.
Furthermore, the authors also demonstrated that DARBO remained effective on superconducting quantum computers despite hardware noise by integrating with proper quantum error mitigation techniques.

In addition to the previously discussed methods, robust control optimization techniques have also been explored in the context of QAOA.
\citet{dongRobustControlOptimization2020} demonstrated that the error of QAOA simulation could be significantly reduced by robust control optimization techniques, specifically by sequential convex programming (SCP).
This approach ensures error suppression in situations where the source of the error is known but not necessarily its magnitude. 
The study showed that robust optimization improves the objective landscape of QAOA and overall circuit fidelity in the presence of coherent errors and errors in initial state preparation.

\paragraph{Gradient-based Approaches}\label{sec:parameter_gradient}

Gradient-based strategies are one of the most commonly used methods for parameter optimization, which come in various flavors.
The gradient descent method is a first-order iterative optimization algorithm for finding a local minimum of a differentiable function.
Gradient descent takes repeated steps in the opposite direction of the gradient because the direction of the gradient is the direction toward the local maximum.

Combining the gradient descent method with machine learning, \citet{crooksPerformanceQuantumApproximate2018} optimized QAOA on a classical computer using automatic differentiation and stochastic gradient descent using QuantumFlow, a quantum circuit simulator implemented with TensorFlow.
Authors amortized the training cost of QAOA circuits by training variational parameters on batches of problem instances (graphs), thus alleviating the training procedure.
\citet{streifTrainingQuantumApproximate2020} also simulated on classical hardware with tensor networks instead of making parameter updates by repetitive calls of the QPU\@.

As a notable variant in the gradient descent family, Stochastic Gradient Descent (SGD) has been widely used in machine learning for training deep neural networks and made it an option for efficient exploration of quantum circuits using classical simulation due to accelerated optimization and ease of use.
However, implementing stochastic gradient descent directly on a quantum computer is demanding, requiring many measurements for each gradient component, which can be computationally costly and complex.
\citet{swekeStochasticGradientDescent2020} makes use of SGD, which replaces the exact partial derivative at each optimization step with an estimator of the partial derivative to tune the value and provide numerical results using a gradient, offering a practical way to utilize SGD in a quantum context.
\citet{sungUsingModelsImprove2020} introduced a surrogate-model-based algorithm called the Model Gradient Descent (MGD), which is also inherently stochastic.
This method incorporates a least-squares quadratic model to estimate the gradient of the objective function, allowing the previously evaluated points to be reused and leading to more efficient optimization.
Empirical comparisons with other popular optimizers (SGD, SPSA, BOBYQA, Nelder-Mead) suggested that stochastic optimizers such as MGD may be advantageous in realistic settings since they are more robust to variations in problems and show good tolerance to noise.

The work by \citet{yaoPolicyGradientBased2020} is focused on the fact that the original QAOA selects initial parameters at random and optimizes them through gradient-based methods, which can be computationally expensive and vulnerable to noise in NISQ devices and may hinder the optimization process.
Noting this, the authors suggested a new method for selecting and optimizing the variational parameters of the model. 
This is achieved by assigning a probability distribution to randomly selected initial parameters and using this distribution as a basis for a policy gradient-based reinforcement learning algorithm to find optimal variational parameters.
It was demonstrated that this policy-gradient-based model (PG-QAOA) does not require derivatives to be computed explicitly and can perform well even if the objective function is not smooth concerning the error.
This probability-based approach also has the advantage of resisting perturbations and noise. 
In this sense, policy-gradient-based reinforcement learning algorithms are well suited for optimizing the variational parameters of QAOA in a noise-robust fashion, opening up the way for developing reinforcement learning techniques for continuous quantum control.
Additionally, similar to the MGD extension for gradient descent, the vanilla policy gradient method~\cite{yaoPolicyGradientBased2020} can also be extended with a surrogate model to reduce the variance in the estimation of the policy gradient, which is dubbed the Model Policy Gradient (MPG) method~\cite{sungUsingModelsImprove2020}.
Like MGD, MPG showed good tolerance to noise and robustness to problem variations.

In the work by \citet{lotshawEmpiricalPerformanceBounds2021}, the effectiveness of circuit optimization using BFGS was assessed against exact optimization software and brute force solutions for several graphs up to $p = 2$. 
Empirical results confirmed that BFGS could return optimal QAOA angles in all the test cases.  
However, it is still unclear how BFGS will behave for larger parameter vectors, i.e., for $p \geq 3$.
The authors also found that optimized angles returned by BFGS reveal various symmetry patterns, where multiple angle solutions gave the same expectation value of the cost function.
In light of this, they suggest that appropriate heuristics may be employed to select parameters more efficiently.


\paragraph{Machine Learning Approaches}\label{sec:parameter_machine}

In addition to gradient-based methods, machine learning has been investigated as a different way to find optimal parameters for QAOA and enhance the optimization process.

\citet{alamAcceleratingQuantumApproximate2020} applied machine learning techniques to accelerate QAOA optimization based on parameters' correlation.
The authors noted a correlation among parameters of the lower-depth and higher-depth QAOA layers, so they exploited it by training machine learning models to predict the variational parameters close to the optimal values.
Then, these quasi-optimal parameters are fine-tuned with classical optimizers to generate the final solution.
The authors trained four machine learning models: Gaussian Process Regression (GPR), Linear Regression (LM), Regression Tree (RTREE), and Support Vector Machine Regression (RSVM).
They found that the proposed machine learning-based approaches can shorten the optimization iterations by 44.9\% on average.

A meta-learning approach based on classical Long Short-Term Memory (LSTM) neural networks was also investigated by \citet{wangQuantumApproximateOptimization2021} and \citet{wilsonOptimizingQuantumHeuristics2021}, where similar conclusions were drawn on the goodness of meta-learner techniques for QAOA optimization.
In~\cite{wangQuantumApproximateOptimization2021}, QAOA with meta-learning (MetaQAOA) for the MaxCut problem was proposed: an LSTM neural network was used as a black-box optimizer in order to help find optimal QAOA parameters.
Numerical simulations showed that MetaQAOA converged faster and to a better value than the vanilla QAOA with local optimization methods such as Nelder–Mead and L-BFGS-B.
Similarly, \citet{wilsonOptimizingQuantumHeuristics2021} compared the performance of an LSTM meta-learner to evolutionary strategies, L-BFGS-B and Nelder-Mead approaches, confirming the results of \citet{wangQuantumApproximateOptimization2021} and showing that the meta-learner applied to a QAOA problem finds the global optima more frequently than all other optimizers tested in the paper, is more resistant to noise and can easily generalize to larger problems even if it was trained on small problems.

\citet{moussaUnsupervisedStrategiesIdentifying2022} made use of unsupervised ML, namely clustering, for setting the QAOA angles without optimization. 
They considered several inputs to the clustering algorithm, including the angle values, instance features, and the output from a variational graph autoencoder.
Their findings demonstrated that such a method is effective in learning how to set QAOA parameters, reducing circuit calls while maintaining a relatively small reduction in the approximation ratio of less than 1--2\%. 
They achieved comparable results to those obtained through exhaustive angle optimization, translating to significant reductions in the required circuit calls.

\citet{deshpandeCapturingSymmetriesQuantum2022} used a Graph Neural Network (GNN) to predict QAOA’s optimal parameters for unseen MaxCut instances, with up to nine vertices and a depth of $p=3$.
The GNN could predict quasi-optimal QAOA’s parameters within 2.7\% of the optimal parameter solution, which could then be used as a warm-start for QAOA.
However, the scalability of this approach is uncertain, as it requires many training instances for the GNN to perform well in the prediction tasks.

Reinforcement Learning (RL) has also been investigated in the VQA framework to improve parameters optimization~\cite{lockwoodReinforcementLearningQuantum2020,beloborodovReinforcementLearningEnhanced2020}. 
On this matter, some papers explored the potential of RL in assisting QAOA's optimization for finding optimal parameters~\cite{khairyLearningOptimizeVariational2020, patelReinforcementLearningAssisted2022}. 
\citet{khairyLearningOptimizeVariational2020} try to find optimal QAOA parameters through two different machine learning methods: an RL technique and a Kernel Density Estimation (KDE) approach. 
The RL framework trains a policy network to optimize QAOA circuits. 
The network exploits geometrical regularities in the QAOA energy landscapes to efficiently find high-quality solutions for unseen test instances with only a few hundred quantum circuit evaluations.
The KDE technique is used to create a generative model of optimal QAOA parameters, which can be employed to generate new parameters and quickly solve test instances.
Extensive simulations with IBM Qiskit Aer demonstrated that both the proposed methods are effective in finding good QAOA parameters, achieving superior approximation ratios compared to other commonly used off-the-shelf optimizers.
Finally, \citet{patelReinforcementLearningAssisted2022} uses RL to enhance the performance of the RQAOA (Section~\ref{subsec:rqaoa}). 
In particular, they compared the original RQAOA~\cite{bravyiObstaclesVariationalQuantum2020} with a proposed RL-enhanced RQAOA variant (RL-RQAOA).
The latter trains the circuit’s parameters via RL and uses correlations between qubits to perform variable elimination at every iteration of the algorithm. 
Through simulations over an ensemble of randomly generated weighted $d$-regular graphs, the authors empirically showed that a $p=1$ RL-RQAOA consistently outperforms RQAOA and simple classical RL agents.

\paragraph{Others}

In \citet{barkoutsosImprovingVariationalQuantum2020}, they proposed using the Conditional Value-at-Risk (CVaR) instead of choosing the sample mean of the expectation values as the objective function. 
CVaR is a risk metric that estimates the expected loss of a portfolio under a specified confidence level by focusing on the tail end of the loss distribution. 
The researchers demonstrated that employing the CVaR aggregation function in QAOA enabled the algorithm to achieve optimal solutions more rapidly in simulations and on real quantum hardware, such as the IBMQ Poughkeepsie 20-qubit device.

To address the difficulty of finding optimal parameters at large $p$ values and to achieve a higher approximation ratio, \citet{leeParametersFixingStrategy2021} proposed a parameter fixing strategy for QAOA training. 
Starting from $p=1$ and continuing to a desired $p$ level, the algorithm initializes QAOA with the optimal parameters from previous layers. 
The algorithm finds the best parameters for each layer, fixes them, and then adds another layer on top of the previous one with new parameter values to optimize.

In their study, \citet{lotshawEmpiricalPerformanceBounds2021} developed a search heuristic for QAOA parameter optimization that is effective for a wide range of graphs, which also significantly decreases computational expenses when compared with the Broyden-Fletcher-Goldfarb-Shanno (BFGS) search with random seeding, which is a widely used optimization algorithm. 
They achieved this by examining how the optimized angle patterns based on a wide range of graphs can help efficiently identify suitable approximate angles.
They demonstrated that identifying consistent patterns among the optimized variational parameters is an efficient heuristic for solving MaxCut problems with up to 9 vertices with up to $p=3$ QAOA.

\subsubsection{Overcoming barren plateaus}\label{sec:parameters_barren_plateaus}

A significant challenge in training parameterized quantum circuits, such as QAOA, is addressing the issue of barren plateaus in the cost function landscape (Section~\ref{sec:barren_plateaus}). 
Barren plateaus are regions where the gradients concerning the trainable parameters vanish exponentially in the number of qubits, causing the optimization process to stall. 
To tackle this problem, several approaches have been proposed in the literature, aiming to improve the optimization process and help QAOA converge to better solutions. 

One promising strategy to overcome this issue is presented in the work of \citet{skolikLayerwiseLearningQuantum2021}, in which a layerwise learning strategy is developed.
During optimization, the circuit depth is grown incrementally, with only subsets of parameters being updated with every training step.
In this approach, the circuit structure and a number of parameters are successively grown while the circuit is trained, and the randomization effects are contained in subsets of the parameters in all the training steps.
This avoids initializing on a plateau and reduces the probability of creeping onto a plateau during training.
The authors demonstrate the success of this strategy in an image classification problem with a general parameterized quantum circuit, which obtained an average generalization error of 8\% lower than in standard learning schemes, but the methodology applies to all VQAs, such as QAOA\@.

The phenomenon of parameter concentration in QAOA circuits has been investigated by \citet{akshayParameterConcentrationsQuantum2021}. 
They found that optimal parameters concentrate as an inverse polynomial concerning the problem size, which is beneficial for improving circuit training. 
Empirical investigation revealed two symmetric branches of optimal parameters, within which parameters concentrated in a fixed range of values as the number of qubits varied. 
This concentration effect allows training a depth-$p$ QAOA on a fraction $w < n$ of qubits and asserting that these parameters are nearly optimal on $n$ qubits and $p$ layers, thereby reducing training time.

\citet{dupontEntanglementPerspectiveQuantum2022} investigated the growth and spread of the entanglement resulting from optimized and randomized QAOA circuits to solve the MaxCut problem on different graphs.
The study finds a volume-law entanglement barrier between the initial and final states.
Entanglement barrier $S \sim N$ must be crossed for large-depth QAOA circuits, making entanglement-based simulation methods challenging.
This study also investigates the entanglement spectrum in connection with random matrix theory.
The results are compared with a quantum annealing protocol, and implications for the simulation of QAOA circuits with tensor network-based methods are discussed.

Finally, a promising general attempt at tackling the problem of barren plateaus stems from the idea of beginning the optimization process closer to the target parameters (Section~\ref{sec:analysis_parameter_optimization}).
This may seem like an obvious solution. However, it is challenging to execute in practice because the solution landscape is almost entirely unknown in most variational problems.

\subsubsection{Parameter transferability and reusability}
\label{sec:transferability_of_params}

One promising approach to finding optimal QAOA parameters lies in the transferability and reusability of optimal parameters across different problem instances. 
This concept builds on the observation that optimal parameters tend to concentrate around specific values and that these values can be transferred from one problem instance to another based on their local characteristics.

\citet{galdaTransferabilityOptimalQAOA2021} provided a theoretical foundation for parameter transferability in QAOA. 
They showed that optimal QAOA parameters converge around specific values, and the transferability of these parameters among different QAOA instances can be predicted and described based on the local characteristics of the subgraphs composing the original graph. 
This observation provides a method for identifying categories of combinatorial optimization problems where QAOA and other VQAs can offer significant speedup. 

Based on this idea, \citet{shaydulinMultistartMethodsQuantum2019} proposed to reuse the optimal QAOA parameters for a given problem as an initial point for similar problem instances.
They demonstrated that this not only improves the quality of the solution by avoiding the local optima but also reduces the number of evaluations required to reach it.
Building on this, \citet{shaydulinQAOAKitToolkitReproducible2021} leveraged the property of optimal parameter transfer across similar graphs and proposed QAOAKit, a Python framework that includes a set of pre-optimized parameters and circuit templates for QAOA\@.
Given an input graph, quasi-optimal parameters are obtained through a graph isomorphism certificate for the input graph, which is then employed as a key to extract the angles from QAOAKit's database.
If optimal angles are not present in the database for a specific graph instance, the system will provide the closest fixed angles instead.

\subsubsection{Leveraging parameter symmetries}\label{sec:parameter_symmetry}
Parameter symmetries can be leveraged to simplify the optimization process and eliminate degeneracies in the parameter space, contributing to more efficient QAOA performance. 
This subsection discusses various works exploring and exploiting QAOA parameters' symmetries for improved optimization, focusing on symmetry in objective functions, cost Hamiltonians, and QAOA parameters.

\paragraph{Symmetry in Objective Functions and Cost Hamiltonians}

\citet{shaydulinExploitingSymmetryReduces2021} established a connection between the symmetries of the objective function and the cost Hamiltonian concerning the QAOA energy.
They showed that excluding terms connected by symmetry could significantly reduce the cost of evaluating the QAOA energy.
They used fast graphs and automorphism solvers to compute the problem's symmetries. 
Although their approach provides a median speedup of 4.06 for $p = 1$ on 71.7\% of the graphs considered, on a benchmark where 62.5\% of the graphs are known to be hard for automorphism solvers, the automorphism calculation could require more time than the one saved by exploiting symmetry in the worst-case scenario.

\citet{shaydulinClassicalSymmetriesQuantum2021} investigated the correlation between QAOA and the inherent symmetries of the target function to be optimized. 
They revealed how the symmetries of the objective function result in invariant measurement outcome probabilities among states connected by those symmetries, regardless of the number of layers or algorithm parameters employed. 
Using machine learning techniques, the authors leveraged these symmetry considerations to predict the QAOA performance accurately---by analyzing a small set of graph symmetry properties, they could predict the minimum QAOA depth required to achieve a desired approximation ratio on the MaxCut problem.

\paragraph{Symmetry in QAOA Parameters}

\citet{akshayParameterConcentrationsQuantum2021} investigated the phenomenon of parameter concentration in QAOA circuits, and they found that the optimal parameters concentrate as an inverse polynomial concerning the problem size, which is beneficial for improving circuit training. 
From empirical investigation, authors identified two symmetric branches of optimal parameters; within these branches, parameters concentrated in a fixed range of values as the number of qubits varied.
The concentration effect allows training a depth-$p$ QAOA on a fraction $w < n$ of qubits and asserting that these parameters are nearly optimal on $n$ qubits and $p$ layers, thereby reducing training time. 
In other words, if parameter concentration occurs, the optimal set of parameters for $n$ qubits is polynomially close to the optimal set of parameters for $n+1$ qubits. 

\citet{shiMultiAngleQAOADoes2022} investigated the connection between symmetries in the input graphs and redundancy in ma-QAOA parameters~\cite{herrmanMultiangleQuantumApproximate2022}, revealing that symmetries can result in a reduction of the number of parameters without decreasing the quality of the solution. 
The authors analyzed all the connected, non-isomorphic graphs with eight nodes, noticing that in over two-thirds of these graphs, the same accuracy ratio on the MaxCut problem can be obtained by reducing the number of parameters by 28.1\%, exploiting the natural symmetries of the graphs. 
Furthermore, they demonstrated that in 35.9\% of the graphs, the aforementioned reduction could be accomplished by utilizing the largest symmetry. 
On the other hand, in the cases where the reduction in the number of parameters led to a decrease in performance, utilizing the largest symmetry led to a mere 6.1\% decrease in the cost while successfully reducing the parameter count by 37.1\%.

\citet{zhouQuantumApproximateOptimization2020} suggested an approach to parameterize QAOA, which could simplify the optimization process by decreasing the dimension of the search space. 
This was achieved by identifying and eliminating degeneracies arising from inherent symmetries in the parameter space, including time-reversal symmetry and $\mathbb Z_2$ symmetry, before searching for patterns in optimal QAOA parameters.
Moreover, for QAOA applied to MaxCut problems on regular undirected (udR) graphs, there was an additional symmetry due to the structure of the problem that created redundancy.
This process of removing degeneracies creates a more navigable parameter space, facilitating a smoother and more effective search for optimal QAOA parameters.

Apart from using classical optimizers in a variational loop, optimal QAOA parameters can sometimes be derived analytically. 
\citet{wangQuantumApproximateOptimization2018} derived analytical expressions to solve optimal parameters for the level-1 QAOA applied to MaxCut on general graphs.
Although the analysis can be extended to higher $p$ values theoretically, the number of terms involved quickly becomes intractable for direct calculation. 
Therefore, the authors proposed a fermionic representation for a specific instance of MaxCut: the ring of disagrees, or the one-dimensional antiferromagnetic ring. 
This approach translates the QAOA-induced evolution of the system into quantum control of a group of independent spins, allowing for the derivation of analytical solutions for any $p$, thus simplifying the search for optimal parameter values. 
By exploring symmetries among parameter values, the authors identified a lower-dimensional sub-manifold that could minimize the search effort. 
Furthermore, they conducted a numerical investigation into the parameter landscape and empirically demonstrated that all minima are global minima.

    \subsection{Computational Resource Efficiency}%
    \label{sec:analysis_performance}

In this section, we delve into the computational resource efficiency of QAOA, focusing mainly on the time it takes to reach desired solutions.
We explore the speedup potential of QAOA over classical algorithms, highlighting instances where it has demonstrated superiority (Section~\ref{sec:speedup_potential}). 
However, obstacles that hinder the realization of quantum speedups with QAOA also exist, such as the challenges in parameter optimization and the overheads of error correction in early fault-tolerant quantum computers (Section~\ref{sec:speedup_obstacles}). 
Moreover, we highlight several strategies aimed at enhancing the runtime performance of QAOA (Section~\ref{sec:performance_improvements}), in addition to the ones that focus on improving parameter optimization discussed in Section~\ref{sec:analysis_parameter_optimization}.
By examining these aspects, we aim to provide a comprehensive analysis of the computational resource efficiency of QAOA, offering insights into its potential for enabling quantum advantage in optimization problems.

\subsubsection{Speedup potential}\label{sec:speedup_potential}

The QAOA is considered one of the leading candidate algorithms to achieve a quantum advantage over classical algorithms and quantum annealing.
One key aspect in determining this advantage is the ability of the QAOA to solve problems faster than other algorithms, thereby achieving a quantum speedup.
This subsection will explore recent research efforts that shed light on the QAOA's potential for achieving quantum speedup in different problem domains.

Based on the amortization of the training cost by optimizing batches of MaxCut problem instances, \citet{crooksPerformanceQuantumApproximate2018} argued that an analysis of the computational complexity of QAOA hinges on the number of gates needed to implement a single instance of the quantum evolution.
Since the required number of QAOA steps for a given performance does not appear to be strongly correlated with the problem size, the primary limiting factor on a gate-based quantum computer is the number of two-qubit gates required to implement a single round of the algorithm.
Based on their analysis, QAOA for MaxCut on an $n$-node graph is expected to require $O(n^2p)$ gates and have a run time of $O(np)$ (assuming $O(n)$ gates can be applied in parallel), where $p$ is the number of layers in the QAOA circuit.
In contrast, the classical Goemans-Williamson (GW) algorithm requires a run time of $O(nm)$ for irregular graphs (ignoring logarithmic factors), where $m$ is the number of edges.
These findings suggest that QAOA, even at modest depths, may offer a speedup over its classical counterpart for dense graphs.

QAOA also exhibits a speedup potential in the Minimum Vertex Cover (MVC) problem, as demonstrated by \citet{zhangApplyingQuantumApproximate2022}.
In their study, they applied QAOA with varying numbers of layers ($2 \leq p \leq 10$) to find the MVC on an undirected graph consisting of 10 vertices and 16 edges.
They found that QAOA was able to effectively solve the MVC problem within a computational time complexity of $O[\text{poly}(k) + \text{poly}(p)]$, where $k$ represents the number of iterations in the optimization process.
In contrast, the time complexity of a competitive decision algorithm for MVC is $O(2n^2 +n^4)$, with $n$ being the total number of vertices in the graph. 
Based on this comparison, the authors concluded that QAOA provides an exponential acceleration for large MVC problems.

In another study, \citet{ebadiQuantumOptimizationMaximum2022} implemented QAOA on 2D Rydberg atom arrays with up to 289 qubits.
In their experiment, QAOA was applied to find the Maximum Independent Set (MIS) on 115 randomly generated graphs of various sizes, with the number of vertices ranging from 80 to 289.
They benchmarked the results of the quantum algorithm against a classical counterpart, Simulated Annealing (SA)~\cite{kirkpatrickOptimizationSimulatedAnnealing1983}.
On graph instances in the deep-circuit regime, where $\delta_\text{min} > 1/T$, that is, the minimum energy gap $\delta_\text{min}$ of a graph is large enough to be resolved in the duration of the quantum evolution $T$, QAOA exhibited a superlinear quantum speedup compared to SA.
Specifically, for SA, the probability of observing an MIS scales as $P_\text{MIS} = 1 - \exp(-C\eta^{-1.03})$, where $\eta$ is a parameter characterizing the difficulty to reach the global optimum for a particular graph.
On the other hand, QAOA gave rise to an improved scaling, $P_\text{MIS} = 1 - \exp(-C\eta^{-0.63})$.
Since the runtime needed to find a solution is proportional to $1/P_\text{MIS}$ by repeating the experiment, the smaller exponent in the scaling for QAOA thus implies a quantum speedup over SA.
However, it should be noted that the observed speedup is specific to graph instances in the deep-circuit regime, and it remains an open question whether this advantage can be extended to more general cases.


QAOA has demonstrated its ability to provide quantum speedup in optimization problems and other domains. 
One notable example is the unstructured search problem, where \citet{jiangNearoptimalQuantumCircuit2017} proposed a novel quantum algorithm by incorporating the QAOA circuit and replacing the original diffusion operator in Grover's algorithm with the transverse field.
Their approach showcased the potential of QAOA in unstructured search, achieving a near-optimal query complexity of $T \sim O(\sqrt{N})$ with an intermediate number of layers ($p \gg 1$), where $N$ represents the search space size. 
This result demonstrates a quadratic quantum speedup similar to Grover's original algorithm.
In a separate study, \citet{niuOptimizingQAOASuccess2019} investigated the dependence of the success probability of the QAOA on its circuit depth $p$ by analyzing its performance for the state transfer problem in a one-dimensional qubit chain of length $N$ using two-qubit XY Hamiltonians and single-qubit Hamiltonians.
The authors derived analytical expressions for QAOA's success probability scaling as a function of the circuit depth.
Interestingly, they established a connection between the state transfer problem and Grover's search algorithm, where the $p$-th Grover iteration for searching the transferred state from an initial state can be represented by the unitary realized by a depth-$p$ QAOA circuit.
In the low-depth limit, $p \to 1$, the total number of steps required to achieve the target state is of order $O(N)$, producing the quadratic Grover-like speedup.
More recently, \citet{anQuantumLinearSystem2022} demonstrated that QAOA could also solve a Quantum Linear System Problem (QLSP) nearly optimally, with $O(\kappa\; \text{poly}(\log(\kappa/\epsilon )))$ runtime, where $\kappa$ is the condition number and $\epsilon$ is the target accuracy.
In contrast, the best classical algorithm, the conjugate gradient method, has a time complexity $O(N \sqrt{\kappa} \log(1/\epsilon))$, where $N = 2^n$ is the size of the system.
QAOA thus has an exponential advantage concerning $N$ over the classical algorithm.
It also outperforms various quantum algorithms for this problem, such as the HHL algorithm~\cite{harrowQuantumAlgorithmLinear2009} and those based on Linear Combination of Unitaries (LCU)~\cite{childsQuantumAlgorithmSystems2017} Quantum Signal Processing (QSP)~\cite{gilyenQuantumSingularValue2019}.


\subsubsection{Obstacles to quantum speedup}\label{sec:speedup_obstacles}

While the QAOA has demonstrated quantum speedup in specific problem domains (see Section~\ref{sec:speedup_potential}), it is crucial to recognize that there are still numerous obstacles to overcome in achieving a quantum advantage in a more general context. 
Challenges related to circuit depth, entanglement, optimization of variational parameters, and noise pose significant hurdles. 

For example, \citet{guerreschiQAOAMaxCutRequires2019} attempted to assess the feasibility of achieving a quantum speedup using QAOA in a more realistic setting that accounts for decoherence and dissipation effects.
They aimed to determine the minimum number of qubits required for QAOA running on a quantum computer to outperform state-of-the-art classical solvers on combinatorial problems.
They concluded that, in such a scenario, QAOA must run in a maximum of a minute to compete with classical solvers on problems with less than 400 variables.
It would only be possible to achieve a quantum speedup for a specific combinatorial problem once several hundreds of qubits are available.
They suggested that the reason for the exponential cost of the QAOA protocol is not due to the complexity of the quantum circuits but rather the challenge of optimizing the variational parameters used in the algorithm.
In this sense, the efficiency of QAOA is strongly tied to its depth, i.e., the number of layers ($p$) chosen, since a larger QAOA depth also implies more parameters to optimize.

The challenge of parameter optimization is further supported by the work of \citet{herrmanMultiangleQuantumApproximate2022}.
In their ma-QAOA (Section~\ref{subsec:ma-qaoa}) proposal, the performance of the standard QAOA is enhanced by adding more parameters to the circuit. 
Even though the approximate solution found by the ma-QAOA is better or equal to that of the standard QAOA, the increased number of variational parameters made the optimization process more challenging.
The authors noted that the time required for each iteration of the optimization algorithm was slower for the ma-QAOA than the standard QAOA, as the number of gradient components increases linearly with the number of variables to optimize.
However, it should be noted that from a theoretical perspective, the number of layers in ma-QAOA is always less than or equal to the number of layers required for the standard QAOA to achieve the same approximation ratio. 
This implies that the ma-QAOA potentially requires fewer samples or shallower circuits than the standard version.
Further studies are needed to compare the performance of the two algorithms with an equal number of parameters to analyze their convergence properties. 
Overall, empirical results indicate that finding good parameters for ma-QAOA typically requires polynomial time.
The challenge of finding optimal parameters was noted by \citet{shaydulinMultistartMethodsQuantum2019}, who conducted a benchmark study comparing six different derivative-free local optimization methods for QAOA.
With a budget of 1000 optimization steps and an optimistic assumption of 1000 measurements needed for obtaining the statistics to calculate the cost function value, the estimated time cost for a complete QAOA optimization is about 16 minutes, which is orders of magnitude greater than the runtime of classical state-of-the-art AKMAXSAT solvers~\cite{guerreschiQAOAMaxCutRequires2019}.
They also observed that as the number of QAOA layers increased, the fraction of problems solved within a given optimization budget significantly decreased across all the derivative-free methods tested. 
These findings suggest that achieving a quantum speedup with QAOA, even at low depths, is challenging.
Moreover, problems such as barren plateaus~\cite{mccleanBarrenPlateausQuantum2018, wangNoiseinducedBarrenPlateaus2021} further exacerbate such situation, and we will need better parameter optimization strategies to improve the runtime of QAOA.
More details on parameter optimization in QAOA can be found in Section~\ref{sec:analysis_parameter_optimization}.

Even though QAOA was a heuristic algorithm designed to be run on NISQ devices, \citet{sandersCompilationFaultTolerantQuantum2020} considered its utility on small fault-tolerant surface code quantum computers with around a million physical qubits or less to search for practical quantum advantages.
They estimated the resources required to implement several heuristic quantum algorithms, including QAOA for the Sherrington-Kirkpatrick (SK) model and the Low Autocorrelation Binary Sequence (LABS) problem on early fault-tolerant quantum processors.
The study revealed that the substantial overhead of state distillation in error correction introduces significant slowness and inefficiency in executing these algorithms. 
As a result, they concluded that any quantum optimization algorithm offering only a quadratic speedup is unlikely to produce any quantum advantage on these processors unless significant improvements in the surface code implementation, such as faster state distillation, are in place.
However, the prospects for an error-corrected quantum advantage on a modest processor are significantly more promising with quartic speedups~\cite{babbushFocusQuadraticSpeedups2021}. 
To this end, \citet{mccleanLowDepthMechanismsQuantum2021} advocated for a better understanding of the structure of classical optimization problems and the underlying physical mechanism of quantum optimization algorithms. 
This is because, for problems lacking structure, quadratic speedups are the best one could hope for.


\subsubsection{Runtime performance improvements}\label{sec:performance_improvements}

Despite the challenge in achieving a quantum speedup with QAOA due to many factors, many methods have been proposed in the recent literature to enhance the efficiency of the algorithm to reach desired solutions.
Most proposals focus on improving the parameter optimization process, extensively surveyed in Section~\ref{sec:analysis_parameter_optimization}.
Below we will highlight a few techniques that target other aspects of the algorithm.

The efficiency of QAOA in reaching optimal solutions has been enhanced through various modifications to its ansatz design.
One such modification is the ADAPT-QAOA (Section~\ref{subsec:adapt-qaoa}) proposed by \citet{zhuAdaptiveQuantumApproximate2020}, which empirically demonstrated a faster convergence than the original QAOA due to shortcuts to adiabaticity.
By incorporating entangling gates in the mixer operator pool, ADAPT-QAOA achieves a reduction in both the number of variational parameters and the number of CNOT gates by approximately 50\%, while simultaneously yielding better results than the original algorithm as the number of layers increases.
Another approach to improving the runtime of QAOA is the introduction of the ab-QAOA (Section~\ref{subsec:ab-qaoa}), as proposed by \citet{yuQuantumApproximateOptimization2022}.
In ab-QAOA, local fields are integrated into the operators themselves.
Numerical simulations conducted on MaxCut problems have shown that ab-QAOA significantly decreases the QAOA runtime compared to the original algorithm for a given level of accuracy.
Significantly, this improvement in runtime increases with problem size, resulting in polynomially shorter computation times than the vanilla QAOA concerning the number of nodes in the graph.
Additionally, ab-QAOA requires the same number of quantum gates and measurements, making it a promising candidate for achieving a quantum advantage in combinatorial optimization problems.

\citet{liHierarchicalImprovementQuantum2020} hierarchically analyzed the effects of influential factors on QAOA's runtime performance and proposed a 3-level improvement of the hybrid quantum-classical optimization for object detection.
They achieved a significant speedup of more than 13 times at the first level by selecting the L-BFGS-B classical optimizer. 
This choice improved the efficiency of the classical optimization process.
For the second level improvement, the authors constrained the QAOA circuit parameters to the range $(0,\pi)$, exploiting the symmetry of these parameters.
This constraint led to a runtime acceleration of up to 5.5 times, with the expectation of further improvements for deeper QAOA circuits.
Moreover, an acceleration of more than 1.23 times was obtained using parameter regression.
Finally, at the third level, they empirically demonstrated that the circuit would achieve better fidelity by optimally rescheduling gate operations, especially for deeper circuits---a shorter critical path would not only make the QAOA's circuit execution faster, it would also mitigate the impacts of noise and decoherence.

\citet{larkinEvaluationQAOABased2022} proposed a new metric for evaluating the runtime performance of QAOA in solving the MaxCut problem. 
They focused on the probability of observing a sample above a certain threshold. Given a desired approximation ratio value, QAOA’s efficiency is measured by the time needed to observe at least one sample with the desired approximation ratio, with a probability of at least 50\%. 
In this sense, the efficiency could also be seen as the number of circuit repetitions before a cut value above a fixed approximation ratio is observed.
Specifically, they considered the approximation ratio $\alpha$ and calculated the likelihood of observing a cut value above $\alpha C_\text{max}$ in the first $K$ samples. 
When this probability exceeded 50\%, they took $K$ as the expected number of repetitions needed to obtain the desired approximation. 
By doing so, in the training phase of the algorithm, the authors managed to reduce the execution time for MaxCut on random 3-regular graphs by two orders of magnitude compared to previous estimates reported in~\cite{guerreschiQAOAMaxCutRequires2019}. 
This was achieved by reducing the number of samples used to calculate the average approximation ratio and also by treating single samples as candidate solutions without waiting for the parameters' optimization to converge.
The performance of QAOA was also evaluated in comparison with some of the best classical alternatives, i.e., an exact solver (AKMAXSAT)~\cite{kuegelImprovedExactSolver}, an approximate solver (GW)~\cite{goemansImprovedApproximationAlgorithms1995}, and a heuristic solver (DESOUSA2013)~\cite{dunningWhatWorksBest2018}.
Thanks to the improved efficiency in the proposed parameter optimization process and approximation ratio calculation, the runtime performance of the QAOA was competitive concerning the aforementioned solvers.
However, further studies are required to assess the effectiveness and scalability of this approach on different types of graphs and larger instance sizes.

    \subsection{Quality of Solution}%
    \label{sec:efficiency}
Since its inception, there has been a persistent pursuit of establishing rigorous mathematical frameworks to analyze the performance of the QAOA and thereby provide specific theoretical bounds.
These theoretical bounds play a crucial role in understanding the capabilities and limitations of this quantum optimization algorithm.
On the one hand, they allow us to compare the performance of QAOA with classical optimization algorithms, providing a basis for evaluating the quantum advantage it may offer. 
This analysis aids in determining the feasibility and potential real-world impact of QAOA in various application domains.
On the other hand, it enables us to assess the algorithm's inherent limitations and identify scenarios where it may be suboptimal or inefficient. 
This understanding helps guide the development of alternative optimization approaches or identify areas where the algorithm can be improved. 
In the following, we will survey previous studies that explore the performance guarantees (Section~\ref{sec:solution_guarantees}) and limitations (Section~\ref{sec:performance_limitations}) of QAOA as well as some classical algorithms on different instances of optimization problems. 
Although most theoretical studies have primarily focused on low-depth QAOA analysis due to the complexity involved in higher depths, results applicable to higher depths, and even any depths, also exist.
These results are summarized in Table~\ref{tab:theoretical_performance}.
In presenting these results, we will focus on whether QAOA can outperform classical algorithms in specific problems.
Furthermore, we will highlight empirical evidence (Section~\ref{sec:empirical_evidence}) that explores the potential advantages of QAOA compared to quantum annealing or classical algorithms.

{
\renewcommand{\arraystretch}{2.5}
\begin{table}[htp!]
    \centering
    \footnotesize
    \begin{tabular}{>{\raggedright}p{0.12\linewidth}p{0.23\linewidth}p{0.3\linewidth}p{0.25\linewidth}}
	\toprule
	\textbf{Problem} & \textbf{Graph/Hypergraph} & \textbf{Algorithm} & \textbf{Performance} \\
	\midrule
	MaxCut & 2-regular & QAOA$_p$~\cite{farhiQuantumApproximateOptimization2014, wangQuantumApproximateOptimization2018, mbengQuantumAnnealingJourney2019} & $\geq \frac{2p+1}{2p+2}$ \\
    \hphantom{} & 3-regular & QAOA$_1$~\cite{farhiQuantumApproximateOptimization2014} & $\geq 0.6924$ \\[-6pt]
    \hphantom{} & \hphantom{} & QAOA$_2$~\cite{wurtzMaxCutQuantumApproximate2021} & $\geq 0.7559$ \\[-6pt]
	\hphantom{} & \hphantom{} & QAOA$_3$~\cite{wurtzMaxCutQuantumApproximate2021} & $\geq 0.7924$ \\
    \hphantom{} & $D$-regular with girth > 3 & QAOA$_1$~\cite{wangQuantumApproximateOptimization2018} & $\geq \frac{1}{2} + \frac{0.303}{\sqrt D}$ \\[-6pt]
    \hphantom{} & \hphantom{} & 1-step threshold algorithm~\cite{hastingsClassicalQuantumBounded2019} & $\geq \frac{1}{2} + \frac{0.335}{\sqrt D}$ \\
    \hphantom{} & $D$-regular with girth > 5 & QAOA$_2$~\cite{marwahaLocalClassicalMAXCUT2021} & $\geq \frac{1}{2} + \frac{0.407}{\sqrt D}$ \\[-6pt]
    \hphantom{} & \hphantom{} & 2-step threshold algorithm~\cite{barakClassicalAlgorithmsQuantum2022} & $\geq \frac{1}{2} + \frac{0.417}{\sqrt D}$ \\[-6pt]
    \hphantom{} & \hphantom{} & Any 1-local algorithm~\cite{marwahaLocalClassicalMAXCUT2021} & $\leq \frac{1}{2} + \frac{1/\sqrt 2}{\sqrt D} \approx \frac{1}{2} + \frac{0.707}{\sqrt D}$ \\
    \hphantom{} & $D$-regular with girth > $2p+1$ & $p$-local ALR algorithm~\cite{barakClassicalAlgorithmsQuantum2022} & $\geq \frac{1}{2} + \frac{2/\pi}{\sqrt D} \approx \frac{1}{2} + \frac{0.6366}{\sqrt D}$ \\[-6pt]
    \hphantom{} & \hphantom{} & QAOA$_p$~\cite{bassoQuantumApproximateOptimization2022} & $\geq \frac{1}{2} + \frac{0.6408}{\sqrt D}$ at $p=11$ \\[-6pt]
    \hphantom{} & \hphantom{} & Gaussian wave process~\cite{hastingsClassicalAlgorithmWhich2021} & $\geq \frac{1}{2} + \frac{c}{\sqrt D}$ with $c > 2/\pi$ not optimized \\
    \hphantom{} & Bipartite $D$-regular & QAOA$_p$ with $p \lesssim \epsilon\log n$~\cite{farhiQuantumApproximateOptimization2020} & $\leq \frac{1}{2} + O(\frac{1}{\sqrt D})$ \\
    SK model & Infinite size & QAOA$_p$~\cite{farhiQuantumApproximateOptimization2022} & $\geq 0.6393$ at $p = 11$ \\[-6pt]
    \hphantom{} & \hphantom{} & SDP algorithms~\cite{aizenmanRigorousResultsSherringtonKirkpatrick1987, montanariSemidefiniteProgramsSparse2016, bandeiraComputationalHardnessCertifying2020} & $\geq 2/\pi \approx 0.6366$ \\[-6pt]
    \hphantom{} & \hphantom{} & AMP algorithm (no OGP)~\cite{montanariOptimizationSherringtonKirkpatrick2021, alaouiOptimizationMeanfieldSpin2021} & $\geq (1-\epsilon)P_\text{opt}$ with $P_\text{opt}\approx 0.7632$ \\
    Max-3XOR & $D$-regular & QAOA$_1$~\cite{farhiQuantumApproximateOptimization2015} & $\geq \frac{1}{2} + O(\frac{1}{\sqrt{D}\ln D})$ \\
    \hphantom{} & $D$-regular with random signs or no overlapping constraints & QAOA$_1$~\cite{farhiQuantumApproximateOptimization2015, linPerformanceQAOATypical2016} & $\geq \frac{1}{2} + O(\frac{1}{\sqrt{D}})$ \\
    Max-$k$XOR & $D$-regular with random signs or no overlapping constraints & QAOA$_1$~\cite{marwahaBoundsApproximatingMax2022} & $\geq \frac{1}{2} + \frac{0.377}{\sqrt{D}}$ at $k = 5$ \\[-6pt]
    \hphantom{} & \hphantom{} & 1-step threshold algorithm~\cite{marwahaBoundsApproximatingMax2022} & $\geq \frac{1}{2} + \frac{0.370}{\sqrt{D}}$ at $k = 5$ \\
    \hphantom{} & $D$-regular with girth $> 2p+1$ & QAOA$_p$~\cite{bassoQuantumApproximateOptimization2022} & $\geq \frac{1}{2} + O(\frac{1}{\sqrt{D}})$ \\
    MIS & Random graphs with bounded average degree & QAOA$_p$ with $p \lesssim \epsilon\log n$~\cite{farhiQuantumApproximateOptimization2020a} & $\leq 0.854$ \\
    Diluted $k$-spin model & Even $k\geq 4$ & QAOA$_p$ with $p \lesssim \epsilon\log n$~\cite{chouLimitationsLocalQuantum2022} & $< P_\text{opt}(k)$ (bounded away from optimality due to OGP) \\
    Fully connected $k$-spin model & Even $k\geq 4$ & QAOA$_p$~\cite{bassoPerformanceLimitationsQAOA2022} & $< P_\text{opt}(k)$ (bounded away from optimality due to OGP) \\
	\bottomrule
    \end{tabular}
    \caption{Summary of established theoretical bounds on the performance of QAOA and selected classical algorithms on optimization problems.}
    \label{tab:theoretical_performance}
\end{table}
}

\subsubsection{Solution guarantees}%
\label{sec:solution_guarantees}

In the original QAOA proposal by \citet{farhiQuantumApproximateOptimization2014}, the authors focused on applying QAOA of fixed depth $p$, denoted as QAOA$_p$ for convenience, to the MaxCut problem on 2-regular and 3-regular graphs. 
On 2-regular graphs, also known as the ring of disagrees, they conjectured that on rings of size $n > 2p$, QAOA$_p$ can achieve $\alpha \geq (2p+1)/(2p+2)$, which was later proved by \citet{mbengQuantumAnnealingJourney2019}.
This means that at large $p$, QAOA can produce solutions arbitrarily close to the true optima on the ring of disagrees. 
On 3-regular graphs, they focused on the single-layer QAOA, QAOA$_1$. 
They proved that with the optimal set of angles, QAOA$_1$ on any 3-regular graph will always produce a cut whose size is at least 0.6924 times the size of the optimal cut, which translates to an approximation ratio $\alpha \geq 0.6924$, an improvement from the average outcome of random guessing, which yields $\alpha = 0.5$.
However, generalizing their analysis to higher values of $p$ becomes challenging due to the double-exponential growth of the complexity of the classical algorithm used to find optimal parameters, which scales as $O(2^{2^p})$.

\citet{wurtzMaxCutQuantumApproximate2021} subsequently extended the worst-case guarantees on 3-regular graphs to QAOA$_2$ and QAOA$_3$.
They found that for $p=2$, $\alpha \geq 0.7559$ on 3-regular graphs with girth $> 5$. 
The girth of a graph is the length of its shortest cycle.
Therefore, girth $> 5$ means a graph contains no triangles, squares, or pentagons.
For $p=3$, $\alpha \geq 0.7924$ on graphs with girth $> 7$.
The girth requirements stem from the large loop conjecture, which was one of the bases of their analysis. 
This conjecture posits that the worst-case graphs for fixed $p$ are $p$-trees, which have no cycles less than $2p+2$.
While this study did confirm the theoretical performance increase with an increasing $p$, it also showed that up to $p=3$, QAOA is not able to outperform the best classical algorithms on 3-regular graphs in the worst-case scenario, such as the GW algorithm~\cite{goemansImprovedApproximationAlgorithms1995} that gives a performance guarantee of 0.8786 on any type of graph, and another classical algorithm based on Semidefinite Programming (SDP)~\cite{halperinMAXCUTCubic2004} that achieves at least 0.9326 on graphs of maximum degree 3.

Another extension to~\cite{farhiQuantumApproximateOptimization2014} was provided by \citet{wangQuantumApproximateOptimization2018}. 
They showed that on any triangle-free $D$-regular graphs, where $D$ denotes the degree, the optimized QAOA$_1$ could provide an approximation ratio of at least $\frac{1}{2} + \frac{c}{\sqrt{D}}$ with $c\approx 0.303$, which outperforms the lower bound of the
best known local classical algorithm at the time, the threshold algorithm, which gives $c \approx 0.281$~\cite{hirvonenLargeCutsLocal2017}.
However, \citet{hastingsClassicalQuantumBounded2019} soon argued that this bound is not tight and showed through a direct calculation that the 1-step threshold algorithm with the optimal parameter exhibits an immense improvement over random assignment compared to QAOA$_1$ in the asymptotic limit as $D\to\infty$, i.e., $\alpha \geq \frac{1}{2} + \frac{c}{\sqrt D}$ with $c\approx 0.335$.
Moreover, it was pointed out that the threshold algorithm belongs to a broader class of local algorithms called the local tensor algorithms.
Numerical calculations revealed that for all $2\leq D < 1000$, a single step of this classical algorithm would outperform QAOA$_1$ on all triangle-free MaxCut instances.
Later, \citet{marwahaLocalClassicalMAXCUT2021} generalized the 1-step threshold algorithm to an $n$-step version with $n$ parameters.
In particular, they first showed by direct calculation that for degrees $2 \leq D < 500$ and every $D$-regular graph $G$ of girth $> 5$, while QAOA$_2$ has a larger expected cut fraction than QAOA$_1$ on $G$, there exists a 2-local classical MaxCut algorithm outperforms QAOA$_2$ for all $G$.
They concluded that this is likely to hold for all $D$ since the coefficient in the improved fraction $c$ for the 2-step threshold algorithm stabilizes at around 0.417 in the asymptotic limit, compared with that of QAOA$_2$, $c \approx 0.407$.

One may wonder if an optimal value exists for the maximum cut on random $D$-regular graphs.
Interestingly, insights into this question emerge from another problem known as the Sherrington-Kirkpatrick (SK) model, which serves as a mean-field model for spin glass in physics.
This model describes a classical spin system with all-to-all couplings. 
For $n$ spins, the Hamiltonian, or the cost function of the SK model, is given by
\begin{equation}
    C(\vb z) = \frac{1}{\sqrt n}\sum_{i < j} J_{ij} z_i z_j,
\end{equation}
where $z_i\in \{-1, 1\}$, $\forall i$, and each $J_{ij}$ is independently chosen from a distribution with mean 0 and
variance 1.
Making use of the so-called replica trick, \citet{parisiInfiniteNumberOrder1979} first computed the ground state energy of the SK model in the infinite-size limit, that is,
\begin{equation} \label{eq:parisi_val}
    \lim_{n\to\infty}\min_{\vb z} \frac{C(\vb z)}{n} \simeq 0.7632.
\end{equation}
\citet{farhiQuantumApproximateOptimization2022} first conducted an extensive study on the performance of QAOA in solving the SK model in the limit of infinite problem size and the comparison with the classical algorithm based on SDP.
Although recently a classical Approximate Message Passing (AMP) algorithm~\cite{montanariOptimizationSherringtonKirkpatrick2021, alaouiOptimizationMeanfieldSpin2021} was proposed to efficiently find an approximate solution that is asymptotically close to the true optimum of the SK model, i.e., the Parisi value in Eq.~\eqref{eq:parisi_val}, the analysis relies on the assumption of no Overlap Gap Property (OGP)~\cite{gamarnikOverlapGapProperty2021a}.
The OGP refers to the geometric structure of near-optimal solutions for a problem.
It implies that the overlap between any two nearly optimal solutions does not take values in a certain nontrivial interval; the overlap is either big or small, and there is no middle ground.
The best-performing classical algorithm known to date without such assumption is the one based on SDP, the ALR algorithm~\cite{aizenmanRigorousResultsSherringtonKirkpatrick1987} and its variants~\cite{montanariSemidefiniteProgramsSparse2016, bandeiraComputationalHardnessCertifying2020}, which yield $C/n = 2/\pi \approx 0.6366$.
In~\cite{farhiQuantumApproximateOptimization2022}, the authors introduced a formula for calculating the typical-instance energy of QAOA applied to the SK model for given parameters at an arbitrary fixed $p$, which is denoted by $V_p(\bs\gamma, \bs\beta)$, where $\bs\gamma$ and $\bs\beta$ are parameters of the QAOA circuit.
This formula can be evaluated on a classical computer with a run time complexity $O(16^p)$.
In this study, they numerically computed the QAOA performance for $1\leq p \leq 12$, from which they found that the performance with optimal parameters, $\bar V_p$, monotonically increases with an increasing $p$.
While for $9 \leq p \leq 12$, the values of $V_p(\bs\gamma, \bs\beta)$ were not optimized but evaluated at some parameters extrapolated from those at lower $p$, which can be seen as lower bounds of $\bar V_p$, it was found that at $p = 11$, QAOA already outperforms the classical SDP algorithm mentioned above, where $\bar V_{11} \geq 0.6393 > 2/\pi$.
At $p = 12$, QAOA produces $\bar V_{12} \geq 0.6466$.
They also demonstrated the concentration property, that is, with probability tending to one in the infinite-size limit, measurements of the QAOA circuit will produce strings whose energies concentrate at the calculated value $V_p(\bs\gamma, \bs\beta)$.
This implies landscape independence~\cite{brandaoFixedControlParameters2018}, meaning that optimal parameters found for one large instance will also be good for other large instances.

Returning to the question of optimal value for the MaxCut problem on random $D$-regular graphs, \citet{demboExtremalCutsSparse2017} addressed it by revealing the connection between MaxCut and the SK model.
They proved that the maximum cut of a random $D$-regular graph tends to $\frac{1}{2} + \frac{P_*}{\sqrt D}$ as $D\to\infty$, where $P_* \approx 0.7632$ is exactly the Parisi value, i.e., the ground state energy of the SK model.
Recently, an enhancement in the algorithmic performance on the MaxCut problem was presented by \citet{barakClassicalAlgorithmsQuantum2022}.
The authors considered classical $p$-local algorithms, where a vertex’s assignment depends on its radius $p$ neighborhood.
It was proved that for every $p$, there exists a polynomial-time classical $p$-local algorithm for all $D$-regular graphs of girth $> 2p+1$ that outputs a cut value lower bounded by $\frac{1}{2} +\frac{c}{\sqrt D} - O(\frac{1}{\sqrt p})$, where $c = 2/\pi\approx 0.6366$.
In the limit of large degree and girth, the cut fraction tends to $\frac{1}{2} +\frac{2/\pi}{\sqrt D}$, therefore surpassing that given by QAOA$_1$ and QAOA$_2$~\cite{hastingsClassicalQuantumBounded2019, marwahaLocalClassicalMAXCUT2021}.
It is worth noting that in this limit, the performance guarantee is the same as that of the SDP algorithms for the SK model~\cite{aizenmanRigorousResultsSherringtonKirkpatrick1987, montanariSemidefiniteProgramsSparse2016, bandeiraComputationalHardnessCertifying2020}. 
Nevertheless, neither of these two classical algorithms achieves the optimal Parisi value.

Can QAOA at larger depths match or outperform this classical algorithm on the MaxCut problem?
\citet{boulebnanePredictingParametersQuantum2021} took an essential step towards addressing this question by establishing an exact expression for the QAOA$_p$ energy on sparse random $D$-regular graphs in the infinite-size limit for any $p$.
In the limit $D\to\infty$, this expression is closely connected to an analogous expression for the SK model~\cite{farhiQuantumApproximateOptimization2022}, allowing parameters obtained for one model to be mapped to the other.
Therefore, in this limiting case, the optimal QAOA parameters can be determined based on the techniques proposed in~\cite{farhiQuantumApproximateOptimization2022} with a time complexity that is exponential in the number of QAOA layers but not in the problem size.
The authors additionally proposed an efficient Monte Carlo algorithm for estimating the QAOA energy on finite-size instances of the SK model. However, its applicability is limited to low depths, i.e., $p \leq 3$.
Around the same time, \citet{bassoQuantumApproximateOptimization2022} also derived an iterative formula for evaluating the QAOA performance on any large-girth $D$-regular graphs for any fixed $p$.
While solving this iteration for any finite $D$ requires a time complexity of $O(p 16^p)$, in the $D\to\infty$ limit, the complexity reduces to $O(p^2 4^p)$.
Therefore, the authors could perform numerical evaluations up to $p = 20$ in the large $D$ limit.
It was discovered that at $p = 11$ and beyond, QAOA achieves a cut fraction better than $\frac{1}{2} + \frac{2/\pi}{\sqrt D}$, beating the classical local algorithm for MaxCut
on large-girth random regular graphs considered by \citet{barakClassicalAlgorithmsQuantum2022}.
This result is reminiscent of the finding made by \citet{farhiQuantumApproximateOptimization2022} for the infinite-size SK model, where QAOA$_{11}$ and beyond also outperform the classical algorithms yielding a ground-state energy estimation of $2/\pi$.
Indeed, \citet{bassoPerformanceLimitationsQAOA2022} showed that the iterative formula for MaxCut on large-girth regular graphs also gives the ensemble-averaged performance of QAOA on the SK model defined on the complete graph.
Hence, these two observations are not coincidental.
The authors also conjectured that as $p \to \infty$, QAOA will be able to achieve the Parisi value on random $D$-regular graphs as $D\to\infty$, and if this were true, it could optimally solve the SK model, too.
However, this remains to be rigorously proven.
On the other hand, \citet{hastingsClassicalAlgorithmWhich2021} demonstrated that a simple modification of the classical algorithm, the Gaussian wave process, can also achieve some other $c > 2/\pi$ in the improved cut fraction, but $c$ was not optimized in this work.
Therefore, it is still unclear if QAOA can outperform any classical algorithm for MaxCut on $D$-regular graphs.

Another class of optimization problem that has garnered significant attention in the field is the Constraint Satisfaction Problems (CSPs).
These problems involve assigning values to variables from a given domain while adhering to constraints.
For example, the combinatorial problem Max-E3LIN2 over $n$ Boolean variables (bits) consists of a set of clauses, each containing exactly three problem variables (E3), where a clause is deemed satisfied if its variables sum to 0 or 1 modulo 2 (LIN2), representing the parity as even or odd.
The objective is to find an assignment of the variables that maximizes the number of satisfied constraints. 
Additionally, if each variable is guaranteed to appear in no more than $D$ clauses, it is an instance of bounded degree $D$.
This problem belongs to a class of CSPs known as Max-$k$XOR, where each clause represents the XOR operation performed on $k$ variables or their negations.
Therefore, Max-E3LIN2 is equivalent to Max-3XOR, while the MaxCut problem can be seen as a special case of Max-2XOR, where all the clauses exhibit even parity.
\citet{farhiQuantumApproximateOptimization2015} considered solving the Max-3XOR problem of bounded degree with QAOA$_1$.
They proved that for any instance, QAOA$_1$ could achieve an average fraction of satisfied constraints of at least $\frac{1}{2} + O(\frac{1}{\sqrt D \ln D})$.
When applied to ``typical'' instances with random signs, i.e., the parity of each constraint is randomly chosen with equal probability, the scaling of the satisfied fraction improves to $\frac{1}{2} + O(\frac{1}{\sqrt D})$.
This matches that of a classical algorithm on both the set of instances with random signs and those with ``triangle-free'' constraints (also known as no overlapping constraints), where any two variables are involved in at most one constraint, and they share no neighbors outside of that specific constraint~\cite{barakBeatingRandomAssignment2015}.
\citet{linPerformanceQAOATypical2016} extended the analysis of the performance of QAOA$_1$ to general CSPs with typicality, including Max-$k$XOR and Max-$k$SAT.
Max-$k$SAT is similar to Max-$k$XOR but differs in that each constraint represents the logical OR operation applied to $k$ variables or their negations.
They demonstrated that for such problems with bounded degree, QAOA$_1$ could efficiently find, on average, an assignment that satisfies $\mu + O(\frac{1}{\sqrt D})$ fractions of constraints, where the constant $\mu$ denotes the expected fraction of constraints satisfied by a random assignment.
Moreover, it was shown that on triangle-free instances, QAOA could also give an advantage of $O(\frac{1}{\sqrt D})$ over a random assignment.

In a subsequent study, \citet{marwahaBoundsApproximatingMax2022} pointed out that a 1-local algorithm such as QAOA$_1$ on CSPs of bounded degree cannot distinguish between the instances of random signs and those of triangle-free constraints, which explains the same performance scaling in these two settings.
Specifically, the authors compared the performance of QAOA$_1$ and a generalization of the classical threshold algorithm~\cite{hirvonenLargeCutsLocal2017} on triangle-free Max-$k$XOR problems of bounded degree $D$.
While the previous studies~\cite{barakBeatingRandomAssignment2015, linPerformanceQAOATypical2016} did not focus on finding the best possible constant $c$ in the improved satisfying fraction $\frac{c}{\sqrt D}$, in this study, they numerically optimized and evaluated that of both algorithms at every $k < 200$, for each $D < 300$ and when $D\to \infty$.
At $k = 2$, the results verified the asymptotic performance for MaxCut (a special case of Max-2XOR)~\cite{wangIntroductionQuantumOptimization2018, hastingsClassicalQuantumBounded2019}, in which the threshold algorithm outperforms QAOA$_1$.
On Max-3XOR, while QAOA$_1$ beats the threshold algorithm for some values of $D \leq 27$, as $D$ increases, the threshold algorithm performs better.
This observation is also reflected in their asymptotic performances as $D \to \infty$.
Importantly, they found that when $k > 4$, QAOA$_1$ starts outperforming the threshold algorithm in the large degree limit, demonstrating a quantum advantage.
However, this does not rule out the possibility that a different local tensor algorithm~\cite{hastingsClassicalQuantumBounded2019} will match or outperform QAOA at larger $k$.

More recently, efforts have been made to analyze the performance of higher-depth QAOA on Max-$k$XOR.
For instance, \citet{bassoQuantumApproximateOptimization2022} showed that the iterative formula they developed to evaluate the performance of QAOA at any $p$ on MaxCut could be easily generalized to Max-$k$XOR instances with large-girth regular hypergraphs.
They effectively employed this formula to identify optimal QAOA parameters and assess performance for $3\leq k \leq 6$ and $1\leq p \leq 14$, showing that QAOA is capable of producing solutions for Max-$k$XOR problems that approach the true optima more closely as $p$ increases.
Building upon this progress, a subsequent study by \citet{bassoPerformanceLimitationsQAOA2022} further extended the established equivalence between MaxCut and the SK model (a 2-spin model)~\cite{demboExtremalCutsSparse2017}. 
They generalized this equivalence to encompass the relationship between Max-$k$XOR and the fully connected $k$-spin model, which characterizes ensembles of combinatorial optimization problems with random all-to-all $k$-body couplings.
They showed that QAOA's performance at any constant depth $p$ for the $k$-spin model matches the ones for Max-$k$XOR asymptotically on random sparse Erd\H{o}s-R\'{e}nyi hypergraphs and large-girth $D$-regular hypergraphs in the $D\to\infty$ limit.
Given such equivalence, it would be interesting to see if the state-of-the-art classical algorithm for the $k$-spin models, which is the AMP algorithm~\cite{montanariOptimizationSherringtonKirkpatrick2021, alaouiOptimizationMeanfieldSpin2021}, can outperform QAOA at larger $k$, primarily when $k\geq 4$ and is even, where the OGP is known to exist and AMP is bounded away from optimality~\cite{gamarnikOverlapGapProperty2021}.

In addition to optimization problems, theoretical analysis was conducted to evaluate QAOA's performance in a Boolean satisfaction problem, random $k$-SAT.
Like Max-$k$SAT, random $k$-SAT comprises a set of clauses, each being a randomly generated Boolean formula involving $k$ variables.
However, the goal for $k$-SAT is to find an assignment that exactly satisfies all the clauses.
Therefore, the performance of QAOA is measured by the success probability that the algorithm outputs a satisfying assignment.
Since QAOA is repeatedly run until satisfiability is reached, a success probability $p_\text{succ}$ is translated to an expected running time $1/p_\text{succ}$.
Based on a technique for estimating the ``generalized multinomial sums'', \citet{boulebnaneSolvingBooleanSatisfiability2022} derived analytical bounds on the average success probability of QAOA on random $k$-SAT in the limit of infinite problem size, which holds for fixed, sufficiently small QAOA parameters and when $k$ is a power of 2.
In particular, the authors showcased the performance of QAOA on random 8-SAT instances.
They computed the analytical bounds for QAOA with up to $p = 10$.
Moreover, numerical calculations of the median, running time, and the inverse of the success probability were performed for up to $p = 60$.
Both analytical and numerical results suggested that QAOA with a depth $p\approx 14$ would match the performance of a random-local-search-based classical algorithm, WalkSATlm~\cite{caiImprovingWalkSATEffective2015}, with an estimated running time $\lesssim 2^{0.33n}$, where $n$ is the number of problem variables.
QAOA is expected to outperform WalkSATlm with larger numbers of layers. However, the extent of the advantage is still unclear because numerical estimates of the median running for small instances ($12 \leq n \leq 20$) starts to deviate significantly from the theoretical and numerical results based on the success probability as $p$ gets large.

\subsubsection{Performance limitations}
\label{sec:performance_limitations}

In order to understand the full capability of QAOA and to determine the circumstances in which it surpasses classical algorithms, it is equally crucial to explore its performance limitations as it is to understand its solution guarantees.

\citet{bravyiObstaclesVariationalQuantum2020} first realized a limitation of fixed-depth QAOA in solving the MaxCut problem on bipartite $D$-regular graphs.
For graphs with $n$ vertices, they proved that the approximation ratio of QAOA$_p$ is at most $\frac{5}{6} + \frac{\sqrt{D-1}}{3D} \sim \frac{5}{6} + O(\frac{1}{\sqrt D})$ for any constant $D \geq 3$, as long as $p<(\frac{1}{3}\log_2n - 4)/(D+1)$. 
As the degree $D$ gets large, this bound approaches $\frac{5}{6}\approx 0.833$, which falls behind the GW bound 0.8786, indicating that QAOA, in this case, cannot outperform the best classical algorithm.
As pointed out in~\cite{marwahaBoundsApproximatingMax2022}, such obstruction in the $O(\log n)$-depth regime is related to the No Low-energy Trivial States (NLTS) conjecture~\cite{freedmanQuantumSystemsNonkhyperfinite2014}.
The NLTS conjecture suggests the existence of families of local Hamiltonians whose low energy states are all nontrivial, whereas a ``trivial'' one can be achieved by evolving a product state $\ket S$ with a low-depth quantum circuit $U$. 
\citet{bravyiObstaclesVariationalQuantum2020} showed that when the local Hamiltonians correspond to the MaxCut instances, both the initial state $\ket{S} = \ket{+}^{\otimes n}$ and the associated QAOA circuit $U$ possess a $\mathbb Z_2$ symmetry, i.e., a global spin flip.
This symmetry property gives rise to the NLTS property, resulting in the nontrivial ground states and leading to the observed obstruction.
\citet{marwahaBoundsApproximatingMax2022} later generalized this obstruction from Max-2XOR (MaxCut) to Max-3XOR instances.
Although the QAOA circuits for Max-$k$XOR at odd $k$ lack the global $\mathbb Z_2$ symmetry, the authors identified partial $\mathbb Z_2$ symmetry within these instances, that is, the corresponding unitary has $\mathbb Z_2$ spin-flip symmetry only concerning some large, fixed subset of vertices $V_+ \varsubsetneq V$.
Notably, such partial symmetry leads to a constant-fraction obstruction for QAOA at sub-logarithmic depth in solving Max-3XOR.
The study established an upper bound on the satisfying fraction of $0.99 + O(\frac{1}{\sqrt{D}})$ in the large $D$ limit. 
However, the authors did not focus on optimizing the constant in this analysis.
Furthermore, it was conjectured that such constant-fraction obstruction exists for some instances of Max-$k$XOR at every $k$ when QAOA operates at sub-logarithmic depths.
Numerically, they evaluated the performance of QAOA$_1$ and the local threshold algorithm on triangle-free, large $D$ Max-$k$XOR instances for $k$ up to 200.
They found that both algorithms are bounded away from the optimum, a satisfying fraction $\frac{1}{2} + \frac{P(k)}{2}\sqrt{\frac{k}{D}}$, with $P(k)$ being the generalized Parisi value for Max-$k$XOR.

Another well-known limitation of QAOA stems from the locality constraint.
For MaxCut on $D$-regular graphs of girth $>5$, \citet{barakClassicalAlgorithmsQuantum2022} proved that every 1-local algorithm, quantum or classical, can only produce a maximum cut of at most $\frac{1}{2} + \frac{c}{\sqrt D}$, where $c = 1/\sqrt 2\approx 0.7071$.
This result falls short of the true optimum, which is $\frac{1}{2} + \frac{P_*}{\sqrt D}$ with $P_* \approx 0.7632$~\cite{demboExtremalCutsSparse2017}.
More generally, when the algorithm is run at a fixed depth $p$, the measurement outcomes of a particular qubit depend solely on the qubits within its $p$-neighborhood, that is, qubits that are within distance $p$ to the given qubit on the graph.
Consequently, if these neighborhoods are small, QAOA does not ``see'' the whole graph and, in some cases, becomes limited in its algorithmic performance.
In light of this, \citet{farhiQuantumApproximateOptimization2020} refined the bound in~\cite{bravyiObstaclesVariationalQuantum2020} by utilizing the property that local neighborhoods of random $D$-regular graphs resemble trees.
It was proved that when QAOA does not see the whole graph, i.e., when $p \lesssim \epsilon\log n$ with $\epsilon$ being a small constant, the upper bound on bipartite random $D$-regular graphs is $\frac{1}{2} + O(\frac{1}{\sqrt D})$.
In another study by \citet{farhiQuantumApproximateOptimization2020a}, the focus was shifted to the Maximum Independent Set (MIS) problem on sparse random graphs with average degree $D$.
They proved that in the low-depth regime $p \lesssim \epsilon\log n$, QAOA fails to produce an independent set larger than 0.854 times the optimal as $D\to\infty$.
The proof uses the OGP exhibited by the large independent sets of random graphs with bounded average degree~\cite{gamarnikLimitsLocalAlgorithms2017}. 
The OGP has been identified as an obstruction for various classical algorithms, ranging from local algorithms~\cite{gamarnikLimitsLocalAlgorithms2017, gamarnikPerformanceSequentialLocal2017, chenSuboptimalityLocalAlgorithms2019} to AMP algorithms~\cite{gamarnikOverlapGapProperty2021}.
This result establishes that the OGP also obstructs QAOA, limiting its performance in generating large independent sets.
Moreover, in the ``worst'' case where QAOA is applied to the MIS on bipartite random $D$-regular graphs, the approximation ratio approaches 0 at large $D$~\cite{farhiQuantumApproximateOptimization2020}.
\citet{chouLimitationsLocalQuantum2022} generalized this obstruction-by-OGP result on the MIS to random sparse instances of CSPs.
In particular, they focused on the diluted $k$-spin glass model, where the interactions are sampled from a random sparse $k$-uniform hypergraph.
They demonstrated that the OGP exhibited by diluted $k$-spin glasses, for every even $k \geq 4$ as previously shown by \citet{chenSuboptimalityLocalAlgorithms2019}, poses an obstacle for QAOA at sub-logarithmic depths.
Building on the relation to the mean-field $k$-spin glasses, they further extended this obstruction to the signed random Max-$k$XOR instances when $k\geq 4$ is even, where each variable in all clauses carries by a random sign.

What happens when QAOA can see the whole graph?
\citet{bassoPerformanceLimitationsQAOA2022} tried to address this question by studying the fully connected $k$-spin models, where the locality-based arguments used in previous studies of sparse instances~\cite{farhiQuantumApproximateOptimization2020, farhiQuantumApproximateOptimization2020a, chouLimitationsLocalQuantum2022} no longer apply.
Exploiting the equivalence of the performance of QAOA on dense and sparse graphs in the asymptotic limit, the authors established that any constant-$p$ QAOA is still bounded away from optimality on dense $k$-spin models for any even $k \geq 4$ due to the obstruction by the OGP.
This finding reveals a hardness of approximation for QAOA in a new regime where the whole graph is seen.
However, it is essential to note that potential quantum advantage is still possible in these problems, particularly when QAOA surpasses sub-logarithmic depths, as classical AMP algorithms have also exhibited suboptimality~\cite{gamarnikOverlapGapProperty2021}.

\subsubsection{Empirical evidence}%
\label{sec:empirical_evidence}


While theoretical analysis of QAOA has shown its potential to outperform classical optimization algorithms in certain problems (Sections~\ref{sec:solution_guarantees} and \ref{sec:performance_limitations}), it is equally important to examine its empirical performance to assess its practical utility. 
Empirical investigations of QAOA's performance become more relevant at higher depths as obtaining rigorous theoretical bounds for a wide range of problem instances becomes challenging.
Moreover, these empirical studies can provide insights into areas of the quantum algorithm that could be improved to enhance performance.

\paragraph{Simulator-based studies}
\citet{crooksPerformanceQuantumApproximate2018} was among the first to conduct an analysis comparing the performance of QAOA on a quantum circuit simulator with a classical solver to look for a quantum advantage. 
The empirical study showed that the QAOA optimized via stochastic gradient descent could achieve an approximation ratio exceeding that of the classical GW algorithm, even with a modest circuit depth. 
By considering 10-node random graphs generated from the Erd\H{o}s-R\'{e}nyi configuration with a 50\% edge probability, the author measured the average approximation ratio of QAOA for up to $p=32$ on the MaxCut problem on these graphs.
Following the theory~\cite{farhiQuantumApproximateOptimization2014}, the average approximation ratio obtained by QAOA monotonically improves as the number of layers increases; at $p=6$, QAOA starts to outperform the classical GW algorithm.
Furthermore, the author also compared the performance of both algorithms at various problem sizes, ranging from $n=6$ to $n=17$, where $n$ is the number of nodes or vertices.
It was observed that both algorithms maintained their performance even as the graph size increased, suggesting that QAOA's solution quality, at a fixed circuit depth, remains insensitive to problem size.
Notably, for $p \geq 8$, the quantum algorithm consistently outperforms the GW algorithm across different problem sizes, demonstrating a potential quantum advantage.
\citet{lotshawEmpiricalPerformanceBounds2021} later conducted a more comprehensive study by investigating an exhaustive set of MaxCut instances on all connected non-isomorphic graphs with $n \leq 9$ vertices.
However, the depth of QAOA was limited to $p \leq 3$.
In addition to the approximation ratio, they introduced another measure of performance---the probability of obtaining the optimal solution.
As the QAOA depth and the graph size increased, they observed a convergence of the approximation ratio across different graph structures, resulting in a narrower distribution.
Moreover, it was found that on most graphs with $n \leq 9$, QAOA exceeds the worst-case GW bound even by $p = 3$, consolidating the viability of modest-depth QAOA to outperform the classical algorithm in many instances.
Interestingly, while the average probability of obtaining an optimal solution increased with larger $p$, the distributions of this probability widened with increasing $p$.
This contrasts the distributions of the approximation ratio, indicating that the probability of success is more sensitive to the graph structure.

On the other hand, as the QAOA depth increases, a significant challenge arises from the growing complexity of the algorithm, particularly the amount of entanglement in the QAOA circuit. 
A recent study on the $p=1$ QAOA indicated that excessive entanglement might hinder the algorithm's performance on problems such as the Hamming ramp and bush of implications~\cite{mccleanLowDepthMechanismsQuantum2021}. 
In another study by \citet{chenHowMuchEntanglement2022}, the authors pointed out that removing excess entanglement introduced by intermediate layers of the QAOA circuit might yield improved outcomes on the MaxCut problem. 
QAOA circuits with large depths can also create an entanglement barrier between the initial and final states, complicating both their classical simulation and subsequent benchmarks \cite{dupontCalibratingClassicalHardness2022,dupontEntanglementPerspectiveQuantum2022}.
To this end, \citet{sreedharQuantumApproximateOptimization2022} performed numerical simulations of QAOA that restrict the allowed entanglement in the circuit by using the Matrix Product States (MPS) with reduced bond dimensions.
The bond dimension acts as a parameter bounding the system entanglement.
Utilizing layer-wise optimization, they could extend the analysis to high depths, up to $p = 100$.
They also employed a deterministic method to sample only one final bitstring of the algorithm.
Under such a restricted simulation, QAOA still provided successful results for small bond dimensions (comparable to the system size), for $p \approx 30$ and up to 60 qubits.
Based on their findings, they concluded that entanglement plays a minor role in solving the MaxCut and Exact Cover 3 problems studied: provided the depth $p$ of the algorithm is sufficiently high, QAOA can solve the optimization problem exactly or approximately even when the bond dimension is small. 
Therefore, even if high-depth QAOAs have an entanglement barrier that inhibits the classical simulability of the algorithm~\cite{chenHowMuchEntanglement2022, dupontEntanglementPerspectiveQuantum2022}, such entanglement might have little impact on QAOA's ability to find optimal solutions.
However, the interplay between entanglement and circuit depth and their impact on the QAOA performance in other problem instances remains an open research problem.

Besides its depth, parameter initialization is also crucial for achieving an advantage with QAOA, as reported by \citet{zhouQuantumApproximateOptimization2020}. 
They proposed heuristic strategies to optimally initialize QAOA's parameters, as discussed in Section~\ref{sec:parameter_initialisation}. 
They discovered that by employing these strategies, quasi-optimal parameters for QAOA at depth $p$ can be determined in $O[\text{poly}(p)]$ time. 
In contrast, random initialization would necessitate $2^{O(p)}$ optimization runs to attain comparable performance.
The researchers evaluated the performance of QAOA using these optimized parameter values for up to $p = 50$. 
They observed that, on average, the approximation ratio obtained by QAOA improved exponentially (or stretched exponentially) when applied to random graphs.
They also compared the performance of QAOA$_3$ initialized with an interpolation-based heuristic strategy INTERP to that of quantum annealing, showing that QAOA was able to converge to a better solution even on difficult instances where adiabatic quantum annealing failed due to small spectral gaps. 
This was attributed to QAOA's ability to use non-adiabatic mechanisms to overcome the challenges of vanishing spectral gaps.

In addition, \citet{akshayReachabilityDeficitsQuantum2020} showed that the density of the graph also plays a vital role in QAOA’s final result. 
Given a constraint satisfiability problem with $n$ variables and $m$ clauses (constraints), its density $\alpha_d$ is defined as the clause to variable ratio, i.e., $\alpha_d = m/n$. 
For any fixed ansatz, there seem to be high-density instances that are inaccessible to the quantum algorithm, thus limiting its performance. 
From empirical investigation, it turned out that higher-depth versions of QAOA are necessary for achieving satisfactory results for densities $\alpha_d > 1$.
Subsequently, \citet{herrmanImpactGraphStructures2021} extensively evaluated the impact of various graph characteristics on the performance of QAOA with up to $p=3$ for the MaxCut problem. 
The authors investigated all connected non-isomorphic graphs with at most eight vertices and identified some exciting predictors of QAOA's success, such as graph symmetries, odd cycles, and density. 
Despite the limited scope of the investigation, the authors found that graphs without odd cycles have a 12\% higher mean probability of obtaining an optimal solution than those with odd cycles. 
Moreover, the number of edges, clique size, and small odd cycles are positively correlated with QAOA's success and bipartite, Eulerian, and distance regular graphs. 
On the other hand, the diameter of the graph has a negative correlation with the expected cost $F_p(\bs{\gamma},\bs{\beta})$ and, therefore, the approximation ratio. 
These correlations between graphs’ structure and QAOA's performance can be used to identify problem instances where the quantum algorithm is likely to exhibit a quantum advantage.

When working with QAOA, as well as any other quantum algorithm, an important question to consider is when it is advantageous to use it over classical algorithms: in an algorithm selection scenario, it is crucial to thoroughly evaluate factors such as problem size, structure, and available quantum resources before deciding whether to use QAOA or a classical solver.
\citet{lykovSamplingFrequencyThresholds2022} aimed to identify under which conditions QAOA could achieve quantum advantage over classical algorithms, in terms of both quality of solution and runtime performance.
Specifically, inspired by parameter transferability, they adopted a fixed-angle approach similar to the ones proposed in Refs.~\cite{galdaTransferabilityOptimalQAOA2021,streifTrainingQuantumApproximate2020,shaydulinQAOAKitToolkitReproducible2021} such that with just one round of circuit sampling, one could speed up QAOA while maintaining good performance compared to slower conventional approaches.
However, their analysis indicated that multi-shot circuit sampling was necessary after all to match the classical solution quality.
They observed that classical heuristic solvers were capable of producing high-quality approximate solutions in linear time complexity, which is very difficult to beat. 
According to their results, the main obstacle to QAOA’s advantage over classical solvers is the exponential sampling time required for large graph sizes.
Therefore, even if an experiment might demonstrate an advantage for intermediate values of $n$ and $p$, such advantage would be lost with larger problems, independently of the rate of quantum circuit sampling. 
They suggested that a QAOA circuit must be implemented with depth $p > 11$ to match the performance of classical algorithms for large graph sizes and hope to see a quantum advantage.
In another study, \citet{moussaQuantumNotQuantum2020} used ML techniques to detect MaxCut problem instances where QAOA is most likely to obtain a quantum advantage over the classical GW algorithm regarding approximation ratio.
The proposed ML model achieved up to 96\% cross-validated accuracy in predictions.
The capacity to predict scenarios where quantum solutions are likely to surpass classical approaches, allows for strategic allocation of computational resources, thus establishing a methodological foundation for quantum advantage. 
It was also shown that QAOA outperformed GW on most instances of 4-regular graphs up to 24 nodes at depth $p = 10$. 
This result partially corroborates the assertions made by \citet{crooksPerformanceQuantumApproximate2018} a few years prior, albeit under different experimental conditions. 
Despite the absolute validity of such a finding, which holds for the specific parameters and scenarios studied, the possibility of divergent outcomes under different conditions or with varying parameters cannot be discounted.
This emphasizes the inherent complexity in analyzing QAOA's performance, which is susceptible to variations in multiple factors. 
Nevertheless, through ML and explainability methods, \citet{moussaQuantumNotQuantum2020} were able to demonstrate that spectral properties of the graph and basic graph density were the most influential graph features affecting the approximation ratio, deepening what was found in~\cite{zhouQuantumApproximateOptimization2020}.
Similarly, \citet{deshpandeCapturingSymmetriesQuantum2022} employed GNNs as a tool to choose whether to use QAOA or a classical solver for MaxCut instances based on their approximation ratios. 
Specifically, they used a GNN trained on \cite{ORNLQuantumComputing2021} to predict QAOA’s performance, with a relative error of less than 19.7\%.
This could allow for a comparison between the performance of QAOA and that of a classical solver, enabling the choice of the most appropriate solver for a particular optimization problem.

\paragraph{Quantum hardware-based evaluations}
In order to outperform classical solvers, the QAOA must demonstrate optimal performance on real quantum hardware, as opposed to just noiseless simulations.
Most of the works in the literature proposed a QAOA implementation on a superconducting hardware platform, the prevalent technology for building quantum computers. 
One of the earliest experimental demonstrations of QAOA on real quantum hardware was conducted by \citet{otterbachUnsupervisedMachineLearning2017}, who translated a clustering task into a MaxCut problem that QAOA$_1$ could solve. 
The study implemented the quantum algorithm on a Rigetti 19Q quantum processor of 19 functional superconducting transmon qubits. 
Bayesian optimization was employed to optimize QAOA’s parameters. 
The results showed that the algorithm reached the optimal solution in significantly fewer steps than expected by drawing cluster assignments uniformly at random. 
Although this study provided an early experimental demonstration of the effectiveness of QAOA, it did not draw a comparison with leading classical solvers or other methods.

QAOA’s performance was benchmarked against quantum annealing by \citet{willschBenchmarkingQuantumApproximate2020} on a set of weighted MaxCut and 2-SAT problems with up to 16 and 18 variables, respectively, executed on the IBM Q 16 Melbourne quantum computer and the D-Wave 2000Q quantum annealer. 
Three different measures were used to evaluate the algorithm's performance: the probability of finding the ground state, the energy expectation value, and a ratio closely related to the approximation ratio.
The IBM Q processor produced poor results when solving a nontrivial 2-SAT problem with eight variables, even though the $p=1$ QAOA on the simulator had yielded good results for the same problem. 
This suggests that the limitations and noise of the QPU may significantly impact QAOA’s performance.
The study also found that for the set of problem instances considered, QAOA with $p=1, \ldots, 5$ could not compete with quantum annealing when no minor embedding was necessary: the D-Wave machine was even able to outperform QAOA executed on a simulator. 
Interestingly, a correlation was also observed between instances hard for quantum annealing and those hard for QAOA.

\citet{bengtssonImprovedSuccessProbability2020} implemented QAOA on their proprietary quantum hardware platform, consisting of two superconducting transmon qubits and one parametrically modulated coupler. 
They applied the QAOA up to a depth of $p=2$ and solved small instances of the NP-complete Exact-Cover problem with a success probability of 96.6\%. 
This high success probability was achieved regardless of the optimizer used, thanks to the high gate fidelities of their quantum hardware.
The authors benchmarked their quantum hardware's measured state probabilities and cost functions against those of an ideal quantum computer without any noise. They found excellent agreement between the two; this indicates low coherent and incoherent error rates. 
However, the authors noted that even with high gate fidelities, high algorithmic fidelity is not guaranteed.
They predicted that increasing the depth of the QAOA up to $p=3$ on their hardware would not yield a higher success probability since it would result in a longer circuit and hence in a reduced total fidelity, which they estimated to be 94.2\%.
Overall, their results demonstrated promising progress toward the practical implementation of QAOA for solving optimization problems on small quantum devices. However, the challenge of scaling up quantum hardware to solve larger and more complex problems still remains a major issue.

\citet{harriganQuantumApproximateOptimization2021} studied three families of graph problems with QAOA up to $p=5$ on a superconducting quantum processor with 23 physical qubits. 
The problems included hardware grid problems whose topology matches the device's, MaxCut on 3-regular graphs, and the SK model; the latter two instances required compilation to be implemented on the hardware.
The study demonstrated a robust performance of QAOA on hardware grid problems, even for the largest instances with 23 qubits. 
Notably, $n$-independent noise effect on the approximation ratios was observed, i.e., the approximation ratio was observed to be more or less independent of the problem size, which is in agreement with the simulation results in~\cite{crooksPerformanceQuantumApproximate2018}. 
Moreover, the performance increased with circuit depth up to $p=3$; however, the noise degraded the performance for $p=4$ and $p=5$.
This is in accordance with an earlier experimental result provided by \citet{alamDesignSpaceExplorationQuantum2020}, who pointed out that noise in real hardware limits the optimal number of layers for any QAOA instance.
On the other hand, it is important to note that most real-world instances of combinatorial optimization problems cannot be mapped to hardware-native topologies. 
Instead, they require compilation, which involves routing qubits with swap networks. 
This additional overhead can significantly impact the performance of the algorithm.
For these two families of problems investigated, i.e., MaxCut on 3-regular graphs and the SK model, QAOA’s performance decreased with problem size but still provided an advantage over random guessing. 
This study highlights the challenge of using near-term quantum computers to optimize problems on graphs that differ from hardware connectivity.

Some experiments on trapped-ion systems were conducted recently. 
For example, \citet{paganoQuantumApproximateOptimization2020} applied QAOA to approximate the ground-state energy of both the quantum and classical Ising model and investigated the algorithm's performance on a trapped ion quantum computer with up to 40 qubits. 
Specifically, they investigated QAOA’s performance as a function of the number of qubits, ranging from 20 to 40. 
They observed that the performance does not degrade significantly as the system size increases, which was consistent with previous findings by \cite{crooksPerformanceQuantumApproximate2018, harriganQuantumApproximateOptimization2021}.
The study also indicated that increasing the number of QAOA layers from $p=1$ to $p=2$ did not significantly improve the performance of QAOA due to the limitations imposed by the hardware, such as decoherence and bit-flip errors.
Further extensive benchmarks of QAOA were conducted by \citet{bakerWassersteinSolutionQuality2022} on various QPUs from IBM (5 different devices), Rigetti (Aspen-10 with 32 qubits), and IonQ (11 qubits). 
The study examined the QAOA performance up to $p=5$ for portfolio optimization, where the quality of solutions was measured using the Wasserstein distance. 
The results showed that the solution quality peaked at $p=5$ for most QPUs with $n=2$ qubits and $p=4$ for the trapped ion QPU with $n=3$ qubits. 
The study also observed an increase in performance with $p$ using variants of the more general QAOAnsatz at $p=2$ for $n=2$ and $n=3$.
Interestingly, among the IBM QPUs, the authors found that a QPU with a lower quantum volume produced higher quality solutions than QPUs with a higher quantum volume and the same qubit topology. 
This highlights the need for application-specific benchmarking, as general benchmarking metrics may not predict application performance. 
Additionally, the study observed that the quality of solutions produced by all QPUs varied at a level much larger than the stochastic noise from the finite number of shots, suggesting that variability should be regarded as a QPU performance benchmark for given applications.

Finally, QAOA performance on photonic quantum hardware was also demonstrated. 
\citet{qiangLargescaleSiliconQuantum2018} realized a fully programmable two-qubit quantum processor with large-scale silicon photonic circuits based on the linear-combination protocol. 
The authors programmed the device to implement 98 different two-qubit unitary operations with an average quantum process fidelity of $93.2 \pm 4.5$, a two-qubit QAOA with $p=1$, and efficient Szegedy directed quantum walks simulation. 
They applied QAOA to three examples of CSPs, with the classical fidelities between experiment and theory of $99.88 \pm 0.10 \%$, $96.98 \pm 0.56 \%$, and $99.48 \pm 0.27 \%$, respectively.


    \subsection{Noise and Errors Considerations}%
    \label{sec:noiseanderrors}

As mentioned in Section~\ref{sec:QAA}, the QAOA is a quantum algorithm inspired by adiabatic quantum computing~\cite{farhiQuantumComputationAdiabatic2000, farhiQuantumAdiabaticEvolution2001, albashAdiabaticQuantumComputation2018} that leverages layering to achieve the desired solution state.
Theoretically, its approximation ratio ($\alpha$) to the true solution increases as the number of alternating cost and mixer layers ($p$) increases.
However, in reality, the performance gain through an increasing number of layers may be seriously challenged by the noise and errors accumulated in a deeper circuit in NISQ-era devices.
Besides circuit depth, increasing system sizes (e.g., for larger graphs) and higher noise rates of hardware can also contribute to an increase in the total amount of noise accumulated and hinder the algorithm's effectiveness in practical settings.

Recent findings have revealed that noise and errors pose substantial challenges to the scalability and
performance of VQAs such as QAOA. 
Both local and correlated noise can adversely affect QAOA, as discussed in Section~\ref{sec:noise_characterization}. Consequently, while QAOA may demonstrate a potential quantum advantage in certain noiseless settings, its performance in near-term hardware can be severely compromised, as noted in Section~\ref{sec:noise_advantage}. 
However, there is hope for overcoming some of these challenges, as various error mitigation techniques have been proposed to improve QAOA's practicality in the near future. 
Section~\ref{sec:noise-mitigation} explores some of these techniques in more detail.

\subsubsection{Characterizing the sources of noise}\label{sec:noise_characterization}

In general, noise can be classified into two types: local (uncorrelated) and correlated.
Effects of local, uncorrelated noise and errors on different VQAs have been under active investigation in recent years~\cite{mccleanTheoryVariationalHybrid2016, mccleanHybridQuantumclassicalHierarchy2017, gentiniNoiseresilientVariationalHybrid2020, sharmaNoiseResilienceVariational2020, fontanaEvaluatingNoiseResilience2021, mohantyAnalysisVehicleRouting2022}.
For example, \citet{kungurtsevIterationComplexityVariational2022} highlight that, due to the noise inherent in near-term quantum devices, the evaluations of the objective function are systematically biased, necessitating a different perspective on the convergence analysis of these classical optimization procedures.

In the case of QAOA, typical local quantum noise channels, including dephasing, bit-flip, and depolarizing channels, were considered.
Theoretical studies showed that the QAOA performance, characterized by the output state fidelity and the approximation ratio, degrades as a power law with the noise strength, with the power being proportional to the system size~\cite{xueEffectsQuantumNoise2021, marshallCharacterizingLocalNoise2020}.
Based on these analytical equations, one can model the trade-off between a deeper QAOA circuit with more layers which gives a better approximation ratio per se, but at the expense of greater performance degradation due to noise.
In other words, an optimal number of layers can be determined for QAOA for any given noise rate.
Moreover, the scalability of QAOA can also be affected by noise.
A noisy implementation of QAOA would be effective if it could produce a measurement result from the intended, noiseless quantum state distribution.
Using a local noise model, \citet{lotshawScalingQuantumApproximate2022} showed that the number of measurements needed to achieve the above goal increases exponentially with the gate error rates, which translates to exponential time complexity, assuming that the number of measurements is proportional to the time taken to reach a solution.
Such a measurement scaling, therefore, significantly limits the scalability of QAOA implementations on near-term devices.
Another factor that affects QAOA's scalability is intimately tied to the variational nature of the QAOA circuit. 
\citet{wangNoiseinducedBarrenPlateaus2021} showed that even local noise, which includes depolarizing noise and certain kinds of Pauli noise, can induce barren plateaus, where the gradient of the cost function landscape vanishes exponentially with an increasing circuit depth.
Such noise-induced barren plateaus (NIBPs) are conceptually distinct from the noise-free barren plateaus first introduced by \citet{mccleanBarrenPlateausQuantum2018} (Section~\ref{sec:barren_plateaus}), as the gradient vanishes with increasing problem sizes at every point on the cost function landscape, rather than probabilistically. 
Furthermore, they cannot be mitigated with strategies that are used to avoid the noise-free ones, such as layerwise training~\cite{skolikLayerwiseLearningQuantum2021, liuLayerVQEVariational2022}.
NIBPs not only pose a challenge to parameter optimization in QAOA even with gradient-free optimizers, inhibiting its near-term scalability but could also destroy any potential quantum speedup.
Conversely, local noise sources may not always be detrimental and, in some cases, may aid the algorithm's optimization process.
One example was shown by \citet{camposTrainingSaturationLayerwise2021}, who employed the layerwise training technique to QAOA, where instead of training all the parameters in a variational circuit at once, layers of the circuit are added to the training routine successively.
They observed that training saturation occurs, that is, the fidelity between the output and the target states stops improving past a certain number of layers.
Interestingly, the authors also showed that local coherent dephasing noise could remove such training saturation, recovering the effectiveness of layerwise learning.
Further investigation is needed to determine if other noise sources beyond the simple noise model can play a similar role.

On the other hand, correlated errors such as crosstalk~\cite{heinzCrosstalkAnalysisSinglequbit2021, niuAnalyzingCrosstalkError2021}, the non-Markovian $1/f$ noise~\cite{burkardNonMarkovianQubitDynamics2009}, and interactions with environmental fluctuators~\cite{schlorCorrelatingDecoherenceTransmon2019}, are shown to be relevant and prevalent in NISQ devices.
The implications of crosstalk noise in measurements for QAOA were studied by \citet{maciejewskiModelingMitigationCrosstalk2021}.
The authors introduced a correlated measurement noise model that can be efficiently characterized by the Diagonal Detector Overlapping Tomography (DDOT) performed on IBM's and Rigetti's superconducting quantum processors.
It allows estimating $k$-local crosstalk noise in an $N$-qubit device using $\mathcal O(k2^k\log N)$ quantum circuits.
Furthermore, through a numerical simulation on random MAX-2-SAT instances and the SK model on an 8-qubit system, the authors demonstrated that while the correlated readout noise has a mild effect on the optimization, it could still alter the energy landscape of QAOA, rendering the solution in sub-optimal regions.
Another correlated error, denoted as the precision errors, resides in the misspecification of QAOA's parameters due to imperfect control and is more prominent as the QAOA order, i.e., the number of alternating layers, grows.
\citet{quirozQuantifyingImpactPrecision2021} found analytically that precision errors lead to an exponential decrease in success probability of QAOA with increasing QAOA order and noise strength, which was also supported by the numerical results of the QAOA variants for Grover’s search and the 1D transverse-field Ising model.
Another study conducted by \citet{karamlouAnalyzingPerformanceVariational2021} on IBM's \verb|ibmq_boeblingen| quantum computer also revealed that a coherent error induced by the residual ZZ-coupling between the transmon qubits is the major bottleneck that limits the performance of QAOA on near-term superconducting quantum processors.
More recently, the effects of both temporally and spatially correlated noise on the performance of QAOA were studied by \citet{kattemolleEffectsCorrelatedErrors2022}, based on a toy error model in which every qubit interacts with a single binary fluctuator that can travel through space or time.
Curiously, it was observed in numerical simulation with the SK model on 6 qubits that the QAOA performance improves as the noise correlation strength increases at fixed local error rates.
This suggests a certain degree of noise resilience of QAOA and that the correlation by itself may not have a negative impact on variational algorithms like QAOA\@.
However, further studies are required to test the generalizability of this result.

\subsubsection{Quantum advantage in the presence of noise}\label{sec:noise_advantage}

As mentioned in Section~\ref{sec:efficiency}, QAOA in the noiseless limit is potentially able to outperform the best classical algorithm in solving certain types of optimization problems, thus achieving a quantum advantage.
However, as alluded to earlier, noise in quantum hardware and environment poses great challenges to its performance.
It is, therefore, unclear if such quantum advantage could still be retained in more realistic settings.
Two approaches have been taken to address the above question: theoretical analysis and empirical studies using real quantum hardware or noisy simulators.

On the theoretical front, recent studies concluded that substantial quantum advantages are unlikely for the QAOA with the noise level in current devices, especially for large and dense problems.
Based on the relative entropy inequalities, \citet{stilckfrancaLimitationsOptimizationAlgorithms2021} provided theoretical bounds on various quantities such as the output energy of noisy quantum circuits and the maximum depth of a quantum circuit can have before efficient classical algorithms outperform it.
Applying these bounds to QAOA on the SK model and random 3-regular graphs, it was found that to match the performance of classical devices, the error rates of current quantum computers would need to improve by a couple of orders of magnitude, reaching the level of the expected fault-tolerance threshold.
Building on these results, \citet{weidenfellerScalingQuantumApproximate2022} performed a detailed analysis on scaling QAOA on superconducting qubit-based devices.
They concluded that even though using appropriate SWAP strategies may help alleviate the conundrum~\cite{hashimOptimizedSWAPNetworks2022}, for dense problems, the required gate error rates would still lie far below the fault-tolerant threshold, based on the entropic argument.
Similarly, \citet{depalmaLimitationsVariationalQuantum2023} used techniques from quantum optimal transport and considered simple noise models such as the one-qubit depolarizing noise with probability $P$.
They proved that with noisy quantum circuits at depths $L\sim \mathcal O(P^{-1})$, the probability of observing a single string with better energy than that output by an efficient classical algorithm is exponentially small in the number of qubits, thus providing a stronger statement than the previous result~\cite{stilckfrancaLimitationsOptimizationAlgorithms2021}, which held only for the expectation of the output.
Moreover, \citet{gonzalez-garciaErrorPropagationNISQ2022} adopted a different approach, i.e., a random circuit model, to study the propagation of arbitrary single-qubit errors in variational quantum circuits and reached a similar conclusion.
Applying such a model to the QAOA circuit, they estimated that the required error rate for a possible quantum advantage scales as $P \sim 1/(nD)$, with $n$ being the number of qubits, and $D$ the circuit depth, which translates to a value lower than $10^{-6}$ (assuming $n = 1000$ for a potential quantum advantage to start becoming practically useful~\cite{guerreschiQAOAMaxCutRequires2019}) with a two-dimensional random circuit architecture.

Such a conclusion is supported by various empirical studies on noisy simulators and different quantum hardware platforms, demonstrating noise's detrimental effects on QAOA's performance.
For example, \citet{alamDesignSpaceExplorationQuantum2020} empirically demonstrated both in simulation and on IBM's quantum computer, IBMQX4, that QAOA's performance is noise-sensitive and its sensitivity is higher for high-depth QAOA. 
Their results thus suggest that it is impractical to gain better QAOA performance in realistic settings through unlimited layering and that the noise in the target hardware limits the optimal number of layers for any QAOA instance.
Another piece of evidence comes from the work of \citet{harriganQuantumApproximateOptimization2021}, where the authors studied various families of problems with QAOA on Google's superconducting quantum processor.
The problems considered include the hardware grid problems, whose topology matches that of the device, and those that are not native to the hardware and require compilation to be implemented, such as MaxCut on 3-regular graphs and the SK model with all-to-all connectivity.
On the one hand, with the hardware grid problems, QAOA showed robust performance measured in terms of the approximation ratio, where the effect of noise showed minimal dependence on the number of qubits $n$.
Furthermore, by averaging 130 hardware grid problems with $n > 10$, it was reported that QAOA reached its performance maximization at $p = 3$ on the current hardware.
On the other hand, for non-hardware-native problems, evident performance degradation was observed as the number of qubits increased.
It is worth noting that similar $n$-independent performance of the $p=1$ QAOA on hardware-native Ising model was also observed on a trapped ion quantum computer with up to 40 qubits~\cite{paganoQuantumApproximateOptimization2020}.
However, at $p=2$, QAOA with 20 qubits showed a similar performance to the $p=1$ circuit, suggesting that decoherence and errors accumulated during longer evolution times already balanced out the $2\%$ expected performance gain of one additional optimization layer.

Overall, these results demonstrate the negative impact of noise and the difficulty in scaling and achieving any quantum advantage from the near-term implementations of QAOA, particularly for non-hardware-native problems.
Heuristically, the challenge in achieving quantum advantages with the QAOA can be attributed to the following facts~\cite{shaydulinErrorMitigationDeep2021}.
On the one hand, in order to achieve a quantum advantage, the depth of QAOA in its original form should grow at least logarithmically with the problem size~\cite{bravyiObstaclesVariationalQuantum2020, farhiQuantumApproximateOptimization2020a, farhiQuantumApproximateOptimization2020}.
At the same time, implementing QAOA on sparsely connected hardware would require routing qubits via SWAP networks that generally incur a linear cost~\cite{harriganQuantumApproximateOptimization2021, stilckfrancaLimitationsOptimizationAlgorithms2021}.
These two observations combined lead to the circuit depth scaling as $O(n\log n)$ with the number of variables $n$.
On the other hand, QAOA's performance, measured in terms of, e.g., output state fidelity, success probability, etc., could suffer an exponential decay with a growing system size in the presence of a constant error rate~\cite{xueEffectsQuantumNoise2021, marshallCharacterizingLocalNoise2020}.
Therefore, it is unlikely that QAOA circuits with more than logarithmic depth could lead to any quantum advantage without significantly improved quantum hardware or error correction.

\subsubsection{Noise mitigation techniques}\label{sec:noise-mitigation}

In the meantime, many mitigation strategies have also been proposed to reduce the negative impact of noise on QAOA and other VQAs.

Gate count is a primary contributor of noise, and different hardware will have different connectivity that will impact the depth of the quantum circuit, in some instances, dramatically.
This is because the circuit must be modified to align with the qubit coupling map of the device, and the gates must be translated into the set of native gates used by the particular device.
This process is often called compiling.
Therefore, a general strategy to mitigate the effects of errors on current quantum devices is to reduce the number of gates in the compiled quantum circuits.
This can be accomplished by, for example, optimizing the SWAP networks, which are needed for running QAOA on non-hardware-native problems~\cite{hashimOptimizedSWAPNetworks2022}.
\citet{hashimOptimizedSWAPNetworks2022} considered two methods to improve the SWAP networks: using an overcomplete two-qubit gate set for decomposing the relevant quantum gates and SWAP networks, as well as a technique called equivalent circuit averaging, which averages over a set of randomized but logically equivalent circuits to mitigate coherent errors.
As a result, around $60\%$ average reduction in error (total variation distance) was observed for depth-1 QAOA on four transmon qubits on a superconducting quantum processor.
Other methods have also been proposed to optimize the compilation process to reduce the gate counts and improve the success probability of the compiled circuit, such as re-ordering of the multi-qubit CPHASE gate and the use of variation-aware compilation policies~\cite{alamCircuitCompilationMethodologies2020,alamNoiseResilientCompilation2020, alamEfficientCircuitCompilation2020}.
Compared with single-qubit gates, two-qubit gates such as the CNOT are typically more erroneous on a quantum computer. 
Therefore, CNOT gate is an important target for noise mitigation.
\citet{majumdarOptimizingAnsatzDesign2021} suggested two hardware-independent methods of reducing the total number of CNOT gates in the QAOA circuit.
The first method is based on edge coloring of the input graph to minimize the depth of the circuit, and it reduces up to $\lfloor\frac{n}{2}\rfloor$ CNOT gates in the first layer of the QAOA circuit, with $n$ being the number of vertices in the graph.
The other method is based on a Depth First Search (DFS) algorithm, which reduces $n - 1$ CNOT gates in the circuit, although it moderately increases the depth of the circuit in doing so.
The depth of the circuit is proportional to the height of the DFS tree, which can be $n-1$ in the worst case.
Therefore, an analytical condition was derived from balancing such trade-offs while maintaining the lowest error probability.
This condition was also satisfied in the experimental implementations on IBM's \verb|ibmq_manhattan|, in which the authors demonstrated that the edge coloring-based method outperformed the standard QAOA, and the DFS method outperformed both of these implementations.
In a successive paper by the same group~\cite{majumdarDepthOptimizedAnsatz2021}, the DFS-based method was improved.
The authors proposed an $O(\Delta\cdot n^2)$ greedy heuristic algorithm, where $\Delta$ is the maximum degree of the graph.
This algorithm finds a spanning tree of lower height, reducing the overall depth of the circuit while still maintaining the $n-1$ reduction in the number of CNOT gates in the ansatz.
It was shown numerically that this new algorithm achieves nearly a factor of 10 increase in the success probability for each iteration of QAOA for the MaxCut problem.

Other strategies that target more specific errors or are more tailored to QAOA also exist.
For example, to combat the crosstalk effects in measurement noise mentioned in Section~\ref{sec:noise_characterization}, a mitigation technique was proposed to correct the marginal probability distributions affected by the correlated measurement noise~\cite{maciejewskiModelingMitigationCrosstalk2021}.
Tested on IBM's (Rigetti's) devices with 15 (23) qubits, such technique led to an average reduction of errors by a factor $> 22$ ($> 5.5$).
Moreover, based on the numerical simulations, it led to a noticeable improvement in QAOA's performance across multiple graph instances.
Likewise, even though the detrimental nature of the precision errors unveiled in~\cite{quirozQuantifyingImpactPrecision2021}, they may also be effectively mitigated via the digitization of the variational parameters in a binary representation at the cost of an increasing circuit depth.

Proponents also exist to take advantage of symmetries possessed by the problem instances.
Specifically, one such error mitigation technique is called symmetry verification, which was initially proposed for quantum simulation~\cite{bonet-monroigLowcostErrorMitigation2018, mcardleErrorMitigatedDigitalQuantum2019}.
\citet{shaydulinErrorMitigationDeep2021} extended this method to QAOA, i.e., by determining the classical symmetries of the objective function preserved by the QAOA ansatz and then projecting the QAOA state into the symmetry-restricted subspace.
By restricting the state from evolving under such subspace (i.e., the eigenspace of the corresponding symmetry operators), an error that pushes the system out of this subspace can be partially corrected by projecting the system back to the correct eigenspace.
This technique was effective in solving four MaxCut instances on graphs with 3 and 4 nodes on IBM's \verb|ibmq_jakarta| processor, leading to improvements in output state fidelity, expected objective function value, and probability of sampling the optimal solution.
A similar approach based on symmetry verification through mid-circuit postselection was later extended to the quantum alternating operator ansatz, a generalization of the original QAOA. It was shown to be effective in solving the TSP~\cite{botelhoErrorMitigationVariational2022}.
Another example was a proposal by \citet{streifQuantumAlgorithmsLocal2021}, which aims to reduce the overhead of a bit-flip error correcting code by leveraging the Local Particle Number Conservation (LPNC) in certain ansatze, such as the XY quantum alternating operator ansatz.
This technique can detect and correct symmetry-breaking errors, and it is advantageous in the case of high error rates and/or deep circuits.
In a more recent study, \citet{weidingerErrorMitigationQuantum2023} introduced a decoding scheme to mitigate the errors that arise in a special variant of QAOA called parity QAOA~\cite{lechnerQuantumApproximateOptimization2020, dlaskaQuantumOptimizationFourBody2022}, which makes use of parity transformation to encode the optimization problem into a 2D local Hamiltonian that is suitable for implementation on current quantum devices with a planar architecture.
This error mitigation technique exploits the redundant information introduced during parity transformation, where each physical qubit is mapped to multiple logical qubits, and was shown to be advantageous in numerical experiments where a depolarizing error on all circuit gates was considered.

Last but not least, recent advances in general exponential error suppression schemes also offer some hope for making QAOA more practical in the near future~\cite{koczorExponentialErrorSuppression2021, hugginsVirtualDistillationQuantum2021, czarnikQubitefficientExponentialSuppression2021}. However, to solve issues such as NIBPs, quantum technologists must work towards more accurate gates and, eventually, fault-tolerant quantum computing.

    \subsection{Hardware-Specific Approaches}%
    \label{sec:hardware_specific_suggestions}

\begin{table}[htp!]
    \centering
    \footnotesize
    \def\arraystretch{1.5}
    \begin{tabular}{>{\raggedright}p{0.1\linewidth}p{0.1\linewidth}p{0.15\linewidth}p{0.05\linewidth}p{0.05\linewidth}p{0.4\linewidth}}
	\toprule
	 \textbf{Problem} & \textbf{Graph} & \textbf{Hardware} & $n$ & $p$ & \textbf{Device and Remarks}\\
	\midrule
 SK model & Complete & Superconducting & 8--72 & 1  & Rigetti Aspen-M-3~\cite{dupontQuantumEnhancedGreedy2023}. Quantum-enhanced iterative algorithm with a truncated 1-layer QAOA ansatz embedded; outperformed classical greedy threshold.\\
 
 & Complete & Trapped ions & 7--32 & 1 & Quantinuum H2~\cite{leonticaExploringNeighborhood1layer2023}. 1-layer QAOA embedded in Instantaneous Quantum Polynomial (IQP) circuits; $\alpha_\text{avg} = 0.985$. \\
 
 Max-3SAT & Random, clause density of 4 & Trapped ions & 6--20 & 1--20 & IonQ Harmony and Quantinuum H1-1~\cite{pelofskeHighRoundQAOAMAX2023}. At fixed $n$, $\alpha_\text{avg}$ degraded beyond certain $p$ due to noise; at fixed $p$, $\alpha_\text{avg}$ remained insensitive to growing $n$. \\
 
MaxCut & Non-planar 3-regular & Trapped ions & 20 & 1--16 & Quantinuum H1-1~\cite{shaydulinQAOACdotGeq2023}. Solution quality increased monotonically with $p$ up to 10 for all problem instances, with largest $\alpha = 0.94$. \\
& Hardware-grid/non-planar 3-regular & Superconducting & 1--23 & 1--5 & Google Sycamore~\cite{harriganQuantumApproximateOptimization2021}. On hardware-grid problems, performance was independent of $n$, and peaked at $p=3$. On 3-regular graphs, performance degraded as $n$ increased. \\
& Heavy-hex & Superconducting & 27 & 1, 2 & IBM \verb|imbq_mumbai|~\cite{weidenfellerScalingQuantumApproximate2022}. $p=2$ QAOA produced better cuts than $p=1$. \\
& 4~nodes~with 3 edges & Neutral atoms & 4 & 1--3 & ColdQuanta~\cite{grahamMultiqubitEntanglementAlgorithms2022}. $\alpha = 0.669\ (p = 1),\ 0.687\ (p=2),\ 0.628\ (p=3)$. \\

Ising
model & Hardware-native with cubic interactions & Superconducting & 127 & 2 & IBM \verb|ibm_washington|~\cite{pelofskeQuantumAnnealingVs2023} and DWave Advantage\_system4.1 \& 6.1. Quantum annealing outperformed QAOA on all instances. With dynamical decoupling~\cite{niuEffectsDynamicalDecoupling2022}, $p=2$ QAOA marginally outperformed $p=1$. \\
& Hardware-native & Trapped ions & 20--40 & 1, 2 & U~Maryland group~\cite{paganoQuantumApproximateOptimization2020}. Performance and runtime were approximately independent of system size. $p=2$ QAOA performed similarly to $p=1$ at 20 qubits. \\

CSPs & \hspace{-1em}\centering{-} &  Photonic & 2 & 1 & U Bristol group~\cite{qiangLargescaleSiliconQuantum2018}. Classical fidelities with theory of $\approx 99.88\%, 96.98\%, 99.48\%$ achieved for the three 2-bit CSPs considered.\\

MIS & Random & Neutral atoms & 39--289 & 1--5 & Harvard U group~\cite{ebadiQuantumOptimizationMaximum2022}. On an 179-vertex graph instance, $\alpha$ improved with increasing $p$ up to $p=4$. On instances with large enough spectral gaps, QAOA achieved speedup over simulated annealing. \\
\bottomrule
    \end{tabular}
    \caption[Summary of selected state-of-the-art experiments on various quantum devices.]{Summary of selected state-of-the-art experiments on various quantum hardware platforms based on different technologies (superconducting, trapped ions, neutral atoms, and photonics). $n$ represents the number of qubits used in the experiment, $p$ is the number of QAOA layers investigated, and $\alpha_\text{(avg)}$ indicates the (average) approximation ratio. Note that in some experiments not all combinations of $n$ and $p$ were investigated.
    }\label{tab:hardware_runs}
\end{table}

{
\color{red}
}

There has been significant research interest in leveraging specific hardware to enhance the performance of QAOA across various platforms, such as trapped ions~\cite{leonticaExploringNeighborhood1layer2023,  pelofskeHighRoundQAOAMAX2023, shaydulinQAOACdotGeq2023, paganoQuantumApproximateOptimization2020}, neutral atoms~\cite{grahamMultiqubitEntanglementAlgorithms2022, dlaskaQuantumOptimizationFourBody2022}, superconducting qubits~\cite{alamAnalysisQuantumApproximate2019, alamCircuitCompilationMethodologies2020, abramsImplementationXYEntangling2020, alamDesignSpaceExplorationQuantum2020, weidenfellerScalingQuantumApproximate2022, mundadaExperimentalBenchmarkingAutomated2023, bengtssonImprovedSuccessProbability2020, lacroixImprovingPerformanceDeep2020, otterbachUnsupervisedMachineLearning2017, harriganQuantumApproximateOptimization2021, willschBenchmarkingQuantumApproximate2020, dupontQuantumEnhancedGreedy2023}, and photonic quantum computers~\cite{proiettiNativeMeasurementbasedQuantum2022, qiangLargescaleSiliconQuantum2018}.  
The goals of these approaches include overcoming hardware connectivity limitations and mitigating noise-related issues to broaden the applicability of QAOA to a wide range of combinatorial optimization problems. 
A selection of the latest experimental realizations of the QAOA is reported in Table~\ref{tab:hardware_runs}. 
Additionally, hardware implementations provide an opportunity to validate the effectiveness of error mitigation techniques, as discussed in Section~\ref{sec:noise-mitigation}. 
However, it is essential to note that different architectures have advantages and disadvantages. 

Different physical models display different natures in the interactions between qubits, each presenting its unique implementation challenges. 
In the case of superconducting qubits, the mapping of interactions is predetermined by the device's architecture. 
When considering neutral atoms, the limitations arise from the decay of qubit couplings. 
\citet{dlaskaQuantumOptimizationFourBody2022} introduced an innovative four-qubit Rydberg parity gate, which facilitates the use of the parity architecture~\cite{lechnerQuantumAnnealingArchitecture2015}. 
This architecture offers a scalable remapping of qubits, enabling more efficient interconnection between them.
By laser-coupling these atoms to highly excited Rydberg states and employing adiabatic laser pulses, they were able to manipulate computational basis states through imprinting a dynamical phase on them.
The gate, fully programmable and adjustable, enables a direct and straightforward implementation of the parity architecture, essential for encoding complex interaction graphs in atomic arrays. 
They numerically demonstrate implementations of QAOA for small-scale test problems, paving the way for experimental investigations beyond numerical simulations.

The underlying interaction graph of the Sherrington-Kirkpatrick (SK) Hamiltonian has all-to-all connectivity, which prevents direct implementation on planar quantum processors.
\citet{rabinovichIonNativeVariational2022} 
utilized ion native Hamiltonians to develop ansatz families that can prepare ground states of general problem Hamiltonians. 
By testing their algorithm on 6-qubit SK Hamiltonians, they demonstrate that overcoming symmetry protection allows for the minimization of arbitrary instances. 
This work highlights the potential of trapped-ion-based quantum processors for solving a broad range of combinatorial optimization problems.

\citet{rajakumarGeneratingTargetGraph2022}
presented a method for constructing arbitrary unweighted and weighted coupling graphs using global entangling operations in quantum spin systems. 
They provided upper bounds on the number of operations required and proposed a mixed-integer program for finding optimal sequences. 
Their approach is less susceptible to noise compared to standard gate-based compilation, suggesting that global entangling operations may be more efficient for dense, unweighted coupling operations.
Further research is needed to establish tighter upper bounds and explore how the complexity of compilation affects quantum algorithms.

In addition to connectivity, another feature that can be leveraged based on the underlying model of the device is the utilization of higher energy levels beyond qubits, known as qudits, which have $d$ energy levels
. 
For instance, in the realm of photonic quantum computers, one can leverage their inherent capabilities to employ qudits rather than qubits.
For a $k$-color graph, choosing $k=d$ allows an efficient mapping of the problem to the hardware. 
In neutral atom-based quantum computers, qudits offer exciting possibilities for solving real-world problems more efficiently compared to classical methods. \citet{dellerQuantumApproximateOptimization2022} have explored the use of qudits in the context of the $k$-graph coloring and electric vehicle charging problems with global power constraints. 
They introduced the use of the momentum operator $L_x$ as a mixing operator for optimization problems with bounded integer variables.
Their numerical simulations compared QAOA solutions with gradient-based and global evolutionary classical optimizers, with the latter showing better results for the instances considered.


Photonic quantum computers operate on a distinct paradigm of quantum computation known as measurement-based quantum computing (MBQC), as opposed to the gate-based model.
\citet{proiettiNativeMeasurementbasedQuantum2022} presented a MBQC QAOA for photonic quantum computers to solve the Max-$k$-Cut problem. 
They developed an MBQC algorithm using diagonal unitary evolution and demonstrated an up-to 30-fold improvement in terms of cluster state dimension when compared to QAOA gate-based circuit algorithms. 
This work highlights the advantages of tailoring algorithms for photonic quantum computing.
One of the biggest technological challenges in photonic quantum computation are the single photon losses. 
To tackle this issue, \citet{vikstalQuantumApproximateOptimization2023} discussed how standard qubits can be replaced with cat qubits. 
Cat qubits are created as a superposition of two distinct coherent states of light with opposite phase. 
Generally, this kind of states become challenging to realize and manipulate. 
Despite this, what is found is that in the context of QAOA, of the Exact Cover problem in particular, they exhibit a performance advantage (i.e., higher fidelity rates at fixed resources).

    \section{Experimental Results}%
    \label{sec:experimental}
    Quantum computing has made significant strides in recent years, and there is growing interest in systematically comparing VQAs, particularly the QAOA, across both simulators and real devices.
Understanding the performance and efficiency of these algorithms is essential for their practical application and the development of quantum computing.
In this section we present preliminary experimental results obtained from our analysis of the QAOA and its variants. 
We focus on assessing their performance (approximation ratio), efficiency (number of optimization iterations times circuit depth), and the effect of noise and errors on real quantum
computers.

\subsection{Related Tools}

Table~\ref{tab:hardware_runs} 
presents a summary of selected state-of-the-art experiments performed on various quantum hardware platforms based on different technologies (superconducting, trapped ions, neutral atoms, and photonics). 
There is significant interest in comparing between a range of combinatorial problems (like SK, MaxCut, Max-3SAT, Ising and CSPs), a wide range of hardware architectures and sizes (from 2 qubits to 289 qubits) and many different aspects of QAOA variants (for example, a varied number of QAOA layers, from 1 to 20).
It is also interesting to note that the product of these last two quantities---number of qubits and number of layers---has been proposed as a benchmark to check the scalability of QAOA on hardware~\cite{shaydulinQAOACdotGeq2023}.

To facilitate these comparisons, tools such as QPack~\cite{donkersQPackScoresQuantitative2022} and QAOAKit~\cite{shaydulinQAOAKitToolkitReproducible2021} have been developed.
QPack is an application-oriented cross-platform benchmarking suite for quantum computers and simulators, which utilizes scalable QAOA and VQE applications to provide a holistic insight into quantum performance.
It collects quantum execution data and transforms it into benchmark scores for application-oriented quantum benchmarking, including sub-scores based on runtime, accuracy, scalability, and capacity performance.
QAOAKit is a Python toolkit for QAOA built for exploratory research.
It serves as a unified repository of preoptimized QAOA parameters and circuit generators for common quantum simulation frameworks, allowing researchers to reproduce, compare, and extend known results from various sources in the literature.

\subsection{Implementation and Systematic Evaluation of QAOA Variants}

Our study focuses on the experimental results obtained using our custom implementations of the QAOA and its variants rather than relying on existing software tools.
As cloud-based quantum computing platforms for VQAs are increasingly prevalent~\cite{karalekasQuantumclassicalCloudPlatform2020}, we utilized IBM Quantum's cloud-based infrastructure to experimentally evaluate the performance of several QAOA variants on different MaxCut problems.

\begin{figure}[htp!]
    \centering
    \includegraphics[width=\textwidth]{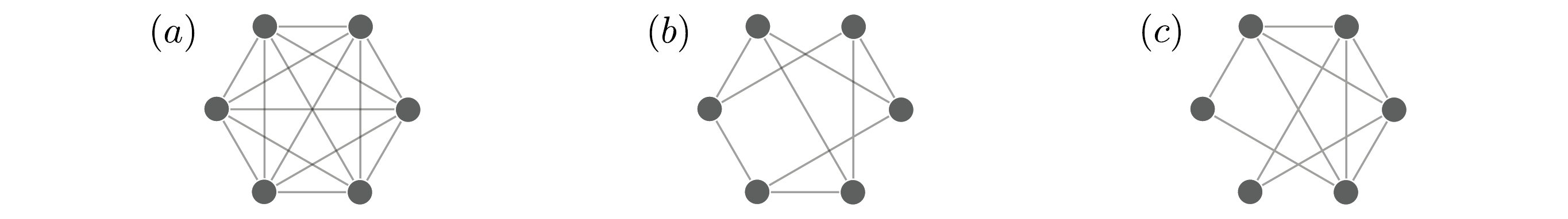}
    \caption[Three different graph types in generated dataset.]{($a$) complete, ($b$) 3-regular, and ($c$) random graph from our generated dataset.}
    \label{fig:graphs}
\end{figure}

\paragraph{Problem types.}
The MaxCut problems used for the experiments were generated beforehand and encompassed a diverse set of graph structures.
This diversity was primarily motivated by the understanding that the performance of the QAOA ansatz is significantly dependent on the specific problem it is applied to.
The chosen graph structures included complete graphs, 3-regular graphs, and random graphs with edge probabilities varying between 0.3 and 0.5 (Figure~\ref{fig:graphs}).
The rationale for choosing these diverse types was to thoroughly evaluate the QAOA variants under various conditions, reflecting the inherent diversity in real-world quantum optimization problems. 
Complete graphs represent densely connected systems where each node is connected to every other node. 
On the other hand, 3-regular graphs provide a more uniform and structured network topology. 
Lastly, random graphs simulate less predictable scenarios, which might present unique challenges to the optimization process.
Moreover, we varied the size of the graphs, with sizes ranging from 4 to 18 nodes. 
This decision is rooted in the understanding that the complexity of the problem---in this case, represented by the number of nodes---could substantially impact the efficacy of the QAOA ansatz. 
By examining smaller graphs (4 nodes), we tested the efficiency of the QAOA variants in solving more straightforward problems. 
We investigated how these variants scale with increasing problem complexity as we increased the number of nodes.
It is crucial to note that while our selection of graph structures and sizes was diverse and aimed to capture everyday use cases, it only covered some potential quantum optimization scenarios.
However, our research is ongoing, and we are actively working on expanding our dataset to encompass additional use cases, including various graph types, sizes, and problem types beyond MaxCut.

To ensure a comprehensive evaluation, we generated multiple instances for each combination of parameters, enabling us to understand better the QAOA variants' performance across different problem instances and graph structures (Table~\ref{tab:problems}).
\begin{table}[htp!]
\centering
\begin{tabular}{lcc}
    \toprule
    Graph Type & Number of nodes & Instances for each combination \\
    \midrule
    Complete & 4 to 18 & 1 \\
    3-regular & 4 to 18 & 3 to 8 \\
    Random (edge $p=0.3 .. 0.5$) & 4 to 18 & 3 to 8 \\
    \bottomrule
\end{tabular}
\caption[Overview of the graph structures and instances used in experiment.]{Overview of the graph structures and instances used for our experiments. A total of 78 graphs where generated for the results we report here.}%
\label{tab:problems}
\end{table}

\paragraph{QAOA variants.}
We evaluated an array of QAOA variants, namely: ``QAOA''~\cite{farhiQuantumApproximateOptimization2014}, ``QAOA+''~\cite{chalupnikAugmentingQAOAAnsatz2022}, ``FALQON+'' (referred to just as ``FALQON'' in this section)~\cite{magannFeedbackBasedQuantumOptimization2022, magannLyapunovcontrolinspiredStrategiesQuantum2022}, ``ModifiedQAOA''~\cite{villalba-diezImprovementQuantumApproximate2021}, ``WS-QAOA''~\cite{eggerWarmstartingQuantumOptimization2021} and ``\maqaoa{}''~\cite{herrmanMultiangleQuantumApproximate2022}.
Each of these variants was chosen to represent a different methodological approach to improving the QAOA's performance. 
They illustrate some of the broad spectrum of ways that researchers have found to enhance the QAOA, such as adding features, speeding up optimization, adapting the ansatz, and utilizing previously computed information for a warm start.

To examine the performance of these variants under various conditions, we used a varying number of layer depth, reported here ranging from 1 to 8 layers.

\paragraph{Optimization.}
For the optimization process, we utilized the COBYLA optimizer provided by the \verb|scipy| library to tune the variational parameters in the QAOA circuits, in line with the results in~\cite{fernandez-pendasStudyPerformanceClassical2022}.
The optimizer iterated until convergence based on the specified tolerance level ($0.0002$) or until 500 iterations, whichever came first.
The maximum number of iterations only influenced \maqaoa{}, which often reached the maximum cap of iterations and needed many thousands of iterations to achieve just a slightly higher approximation ratio.
We observed that this did not affect the overall trends that we discuss below.
However, we will include the unconstrained versions in the future release of our results.

The same optimization settings were used across all variants to ensure a fair comparison. 
Except for ``WS-QAOA'', which provides a specific initialization of the parameters, we randomly initialized the parameters from a uniform distribution $\gamma \to (0..2\pi)$ and  $\beta\to (0..\pi)$.

\paragraph{Implementation \& Hardware}

We implemented the QAOA and its variants in Python using the Qiskit quantum computing framework. 
Each QAOA variant was implemented as a separate class, inheriting a base QAOA class containing the shared functionalities. 
This modular design allows for easy extension and comparison of different QAOA variants. 

Our implementations were executed on both quantum simulator (qiskit) and real quantum devices provided by the IBM Quantum Experience.
To assess the noise resilience of the proposed QAOA variants, we conducted experiments on actual IBM quantum devices, specifically the IBM Quantum Falcon r5.11H quantum processors, which include the \verb|ibm_oslo|, \verb|ibm_lagos|, \verb|ibm_nairobi| and \verb|ibm_perth|.
The results presented here are part of an ongoing study, and the comprehensive details, source code, and detailed outcomes will be disclosed in a forthcoming research paper.



\subsection{Experimental Results and Discussion}

\begin{figure}[ht!]
	\centering
	\includegraphics[width=1\linewidth]{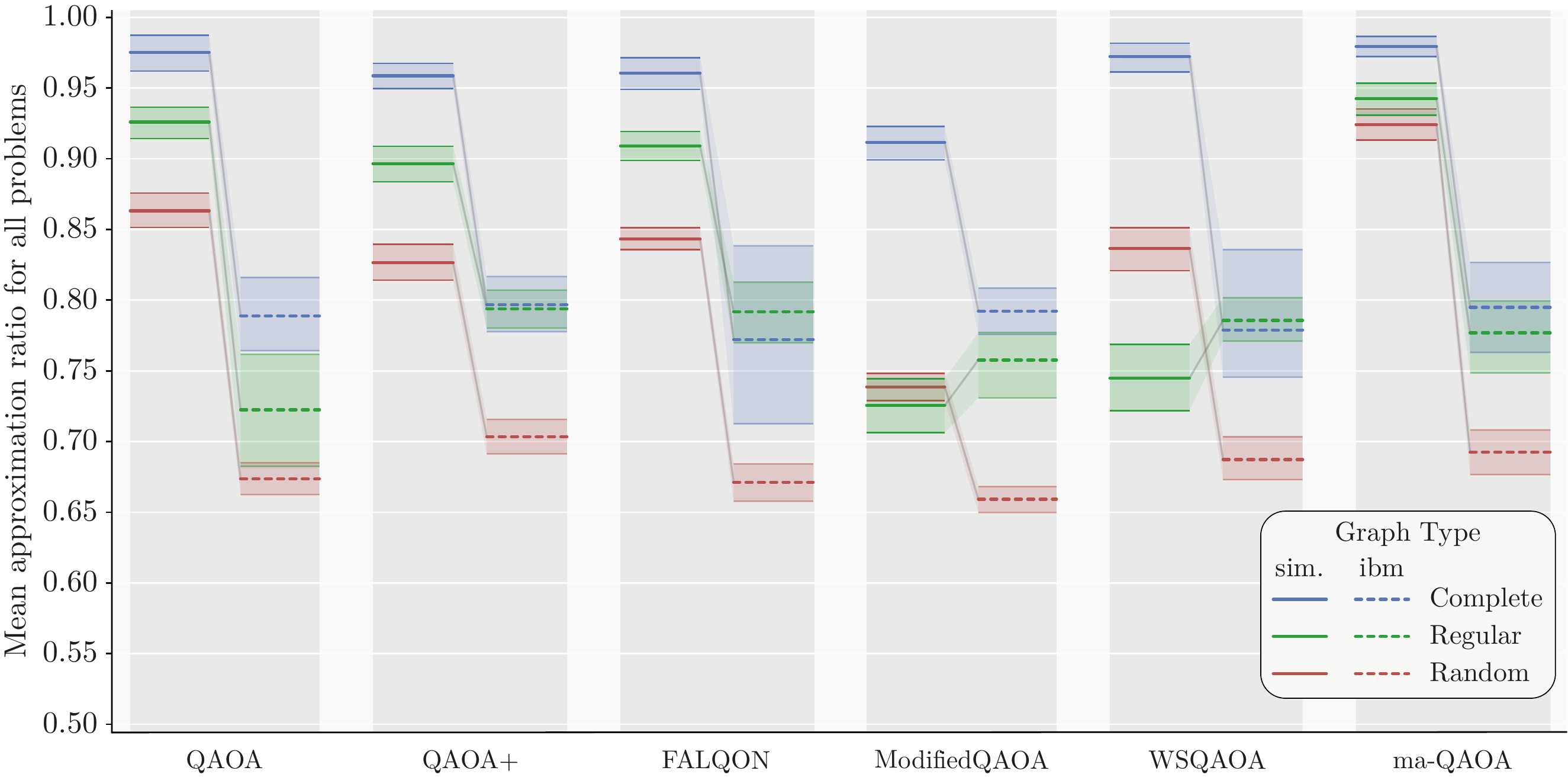}
	\caption[Performance of QAOA variants]{Mean approximation ratios of the QAOA and its variants on different MaxCut problems. 
 For each variant: on the left we report the results from noiseless simulations and on the right the ones from real hardware.} \label{fig:Approximation_ratio_vs_graph_type}
\end{figure}

\begin{figure}[ht!]
	\centering
	\includegraphics[width=1\linewidth]{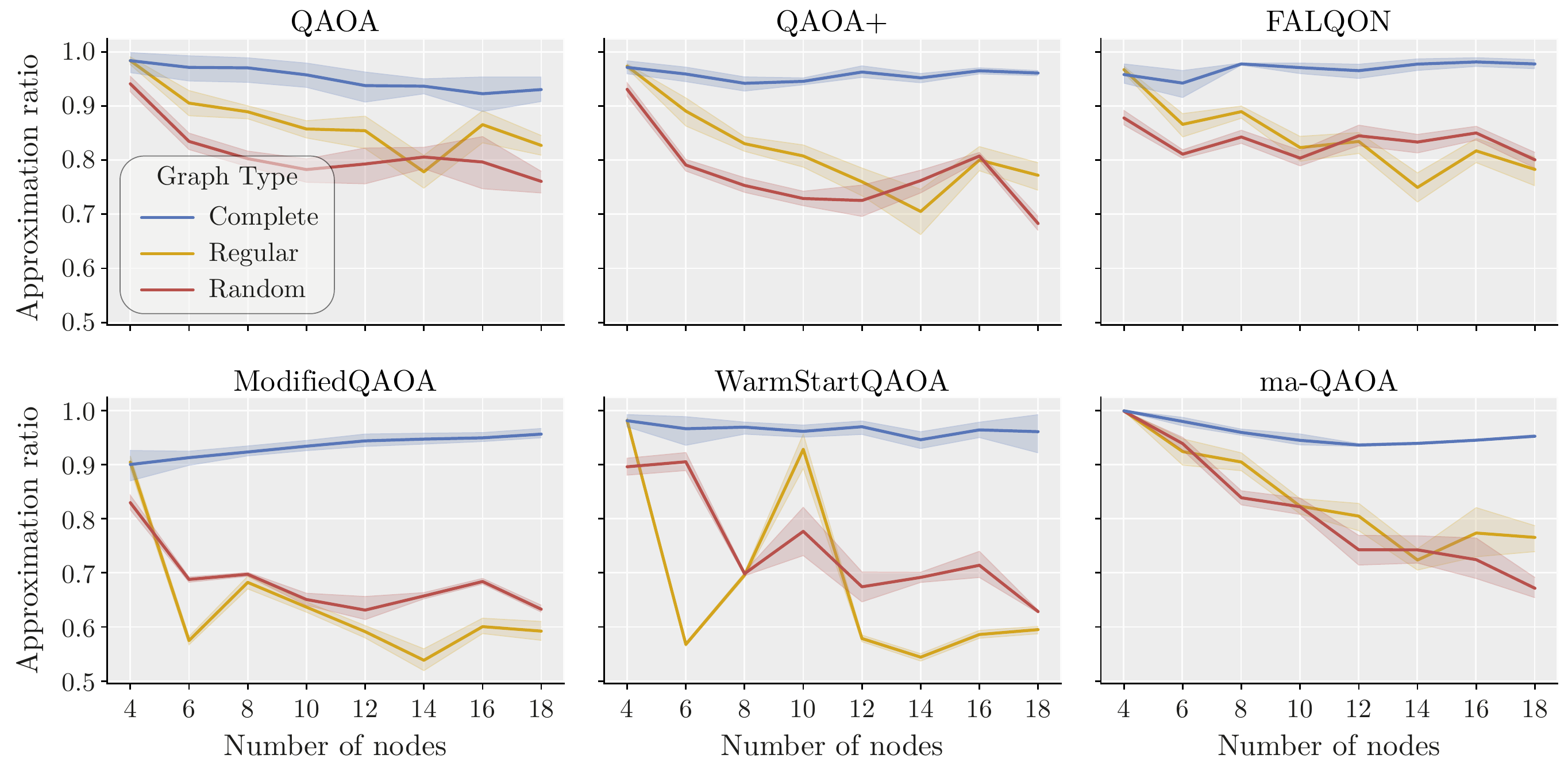}
	\caption[Performance of QAOA variants on a simulator]{Impact of graph size and type on approximation ratios by QAOA variant. 
 A declining trend of mean approximation ratios with increasing graph size is observed. 
 A strong dependence on graph type is evident.
 (Results from simulations)}%
	\label{fig:sim_Approximation_ratio_vs_num_nodes}%
\end{figure}

\begin{figure}[ht!]
	\centering
	\includegraphics[width=1\linewidth]{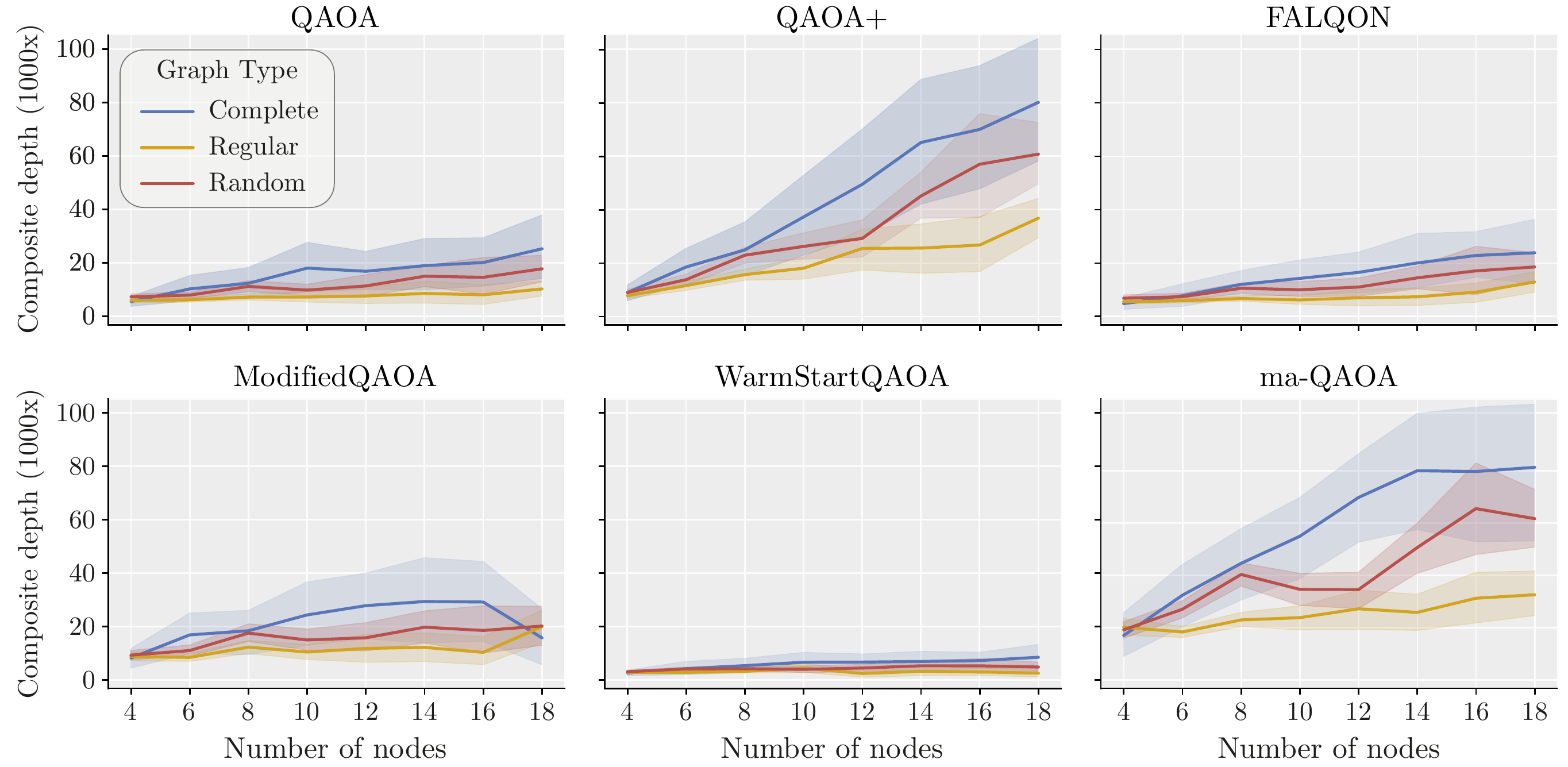}
	\caption[Amount of resources employed by different QAOA variants]{Composite depth of the QAOA and its variants on different MaxCut problems using simulators, namely the depth of the circuit times the number of iterations required. This measure provides an agnostic indication of the computational time required on the quantum device. 
    }%
\label{fig:sim_Gates_used_vs_num_nodes}%
\end{figure}

The experimental results provide insights into the performance of different QAOA variants when applied to various graph types and sizes.
Using the mean approximation ratios as a key performance metric (Figures~\ref{fig:Approximation_ratio_vs_graph_type},~\ref{fig:sim_Approximation_ratio_vs_num_nodes} and~\ref{fig:sim_Gates_used_vs_num_nodes}; Tables~\ref{tab:stats_summary},~\ref{tab:approx_to_layer},~\ref{tab:sim_to_ibm} and~\ref{tab:param_similarity}), we identify some general trends and patterns and discuss the trade-offs between approximation performance, resource consumption, and the consequent implications for problem-solving efficacy. 

\paragraph{Variation of approximation ratio across graph types:} As illustrated in Figure~\ref{fig:Approximation_ratio_vs_graph_type}, the graph type---complete, regular or random---significantly influences the approximation ratio achieved by a QAOA variant.
On both simulations and on real quantum hardware, all variants demonstrate superior approximation ratios when applied to complete and regular graphs, achieving mean ratios as high as 0.98, for complete graphs, and as high as 0.94 for regular graphs on simulations. 
Conversely, when applied to random graphs, the approximation ratio drops to at most 0.92. 

This trend gets even more pronounced as the number of nodes increases (Figure~\ref{fig:sim_Approximation_ratio_vs_num_nodes}).
In complete graphs most variants manage to retain a rather high mean approximation ratio of over 0.9 for at least up to 18 nodes.
On the other hand, in random graphs of 18 nodes the approximation ratio typically lies between 0.6 and 0.8.

This stark variance emphasizes the fundamental influence of the underlying graph structure on a QAOA variant's performance.
It substantiates the observations discussed in Section~\ref{sec:empirical_evidence} and underscores the need for a tailored selection of a QAOA variant keeping in mind the specific graph structure at hand.
A clear understanding of the interplay between graph types and QAOA variants is essential for achieving optimal quantum optimization results.

\paragraph{Degradation in approximation ratio with increasing graph size:}
Figure~\ref{fig:sim_Approximation_ratio_vs_num_nodes} highlights a tendency for mean approximation ratios to decrease as the number of graph nodes increases in regular and random graphs, and in a lesser degree in complete graphs. 
However, this downtrend is not universally applicable to all QAOA variants and is particularly noticeable when scaling from a minimal number of nodes (specifically, 4). 
Beyond this point, the descending trend can be observed, albeit with less pronounced impact. 
Notably, FALQON and vanilla QAOA appear to exhibit greater robustness in the face of this trend, suggesting a relative resilience against the deteriorating effect of increasing graph sizes.

This trends holds true even in the absence of noise. 
For instance, for QAOA, the approximation ratio falls from 0.98 and 0.94 (regular and random, respectively) for 4 nodes to 0.83 and 0.76 for 18 nodes. 
Similarly, FALQON's approximation ratio falls from 0.97 and 0.88 (regular and random, respectively) for 4 nodes to 0.78 and 0.80 for 18 nodes. 
These results indicate that larger graph sizes might make the optimization problem more challenging, and the effectiveness of QAOA variants diminishes as the graph size grows.

\paragraph{Relative efficiency of QAOA variants:} 
Our experimental findings highlight the variation in the performance of different QAOA variants across varying graph types and sizes.
For instance, FALQON and the standard QAOA show the most steady performance across varying graph sizes, as shown in Figure~\ref{fig:sim_Approximation_ratio_vs_num_nodes}.
As another example, the performance of WS-QAOA reveals surges in approximation ratios.
This variability, especially apparent in 10-node regular graphs, stems from its warm-starting algorithm that occasionally produces near-optimal parameters early in the optimization process.

These observations underscore the necessity of choosing an appropriate QAOA variant to achieve optimal results as some variants may be better suited to specific problem instances or graph structures.
However, it is important to note that the relative efficiency of a QAOA variant depends not only on its performance in isolation but also on how it compares to other variants under the same conditions. 
For example, while \maqaoa{} might offer high approximation ratios, it also require significantly more gates, especially as the number of nodes increases (Figure~\ref{fig:sim_Gates_used_vs_num_nodes}). 
Therefore, the relative efficiency of a QAOA variant should be evaluated considering both its performance and the computational resources it requires.

\paragraph{Balancing approximation ratio and resource use:}
The trade-off between the mean approximation ratio and the computational resources required by each QAOA variant is a vital metric to consider.
As illustrated in Figure~\ref{fig:sim_Gates_used_vs_num_nodes} and summarized in Table~\ref{tab:stats_summary} variants present unique trade-offs between their approximation capabilities and the amount of computational resources they demand. 
Some variants achieve higher approximation ratios but require more gates, have higher circuit depth, or need more circuit evaluations, resulting in increased computation time and resource usage.
For example, comparing the \maqaoa{} and WS-QAOA variants, \maqaoa{} might provide higher approximation ratios on larger graphs, however, it also requires significantly more resources in terms of effective circuit depth (circuit depth $\times$ circuit calls). 
On the other hand, WS-QAOA could be less resource-intensive while still delivering reasonably high approximation ratios.
However, it's crucial to note, further complicating the matter, that WS-QAOA includes a relatively expensive classical warm-starting step in its initialization, which contributes to its total resource consumption but not the quantum resource consumption.

The choice between these two QAOA variants would depend on the specific problem instance, available computational resources, and the desired balance between performance and resource usage.
\begin{table}[ht]
\centering
\resizebox{\textwidth}{!}{
\begin{tabular}{@{}lrrrrrr@{}}
\toprule
& \textbf{QAOA} & \textbf{QAOA+} & \textbf{FALQON} & \textbf{ModifiedQAOA} & \textbf{WS-QAOA} & \textbf{ma-QAOA} \\
\midrule
\textbf{Mean Approximation Ratio} & 0.87 & 0.84 & 0.87 & 0.72 & 0.79 & 0.88 \\
\textbf{Average Circuit Depth} & 96.46 & 97.31 & 104.73 & 167.70 & 44.79 & 90.73 \\
\textbf{Number of Circuit Calls} & 108.25 & 287.16 & 104.31 & 95.72 & 99.28 & 450.97 \\
\bottomrule
\end{tabular}
}
\caption{Summary statistics of the selected QAOA variants across all implementation combinations (simulation results).}
\label{tab:stats_summary}
\end{table}   

\paragraph{Effect of circuit layer depth:} 
Based on our experiments (Table~\ref{tab:approx_to_layer}), it's clear that the circuit layer depth has a varying impact on the approximation ratio achieved.
While it plays a significant role it seems to not be as significant as the other factors discussed, at least for depths up to $p=8$.

For instance, the QAOA variant shows an overall increase in the approximation ratio as the layer depth progresses from 1 to 8. 
There are minor fluctuations in this trend at depths 5 and 6, but the general upward pattern persists. 
This suggests that the QAOA variant tends to perform better with an increased layer depth.
Similarly, the FALQON variant shows a consistent improvement in the approximation ratio with increasing layer depth.
In general, the ma-QAOA exhibits the greatest improvement with increased depth, other variants show either a moderate rise (such as QAOA and WS-QAOA) or a more limited response (such as QAOA+ and ModifiedQAOA).

\begin{table}[htp!]
\centering
\begin{tabular}{lrrrrrrrr}
\toprule
\textbf{depth} ($p$)	&  1         &  2         &  3         &  4         &  5         &  6         &  7         &  8		\\
\midrule
\textbf{QAOA}            &  0.810  &  0.854  &  0.860  &  0.865  &  0.901  &  0.892  &  0.917  &  0.909		\\
\textbf{QAOA+}        &  0.789  &  0.856  &  0.855  &  0.858  &  0.855  &  0.853  &  0.838  &  0.889		\\
\textbf{FALQON}          &  0.807  &  0.849  &  0.865  &  0.890  &  0.890  &  0.885  &  0.899  &         -		\\
\textbf{ModifiedQAOA}    &  0.684  &  0.746  &  0.726  &  0.734  &  0.733  &  0.724  &  0.725  &  0.735		\\
\textbf{WS-QAOA}   &  0.785  &  0.756  &  0.770  &  0.806  &  0.804  &  0.796  &  0.790  &  0.817		\\
\textbf{ma-QAOA}  &  0.846  &  0.990  &  0.999  &  0.994  &  0.998  &  0.948  &  0.960  &  0.992		\\
\bottomrule
\end{tabular}
\caption{Mean approximation ratio achieved in relation to circuit layer depth ($p$).}%
\label{tab:approx_to_layer}
\end{table}

\paragraph{Noise-free simulations vs. real quantum hardware:}
The detrimental effect of running on noisy hardware is clearly evidenced when comparing the performance of the various QAOA variants in noise-free simulations versus real quantum hardware (Table~\ref{tab:sim_to_ibm}). 
The real quantum hardware results show a significant reduction in the mean approximation ratios across all the variants as was expected and discussed in Section~\ref{sec:noiseanderrors}.
It is interesting to observe is that in most instances, the general trends associated with different graph types typically persist (Figure~\ref{fig:Approximation_ratio_vs_graph_type}).
This highlights that the underlying structure of a graph, whether it's complete, regular, or random, plays a pivotal role in determining the performance of each QAOA variant.

\begin{table}[htp!]
\centering
\resizebox{\textwidth}{!}{
\begin{tabular}{rcccccc}
    \toprule
        & \textbf{QAOA} & \textbf{QAOA+} & \textbf{FALQON} & \textbf{ModifiedQAOA} & \textbf{WS-QAOA} & \textbf{ma-QAOA} \\
    \midrule
    \textbf{Simulation}    & $0.899 \pm 0.008$ & $0.867 \pm 0.009$ & $0.881 \pm 0.007$ & $0.743 \pm 0.011$ & $0.802 \pm 0.014$ & $0.936 \pm 0.008$ \\
    \textbf{Real hardware} & $0.702 \pm 0.019$ & $0.710 \pm 0.014$ & $0.716 \pm 0.020$ & $0.705 \pm 0.016$ & $0.727 \pm 0.016$ & $0.732 \pm 0.016$ \\
    \bottomrule
\end{tabular}}
\caption[Mean approximation achieved for each variant for all problem types and sizes.]{Mean approximation achieved for each variant for all problem types and sizes. Comparison between real hardware runs (IBMQ) and noise-free simulations.}%
\label{tab:sim_to_ibm}
\end{table}

\paragraph{Proximity of optimal parameters to initial random guess:} The cosine similarity metrics presented in Table~\ref{tab:param_similarity} provide insight into the effectiveness of the COBYLA optimization method in exploring the parameter space for each QAOA variant. 
The values indicate how close the optimal parameters are to their initial random guesses. 
For most variants, such as QAOA, QAOA+, ModifiedQAOA, WS-QAOA, and ma-QAOA, high mean cosine similarity values (greater than 0.96) suggest that the optimization process may not be thoroughly exploring the parameter space. 
This is further corroborated by the low standard deviations, indicating less deviation from the initial parameters on all problem instances. 
On the other hand, FALQON's significantly lower mean cosine similarity (approximately 0.51) and higher standard deviation hint at a more extensive exploration of the parameter space. 
In the case of WS-QAOA, the high cosine similarity is expected due to its design, which encourages starting near optimal parameters. 
This analysis underscores the influence of the optimization process on the performance of QAOA variants, and highlights the need for more effective parameter exploration strategies.
\begin{table}[htp!]
\centering
\resizebox{\textwidth}{!}{
\begin{tabular}{lcccccc}
\toprule
 & \textbf{QAOA} & \textbf{QAOA+} & \textbf{FALQON} & \textbf{ModifiedQAOA} & \textbf{WS-QAOA} & \textbf{ma-QAOA} \\
\midrule
\textbf{Mean cosine similarity}    & 0.988  & 0.963  & 0.511  & 0.986  & 0.978  & 0.976\\
\textbf{Standard Deviation}        & 0.009  & 0.037  & 0.257  & 0.016  & 0.067  & 0.015\\
\bottomrule
\end{tabular}
} 
\caption[Summary of cosine similarity metrics across QAOA variants.]{A summary of cosine similarity metrics across QAOA variants, indicating the proximity of optimal parameters to the initial random guess. High similarity suggests that the COBYLA optimization method may not be exploring the parameter space effectively. The exception is FALQON, which shows substantial divergence from the initial parameters, indicating more extensive exploration. WS-QAOA is expected to start near the optimal parameters due to its design.}%
\label{tab:param_similarity}
\end{table}


\subsection{Limitations and Future Work}
We have presented a curated set of experimental results and observations from our analysis of the QAOA and its variants.  
While this account provides a comprehensive overview, the extensive nature of the QAOA landscape means that numerous findings and results could not be included in this review paper.  
We are currently compiling a more extensive research paper that will delve deeper into the experiments, source code, and outcomes, providing a more granular examination of the performance of various QAOA variants including newer variants such as the ADAPT-QAOA and RecursiveQAOA.
We are also expanding our investigation's graph sizes and graph types to include an array of different MaxCut, weighted-MaxCut, and other problems, thereby offering a more robust comparison across different graph types and sizes.

Moreover, we are broadening our exploration to include a more diverse range of optimizers, enhancing our understanding of the efficiency of various optimization techniques in the context of QAOA, while our experimental setup has also been increased by expanding the number of nodes.
 
The released code will also provide an interactive interface for the implemented QAOA variants, allowing for greater exploration and comparison of their performance across a broader range of problem instances and graph structures.

    \section{Discussion}%
    \label{sec:discussion}
    In this section we provide a comprehensive discussion of our findings.
We begin our discussion with a brief overview of all the results followed by an exploration of proposed applications for the QAOA.
We then offer some insights on the quantum advantage of QAOA and finally conclude with suggestions for future research directions.

\subsection{Summary of Analysis}
\paragraph{QAOA ansatz improvements}

Ansatz selection in QAOA is crucial for solving problems efficiently and balancing specificity and generality to avoid overfitting. 
Multiple approaches have been proposed, including ma-QAOA~\cite{herrmanMultiangleQuantumApproximate2022}, QAOA+~\cite{chalupnikAugmentingQAOAAnsatz2022}, ab-QAOA~\cite{yuQuantumApproximateOptimization2022}, and PG-QAOA~\cite{yaoPolicyGradientBased2020}, which introduce new parameters, layers, and optimization techniques to enhance the traditional QAOA ansatz.

ADAPT-QAOA~\cite{zhuAdaptiveQuantumApproximate2020} provides an iterative method that selects QAOA mixers from a pool of operators, leading to faster convergence and reduced resource requirements. 
RQAOA~\cite{bravyiObstaclesVariationalQuantum2020, patelReinforcementLearningAssisted2022, bravyiHybridQuantumclassicalAlgorithms2022} is a non-local variant that iteratively reduces problem size and employs classical methods for the remaining problem. 
WS-QAOA~\cite{eggerWarmstartingQuantumOptimization2021} attempts to initialize QAOA based on the solution to relaxed QUBO problems and demonstrates the ability to retain the performance guarantees of classical algorithms, such as the GW bound when solving the MaxCut problem.
Quantum Alternating Operator Ansatzes~\cite{hadfieldAnalyticalFrameworkQuantum2022,hadfieldQuantumApproximateOptimization2019,ruanQuantumApproximateOptimization2023,bartschiGroverMixersQAOA2020,goldenThresholdBasedQuantumOptimization2021,goviaFreedomMixerRotation2021} offer a more flexible framework by alternating between general sets of operators, making them suitable for a broader range of problems with hard and soft constraints.
FALQON~\cite{magannFeedbackBasedQuantumOptimization2022,magannLyapunovcontrolinspiredStrategiesQuantum2022} and FQAOA~\cite{yoshiokaFermionicQuantumApproximate2023} tackle quantum optimization from different angles, with FALQON focusing on feedback-based optimization without classical intervention and FQAOA using fermion particle number preservation to impose constraints intrinsically. 
Both methods show potential performance advantages over existing techniques. 
Lastly, Quantum Dropout~\cite{wangQuantumDropoutEfficient2022} offers an innovative approach to handle hard cases in combinatorial optimization problems, selectively modifying the quantum circuit to improve QAOA performance.


\paragraph{Parameter optimization}
Parameter optimization is a crucial aspect of QAOA, especially as the depth and complexity of the algorithm increase~\cite{zhouQuantumApproximateOptimization2020}.
Considering the simultaneous increase in complexity with the increase in depth, methods can be developed to reduce the ansatz depth.
The parameters of the QAOA ansatz create a parameter space, and these parameters should be effectively chosen to govern a proper run of the QAOA\@.
Various techniques have been proposed to address the challenge of finding good initial parameters, such as heuristic strategies~\cite{zhouQuantumApproximateOptimization2020}, parameter fixing~\cite{leeParametersFixingStrategy2021}, graph neural networks~\cite{jainGraphNeuralNetwork2021}, and parameter transferability across graphs~\cite{galdaTransferabilityOptimalQAOA2021,shaydulinMultistartMethodsQuantum2019}. 

Classical optimization algorithms such as gradient-based methods~\cite{crooksPerformanceQuantumApproximate2018,yaoPolicyGradientBased2020} and gradient-free methods~\cite{fernandez-pendasStudyPerformanceClassical2022} have been utilized to explore the QAOA parameter space.
Gradient-based approaches like gradient descent and policy gradient are widely used for optimizing the parameters of quantum circuits like QAOA. However, they can have high computational times when compared to gradient-free methods~\cite{fernandez-pendasStudyPerformanceClassical2022}. 

Various machine learning techniques can be experimented with to provide more effective parametrization that automatically learns and employs the ideal parameter patterns.
Machine learning approaches, including reinforcement learning~\cite{patelReinforcementLearningAssisted2022,khairyLearningOptimizeVariational2020} and meta-learning~\cite{wangQuantumApproximateOptimization2021}, have been applied to improve parameter optimization in QAOA.
These Machine Learning approaches have been applied to train models and find optimal parameters for QAOA, often resulting in faster convergence and better performance.

Overcoming barren plateaus in the cost function landscape is a significant challenge in training parameterized quantum circuits like QAOA~\cite{mccleanBarrenPlateausQuantum2018}. 
Strategies such as layerwise learning and warm-start techniques have been proposed to mitigate this issue~\cite{skolikLayerwiseLearningQuantum2021,eggerWarmstartingQuantumOptimization2021,trugerWarmStartingQuantumComputing2023}. 
Furthermore, the ability to transfer and reuse optimal parameters across different problem instances can potentially enhance the optimization process~\cite{galdaTransferabilityOptimalQAOA2021,shaydulinMultistartMethodsQuantum2019}.



Parameter symmetry is also explored in quantum algorithms like QAOA.
Utilizing parameter symmetries can significantly enhance the optimization process in QAOA by reducing degeneracies and enabling more efficient performance~\cite{zhouQuantumApproximateOptimization2020,shiMultiAngleQAOADoes2022,shaydulinClassicalSymmetriesQuantum2021,akshayParameterConcentrationsQuantum2021,shaydulinExploitingSymmetryReduces2021}.
This approach can provide several advantages, including reducing the QAOA energy evaluation cost, accurately predicting QAOA performance, and improving circuit training by concentrating optimal parameters as an inverse polynomial relative to problem size~\cite{akshayParameterConcentrationsQuantum2021,shaydulinClassicalSymmetriesQuantum2021,shaydulinExploitingSymmetryReduces2021}.
Research also indicates that symmetries can minimize the number of parameters without compromising solution quality, and inherent symmetries can aid in the identification and elimination of degeneracies in the parameter space, leading to a more effective search for optimal parameters~\cite{zhouQuantumApproximateOptimization2020,shiMultiAngleQAOADoes2022}.
Certain studies even suggest that optimal QAOA parameters can sometimes be derived analytically, further simplifying the search process~\cite{wangQuantumApproximateOptimization2018}.

\paragraph{Efficiency and performance}

QAOA is considered a leading candidate for achieving quantum advantage, both in terms of runtime efficiency and quality of the solution.
Several works have already highlighted its speedup potential compared to classical algorithms in optimization~\cite{crooksPerformanceQuantumApproximate2018, zhangApplyingQuantumApproximate2022, ebadiQuantumOptimizationMaximum2022} and other problem domains~\cite{jiangNearoptimalQuantumCircuit2017, niuOptimizingQAOASuccess2019, anQuantumLinearSystem2022}. 
Despite this, quantum speedup with QAOA on near-term devices is unlikely~\cite{guerreschiQAOAMaxCutRequires2019}, primarily due to challenges in optimizing the variational parameters in large QAOA circuits~\cite{herrmanMultiangleQuantumApproximate2022, shaydulinMultistartMethodsQuantum2019, mccleanBarrenPlateausQuantum2018, wangNoiseinducedBarrenPlateaus2021}.
The prospects for achieving quantum advantage on early fault-tolerant quantum computers are also limited for quadratic speedups due to substantial overheads of error correction, which slows down the algorithm significantly~\cite{sandersCompilationFaultTolerantQuantum2020}.
Problem structure should be exploited to achieve high-order speedups when implementing QAOA and other quantum algorithms~\cite{mccleanLowDepthMechanismsQuantum2021}.
Meanwhile, many approaches to improve the runtime efficiency of QAOA have been proposed, primarily targeting the parameter optimization process.
Other improvement strategies have also been explored, including modifying the QAOA ansatz~\cite{zhuAdaptiveQuantumApproximate2020, yuQuantumApproximateOptimization2022}, optimizing gate operations~\cite{liHierarchicalImprovementQuantum2020}, reducing the number of samplings~\cite{larkinEvaluationQAOABased2022}, etc. 

On the other hand, the solution quality, typically characterized by the approximation ratio in solving optimization problems, is another important metric for evaluating the effectiveness of QAOA.
Extensive efforts have been made to understand its theoretical solution guarantees, that is, the lower bounds on QAOA's performance, in asymptotic limits when applied to various problems and the comparison to state-of-the-art classical algorithms (see Table~\ref{tab:theoretical_performance}).
Most notably, QAOA at $p \geq 11$ could outperform the assumption-free classical SDP algorithms in solving the MaxCut problem on $D$-regular graphs with high girth ($> 2p+1$)~\cite{bassoQuantumApproximateOptimization2022} and the SK model~\cite{farhiQuantumApproximateOptimization2022}, which are intimately related~\cite{demboExtremalCutsSparse2017}.
At $p = 1$, QAOA was also demonstrated to surpass the threshold algorithm in the Max-$k$XOR problem on bounded-degree hypergraphs with random signs or no overlapping constraints, for $k > 4$~\cite{marwahaBoundsApproximatingMax2022}.
However, there are known obstacles arising from symmetry~\cite{bravyiObstaclesVariationalQuantum2020, marwahaBoundsApproximatingMax2022} and locality~\cite{barakClassicalAlgorithmsQuantum2022, farhiQuantumApproximateOptimization2020, farhiQuantumApproximateOptimization2020a, chouLimitationsLocalQuantum2022}, which prevent QAOA at modest depths from achieving optimal solutions.
Even for problems where QAOA is not constrained by locality, such as the fully connected $k$-spin models, constant-depth QAOA was shown to be bounded away from optimality~\cite{bassoPerformanceLimitationsQAOA2022}.
Empirically, studies have also been conducted to benchmark the performance of QAOA against classical solvers, using both simulators and real hardware.
Several factors were identified to have significant impacts on the quality of solution provided by QAOA, including the circuit depth~\cite{crooksPerformanceQuantumApproximate2018}, entanglement~\cite{mccleanLowDepthMechanismsQuantum2021, chenHowMuchEntanglement2022, dupontEntanglementPerspectiveQuantum2022, dupontCalibratingClassicalHardness2022}, parameter optimization~\cite{zhouQuantumApproximateOptimization2020}, properties of the underlying graph~\cite{akshayReachabilityDeficitsQuantum2020, herrmanImpactGraphStructures2021}, etc.
Moreover, multiple strategies have been employed to assess the conditions at which it is advantageous to use QAOA over classical algorithms~\cite{lykovSamplingFrequencyThresholds2022, moussaQuantumNotQuantum2020, deshpandeCapturingSymmetriesQuantum2022}.

In summary, while research on QAOA's efficiency and performance has yielded promising results in certain problem instances, classical solvers remain highly competitive in solving a wide range of optimization problems, making it challenging to achieve quantum advantage with QAOA.
This challenge is further compounded by the detrimental effect of noise in current quantum computers.
Therefore, attaining a general and definitive quantum advantage with QAOA remains an open question. 
However, progress has been made in understanding the key aspects that influence the effectiveness of QAOA, such as circuit depth, parameter optimization, graph structures, and more, and corresponding strategies have been proposed to improve these aspects.
Future research and experimental method standardization are needed to assess the extent to which these proposals could make QAOA more advantageous than its classical counterparts.

\paragraph{Hardware considerations}

Both local and correlated errors in current hardware pose significant challenges to QAOA's scalability and performance.
In the case of local errors, theoretical studies suggested that the QAOA performance suffers from an exponential degradation with an increasing noise strength~\cite{xueEffectsQuantumNoise2021,marshallCharacterizingLocalNoise2020}.  
This translates to an exponential time complexity to make QAOA effective, severely limiting its scalability~\cite{lotshawScalingQuantumApproximate2022}.
Another major challenge is that even with just local noise, the parameter optimization of QAOA suffers from the noise-induced barren plateaus~\cite{wangNoiseinducedBarrenPlateaus2021}, which cannot be mitigated with techniques that are effective for the noise-free ones~\cite{mccleanBarrenPlateausQuantum2018}.
The adverse effects of correlated errors have also been investigated, including crosstalk noise~\cite{maciejewskiModelingMitigationCrosstalk2021}, precision errors~\cite{quirozQuantifyingImpactPrecision2021}, the coherent error induced by residual ZZ-couplings~\cite{karamlouAnalyzingPerformanceVariational2021}, etc.
As a result, both theoretical~\cite{stilckfrancaLimitationsOptimizationAlgorithms2021, weidenfellerScalingQuantumApproximate2022, depalmaLimitationsVariationalQuantum2023, gonzalez-garciaErrorPropagationNISQ2022} and empirical studies~\cite{alamDesignSpaceExplorationQuantum2020, harriganQuantumApproximateOptimization2021, paganoQuantumApproximateOptimization2020} suggested that quantum advantage is unlikely with the noise level in current devices. 
To match the performance of classical devices, the error rates in quantum computers would need to improve significantly, even reaching below the fault-tolerant threshold for dense and non-hardware-native graphs.
Despite this, error mitigation techniques, such as optimizing SWAP networks~\cite{hashimOptimizedSWAPNetworks2022}, gate reduction strategies~\cite{alamCircuitCompilationMethodologies2020, alamNoiseResilientCompilation2020, alamEfficientCircuitCompilation2020, majumdarOptimizingAnsatzDesign2021, majumdarDepthOptimizedAnsatz2021}, exploiting problem symmetries~\cite{shaydulinErrorMitigationDeep2021, streifQuantumAlgorithmsLocal2021, weidingerErrorMitigationQuantum2023}, and other error-specific strategies~\cite{maciejewskiMitigationReadoutNoise2020, maciejewskiModelingMitigationCrosstalk2021, quirozQuantifyingImpactPrecision2021, mundadaExperimentalBenchmarkingAutomated2023}, have been proposed for NISQ devices.
Furthermore, over the past few years, devices, especially those utilizing trapped ions, have achieved significant advancements in terms of maturity. 
This progress has allowed for a monotonic improvement in the performance of QAOA with up to 10 layers for graphs consisting of 20 qubits~\cite{shaydulinQAOACdotGeq2023}.

Hardware-specific approaches focused on leveraging platform capabilities like trapped ions, neutral atoms, superconducting qubits, and photonic quantum computers to enhance QAOA's performance. 
For example, research into qudits and long-range interaction platforms has demonstrated potential improvements in optimization problems on neutral atom devices~\cite{dellerQuantumApproximateOptimization2022}. In order to address the primary issue with photonic devices~\cite{vikstalQuantumApproximateOptimization2023}, cat qubits have been employed instead of standard qubits, resulting in a more noise-resilient QAOA. In the context of this particular quantum computation model, significant efforts have also been devoted to fully utilizing the MBQC capabilities of photonic devices~\cite{proiettiNativeMeasurementbasedQuantum2022, zhangCompilationFrameworkPhotonic2022}. 
To achieve improved QAOA performances, the utilization of pulse-level access on IBM devices has been proposed~\cite{niuEffectsDynamicalDecoupling2022, jiOptimizingQuantumAlgorithms2023}.
A four-qubit Rydberg parity gate was introduced for encoding arbitrarily connected graphs in trapped neutral atom systems~\cite{dlaskaQuantumOptimizationFourBody2022}.
Ion native Hamiltonians were used to develop ansatz families for encoding all-to-all connectivity in planar quantum processors~\cite{rabinovichIonNativeVariational2022}.
Lastly, methods for constructing arbitrary coupling graphs using global entangling operations in quantum spin systems were also presented~\cite{rajakumarGeneratingTargetGraph2022}.

\subsection{Potential Applications and Use Cases}


Having discussed various aspects of QAOA in-depth, it is important to explore where the algorithm can and has been applied to modern problems. 
The implementation of quantum optimization algorithms is naturally limited by the capabilities of near-term quantum hardware, as discussed in previous sections, which makes empirical demonstrations valuable as benchmarks for expected outcomes. 
An example of such an empirical benchmark was provided by \citet{lotshawEmpiricalPerformanceBounds2021}. 

Near-term applications of these algorithms are exhibited by results such as those given by \citet{niroulaConstrainedQuantumOptimization2022}, who demonstrated the potential of near-term quantum computers, through QAOA, in addressing industry-relevant \textbf{constrained-optimization problems} by considering an extractive summarization problem. 
They used the Quantum Alternating Operator Ansatz algorithm with a Hamming-weight-preserving XY mixer (XY-QAOA) on a trapped-ion quantum computer. They effectively implemented XY-QAOA circuits on up to 20 qubits, emphasizing the necessity of direct constraint encoding into the quantum circuit.
Their results show that the right choice of algorithm and compatible NISQ hardware can already solve constrained optimization problems relevant to the modern industry and that quantum advantage is near for variational quantum algorithms. 

A similar demonstration of the right choice of quantum algorithm for a problem showing potential quantum advantage is the work of \citet{bravyiHybridQuantumclassicalAlgorithms2022} using RQAOA for approximate vertex $k$-coloring of a graph.
More specifically, RQAOA was applied to tackle the MAX-$k$-Cut problem, a notoriously tricky task in graph theory that involves finding an approximate $k$-vertex coloring of a graph.
Vertex coloring is a crucial problem that finds applications in many fields, such as scheduling, mobile radio frequency assignment, register allocation, etc., making it a pertinent issue in theoretical computer science and various industries.
Their study discovered that level-1 RQAOA is surprisingly competitive, often achieving higher approximation ratios than the best-known generic classical algorithm based on rounding an SDP relaxation for ensembles of randomly generated 3-colorable constant-degree graphs (Section~\ref{subsec:rqaoa}). 
This suggests that the RQAOA could be a powerful tool for NISQ devices and points towards its potential for outperforming classical solutions in certain instances of the graph coloring problem.

\citet{tabiQuantumOptimizationGraph2020} proposed a space-efficient quantum optimization algorithm tailored explicitly for the graph coloring problem.
Given current quantum computing architectures' varying strengths and limitations, they underscored the necessity for flexible circuit design approaches. 
They introduced a space-efficient quantum optimization algorithm that drastically reduces the required qubits for problem encoding. Also, they simultaneously lowered the required layers and optimization iteration steps to reach the optimal solution.
Though their proposed circuits are inherently deeper than traditional methodologies, the exponential reduction in qubit usage makes this an intriguing prospect for real-world applications, particularly given the constraints of current quantum hardware. 
The authors validated the performance of their approach through numerous numerical simulations. They concluded that analogous space-efficient embedding techniques could enhance other graph-related quantum optimization methods, which they earmarked for future exploration.

In a related application, \citet{cookQuantumAlternatingOperator2020} investigated the use of the QAOAnsatz (Section~\ref{subsec:qaltop}) on the Maximum $k$-Vertex Cover problem, a complex task with notable real-world implications such as network security and social network analysis.
Their study revealed that improved performance was achieved by using Dicke states and the complete graph mixer. 
Though challenged by the increasing complexity in subsequent rounds, the results exhibit a promising trend that aligns with the Quantum Adiabatic Algorithm, highlighting the feasibility and potential of QAOA in addressing complex optimization problems.

Another example of a real-world application of QAOA, and variational quantum algorithms more generally, is given by the works of \citet{azadSolvingVehicleRouting2022} and \citet{mohantyAnalysisVehicleRouting2022}, which tackled the Vehicle Routing Problem (VRP), an NP-hard optimization problem of considerable interest to both science and industry. 
It generalizes the traveling salesperson problem and consists of finding the most efficient vehicle route to a set of customers. 
These works showed the comparable performance of VQAs against classical solvers even in the presence of noise, making them relevant for near-term use of quantum algorithms for industry-relevant problems of supply chain management and scheduling. 
A similar result has been achieved by \citet{vikstalApplyingQuantumApproximate2020}, in which the QAOA was applied to the Tail Assignment Problem (TAP), which is the task of assigning individual aircraft to a given set of flights, minimizing overall cost.

In \textbf{financial industry}, \citet{bakerWassersteinSolutionQuality2022} benchmarked the performance of QAOA in portfolio optimization, offering another practical application. 
Their study focused on the solution quality determined by the normalized and complementary Wasserstein distance, suggesting that this measurement can serve as an application-specific performance benchmark.
Their findings indicated that the solution quality exhibited improvement as the QAOA circuit depth increased, reaching its peak at $p=5$ with 2 qubits for most tested systems and at $p=4$ with 3 qubits on a trapped ion processor. 
These results suggest the potential of QAOA and its variants in addressing financial optimization tasks.
A study by \citet{hodsonPortfolioRebalancingExperiments2019} further corroborates the application of QAOA in portfolio optimization. 
They experimentally analyzed the performance of a discrete portfolio optimization problem relevant to the financial industry on an idealized simulator of a gate-model quantum computer. 
Their study demonstrated the potential tractability of their application on NISQ hardware, with portfolios identified within 5\% of the optimal adjusted returns and optimal risk for a small eight-stock portfolio.
Their work also involved designing novel problem encoding and hard constraint mixers for the QAOAnsatz, demonstrating a method to tailor quantum algorithms to specific industry use cases.
Along the same lines, \citet{hegadePortfolioOptimizationDigitized2022} investigated the complex task of discrete mean-variance portfolio optimization by using the variant DC-QAOA (Section~\ref{subsec:dc-qaoa}).

In the domain of \textbf{Hamiltonian simulation}, another promising application of QAOA has emerged. 
\citet{lotshawSimulationsFrustratedIsing2023} explored the use of QAOA in simulating the frustrated Ising Hamiltonians, which serve as toy models crucial for understanding novel magnetic materials. 
The ground state properties of these materials are computationally intensive to calculate with traditional methods due to their inherent complexity.
Using QAOA, the authors examined the Ising spin models on unit cells of square, Shastry-Sutherland, and triangular lattices, finding that a modest number of measurements were sufficient to identify the ground states of the 9-spin Hamiltonians. 
Remarkably, this efficiency persisted even when the Hamiltonians induced frustration in the system, showcasing the potential of QAOA in physical simulations.
In more hybrid approaches, \citet{bradyOptimalProtocolsQuantum2021} combined Quantum Annealing with QAOA to prepare the ground state of a quantum system.
They discovered that the combination of both methods worked best most of the time and concluded that the optimal protocol for minimizing the energy of a quantum state under time constraints is often of the ``bang-anneal-bang'' form rather than the previously thought ``bang-bang'' pulse structure based on Pontryagin's principle~\cite{liangInvestigatingQuantumApproximate2020}. 

In \textbf{communication}, \citet{cuiQuantumApproximateOptimization2022} applied QAOA to the problem of maximum likelihood (ML) detection of binary symbols transmitted over a multiple-input and multiple-output (MIMO) channel. 
They demonstrated that a QAOA-based ML detector could approach the performance of a classical ML detector, revealing the potential for large-scale classical optimization problems to be effectively addressed on NISQ computers.

In the work of \citet{chandaranaDigitizedcounterdiabaticQuantumAlgorithm2022}, a hybrid classical-quantum ansatz was proposed, which uses counterdiabatic protocols to solve the \textbf{protein folding} problem on a tetrahedral lattice. 
The ansatz is inspired by digitized-counterdiabatic quantum computation, which can accelerate adiabatic quantum algorithms and compress quantum circuits~\cite{chandaranaDigitizedcounterdiabaticQuantumApproximate2021, wurtzCounterdiabaticityQuantumApproximate2022}. 
The authors applied this algorithm to various proteins with different numbers of amino acids and qubits, and it was shown to outperform state-of-the-art quantum algorithms, such as the original QAOA, in terms of convergence and circuit depth. 
These demonstrations were performed on several quantum hardware platforms, including trapped-ions and superconducting systems, achieving high success probabilities and efficient ground state convergence. 
The authors noted the remaining difficulties for such an approach to solving complex optimization problems with quantum computers. However, the proposed digitized counterdiabatic protocol opened an investigation into implementing problem-inspired ansatz to industrial use cases using NISQ devices.

In \textbf{computer vision}, \citet{liHierarchicalImprovementQuantum2020} investigated the application of partially occluded object detection under the broader framework of QUBO\@.
They proposed a three-tiered improvement approach for a hybrid quantum-classical optimization for object detection, resulting in significant execution speedup---an over 13-fold speedup was achieved by selecting L-BFGS-B as the classical optimizer.
They also demonstrated that optimally rescheduling gate operations, especially in deeper circuits, resulted in better circuit fidelity at the third level. 
The findings of this study shed light on the potential benefits of QAOA for object detection tasks.

Very recently, \citet{dateQUBOFormulationsTraining2021} proposed using quantum computers to accelerate \textbf{training of machine learning models}.
They formulated three machine learning problems (linear regression, support vector machine, and balanced $k$-means clustering) as QUBO problems and suggested solving them with adiabatic quantum computing.
In this context, QAOA can be employed as an alternative method to find good approximated solutions to such problems.
This could potentially pave the way for using QAOA in the deep learning realm for neural networks' training in the future.
Some other potential future scopes from~\cite{killoranContinuousvariableQuantumNeural2019} include experimental implementation of non-Gaussian gates, exploration of quantum advantages in the presence of decoherence, development of specialized QNNs, investigation of joint architectures, and deeper exploration of fundamental quantum physics concepts in QNNs.

\subsection{The Quantum Advantage of QAOA}
The QAOA shows promising potential in realizing quantum advantage over classical algorithms in specific problem instances and under certain conditions.
This approach targets classical optimization problems, providing increased efficiency and improved performance solutions.
However, this advantage is not yet fully realized due to various challenges, including noise and hardware limitations in near-term quantum devices and the competitiveness of state-of-the-art classical solvers.


In terms of computational runtime efficiency, instances where QAOA has demonstrated superiority include solving the MaxCut problem on dense graphs, offering exponential acceleration for large Minimum Vertex Cover (MVC) problems, and achieving a superlinear quantum speedup compared to Simulated Annealing (SA) for the Maximum Independent Set (MIS) problem on specific graph instances. 
Moreover, QAOA has shown potential in other areas, such as unstructured search problems and Quantum Linear System Problems (QLSP), where it outperforms various classical and quantum algorithms.

In terms of solution quality, multiple studies comparing QAOA to classical algorithms on various optimization problems, including MaxCut, Max-$k$XOR, and other Constraint Satisfaction Problems (CSPs), have revealed that QAOA can outperform classical algorithms under specific conditions or for certain problems. 
For instance, QAOA surpasses classical threshold algorithms for Max-$k$XOR problems when $k > 4$ in the large degree limit. 
Furthermore, compared to classical local algorithms, QAOA provides superior solutions for MaxCut on large-girth random regular graphs at depth $p = 11$ and beyond.

Despite these encouraging results, other studies have found that classical algorithms can still match or surpass QAOA's performance in many scenarios.
A simple modification of the classical algorithm, the Gaussian wave process, has achieved a larger improvement over random assignment compared to QAOA$_1$ in the asymptotic limit as $D \to\infty$ for MaxCut on triangle-free $D$-regular graphs. 
There also exists a 2-local classical MaxCut algorithm that consistently outperforms QAOA$_2$ for all $D$-regular graphs of girth $> 5$.
Furthermore, QAOA can encounter limitations in scenarios where it cannot outperform the best classical algorithm, such as solving the MaxCut problem on bipartite $D$-regular graphs. 
It can also face restrictions due to the locality constraint, which limits its algorithmic performance when the entire graph is not visible to the algorithm. 
When applied to problems with the Overlap Gap Property (OGP), such as the MIS problem on sparse random graphs or Max-$k$XOR with even $k \geq 4$, QAOA can face obstructions limiting its performance.
Despite these challenges, a potential quantum advantage is still achievable in some problems, particularly when QAOA surpasses sub-logarithmic depths, as classical Approximate Message Passing (AMP) algorithms exhibit suboptimality. 
Consequently, more research is needed to understand the extent of QAOA's capabilities and limitations and identify improvement areas or alternative optimization approaches.

One area that can be improved to bolster the QAOA's efficacy is its ansatz design.
Recent advancements in ansatz variants such as ab-QAOA, ADAPT-QAOA, and QAOAnsatz have significantly improved over the standard QAOA\@. 
The ab-QAOA drastically reduces computation time and achieves faster convergence, with improvements increasing proportionally with problem size. 
The ADAPT-QAOA, applying the principle of shortcuts to adiabaticity, exhibits enhanced convergence speed and reduces the number of CNOT gates and optimization parameters by about half, streamlining quantum computation. 
QAOAnsatz introduces more flexibility in defining parts of the ansatz, thereby expanding the range of solvable problems and enabling a larger and potentially more useful set of states to be represented than possible with the original formulation. 
These advancements could accelerate the realization of quantum advantage in solving combinatorial optimization problems.
Other ansatz variants, such as WS-QAOA, have also shown potential advantages over the standard QAOA at low depth, critical for implementation on NISQ devices. 
Likewise, the FALQON+ algorithm has illustrated a significant increase in approximation ratio and success probability with a minor increase in circuit depth and noise degradation, making it suitable for NISQ devices. 
Furthermore, FQAOA has demonstrated substantial performance advantages in portfolio optimization problems by effectively handling constraints. 
Other approaches like Quantum Dropout and ST-QAOA have also indicated improvements in performance for specific types of combinatorial optimization problems.

The performance of QAOA is also strongly tied to its depth (i.e., the number of layers), as a larger depth implies more parameters to optimize. 
Finding optimal parameters often requires polynomial time, making it difficult to achieve a quantum speedup with QAOA, even at low depths. 
Additional complications, such as barren plateaus, aggravate this challenge and necessitate improved parameter optimization strategies.
Therefore, parameter optimization is crucial for QAOA's quantum advantage.
Various techniques have been proposed to effectively initialize QAOA parameters, achieving better results than random initializations.
Many works have also shown that when coupled with diverse optimization strategies such as gradient-based methods, gradient-free techniques, and certain machine learning approaches, QAOA often displays rapid convergence to optimal solutions and better solution quality. 
Moreover, parameter concentration in certain problem instances and symmetries can be exploited to increase the efficiency of parameter optimization further.

There remain, however, substantial obstacles to achieving a quantum advantage in more practical settings, i.e., when running QAOA on real quantum computers. 
Challenges related to various noise sources, including local and correlated errors, present significant hurdles. 
While strategies to mitigate some of these errors have been proposed, quantum devices' error rates need to continue to improve in order to address issues such as noise-induced barren plateaus.
Therefore, work towards achieving consistent quantum advantage with QAOA remains an ongoing and active area of research.

Nonetheless, QAOA exhibits strong compatibility with NISQ devices, enhancing its near-term applicability. 
Some QAOA variants display resilience to noise, making them well-suited for NISQ devices. 
Empirical evidence suggests that QAOA could effectively manage industry-relevant constrained-optimization problems on these quantum platforms, efficiently addressing scalability and compatibility issues. 
This versatility, combined with the successful application of QAOA to various real-world problems, underscores its adaptive nature and a broad range of use cases.









\subsection{Open Questions and Future Directions}

The QAOA represents a noteworthy milestone in quantum computing with its potential for addressing various optimization problems.
Despite significant progress, numerous research opportunities remain in the realm of QAOA. 
Unlocking the full potential of this quantum optimization algorithm may require a combination of theoretical understanding, empirical studies, algorithmic improvements, parameter optimization, and a thorough exploration of entanglement and problem structures, among other factors.




\paragraph{Theoretical insights \& mathematical frameworks:} Enhancing our theoretical understanding of QAOA, primarily at higher depths, is fundamental. 
This includes developing rigorous theoretical frameworks that provide a comprehensive understanding of QAOA's behavior, performance, and inherent limitations compared to classical optimization algorithms.
Further research is needed to understand the interplay between entanglement and circuit depth, particularly how they collectively impact QAOA's performance at higher depths.
This will also facilitate the development of problem-specific algorithms and efficient ansatz designs.
Research on how different problem structures interact with QAOA's functionality and performance is also essential.
This understanding can offer valuable insights into improving performance and applicability across diverse problem sets.

Some specific questions to be addressed in this regard include:

\begin{itemize}
    \item  A rigorous proof of QAOA achieving the Parisi value would be an important result.
\citet{bassoPerformanceLimitationsQAOA2022} conjectured that as $p\to\infty$, QAOA would be able to achieve the Parisi value on random $D$-regular graphs as $D\to\infty$.
If true, this would mean QAOA could also optimally solve the SK model, as well as the MaxCut problem~\cite{demboExtremalCutsSparse2017}.

    \item The Overlap Gap Property (OGP) exhibited by certain problem instances has been shown to pose obstacles for QAOA in achieving optimal solutions.
    These challenge arise when QAOA does not see the whole graph (e.g., when operating at sub-logarithmic depths)~\cite{chenSuboptimalityLocalAlgorithms2019}, as well as when it is able to see the whole graph (e.g., on fully-connected $k$-spin models)~\cite{bassoPerformanceLimitationsQAOA2022}. 
    In comparison, the classical Approximate Message Passing (AMP) algorithm can provably find solutions arbitrarily close to the true optima of $k$-spin models in tha absence of OGP (conjectured for $k=2$), but it also faces suboptimality when $k\geq 4$ is even.
    It therefore remains an open question whether QAOA can achieve a quantum advantage in scenarios involving the OGP.
    
    \item For instances with exceedingly small spectral gaps, QAOA can overcome the adiabatic limitations. \citet{zhouQuantumApproximateOptimization2020} noticed it for small problem sizes, but how this tendency could scale and allow to solve problems out of reach for adiabatic evolution is still unknown.
    
    \item \citet{rajakumarGeneratingTargetGraph2022} proposed a method enabling the construction of arbitrary coupling operations on quantum spin-systems, using global Ising operations and single qubit bit flips.
    This method exhibits promising scaling properties, suggesting potential advantages for dense, unweighted coupling operations like those in QAOA. However, an optimal operation sequence for arbitrary graphs remains elusive, likely pointing to an NP-hard problem. 
    Future work may focus on developing efficient solutions for specific types of graphs, seeking tighter upper bounds on construction, and formally establishing this problem's NP-hard nature. 
    Further research is also needed to explore how the complexity of compilation affects quantum algorithms when constructing arbitrary unweighted and weighted coupling graphs using global entangling operations in quantum spin systems.
    
    \item \citet{boulebnaneSolvingBooleanSatisfiability2022} derived analytical bounds on the average success probability of QAOA on random $k$-SAT in the limit of infinite problem size. 
    Further research could be done to explore QAOA's performance on other Boolean satisfaction problems and instances and investigate the relationship between QAOA's success probability and running time.
   
   \item \citet{dupontEntanglementPerspectiveQuantum2022} studied the growth and spread of entanglement resulting from optimized and randomized QAOA circuits, which could have implications for the simulation of QAOA circuits with tensor network-based methods. 
   Further research could be conducted on these connections and their impact on overcoming barren plateaus.

\end{itemize}

\paragraph{Parameter optimization strategies and initialization techniques:} Efficient methods for finding optimal parameters and initializing the algorithm are crucial for improving QAOA's performance, especially in noise and hardware limitations. 
Strategies include overcoming barren plateaus, exploiting parameter symmetries, and leveraging machine learning or reinforcement learning techniques.
Furthermore, the process of parameter initialization significantly influences QAOA's overall performance. 
Enhancing strategies for optimal parameter initialization and understanding their impact on performance at higher depths are critical future research directions.

In this direction, some promising avenues for future investigation include:

\begin{itemize}
    \item  It has been noticed that for similar graphs, the parameters of the QAOA ansatze share some similarities~\cite{brandaoFixedControlParameters2018}. 
However, it is unclear whether this phenomenon is due to the small graphs that have been investigated or whether this is more of a universal property that connects the ansatze to closely related graphs.

   \item The performance of BFGS for larger parameter vectors (i.e., for $p \geq 3$) in QAOA optimization is unclear~\cite{lotshawEmpiricalPerformanceBounds2021}.
    It would be interesting to evaluate the behavior and effectiveness of BFGS for parameter optimization in QAOA circuits with larger depths.

    \item How can parameter transferability and reusability be further utilized for optimizing QAOA performance?
    \citet{galdaTransferabilityOptimalQAOA2021} provided a theoretical foundation for parameter transferability in QAOA. However, more research is needed to understand how this can be applied across a wide range of problem instances and optimization tasks.

    \item Explore additional ways to leverage parameter symmetries for more efficient QAOA performance.
    \citet{shaydulinClassicalSymmetriesQuantum2021} investigated the correlation between QAOA and the inherent symmetries of the target function, using machine learning techniques to predict the QAOA performance accurately. 
    Additional research could be conducted to improve the understanding of symmetry in objective functions, cost Hamiltonians, and QAOA parameters themselves, leading to more efficient optimization processes.

    \item Can optimal QAOA parameters be derived analytically?
    \citet{wangQuantumApproximateOptimization2018}  derived analytical expressions for solving optimal parameters for the level-1 QAOA applied to MaxCut on general graphs. However, it is unclear whether this approach can be extended to higher values of p or other problem instances.
    Investigate the possibility of deriving analytical expressions for optimal QAOA parameters in various problem instances and optimization tasks. 
    This could potentially simplify the search for optimal parameter values and improve the overall efficiency of the QAOA algorithm.

    \item The effectiveness of unsupervised machine learning approaches, such as clustering,  in setting QAOA parameters has been demonstrated~\cite{moussaUnsupervisedStrategiesIdentifying2022}. However, how well these methods can generalize to other problems and larger instances remains to be seen.
    The same goes for reinforcement learning~\cite{khairyLearningOptimizeVariational2020,headleyApproximatingQuantumApproximate2022}.
    Investigate the applicability and generalization of unsupervised learning and RL methods for QAOA parameter optimization in a broader range of problem settings and larger problem instances.

    \item Can better parameter optimization strategies mitigate the challenge of achieving a quantum speedup with QAOA, even at low depths, due to factors such as barren plateaus~\cite{mccleanBarrenPlateausQuantum2018,wangNoiseinducedBarrenPlateaus2021}?

\end{itemize}



\paragraph{Understanding performance, benchmarking and comparison with classical algorithms:} Developing application-specific benchmarking methods and evaluating QAOA's performance across various problem instances will help determine its advantages over classical algorithms and guide researchers in choosing the most appropriate solver for a given optimization problem.
\textbf{Benchmarking against classical algorithms} will help identify the conditions under which QAOA can consistently surpass classical benchmarks, which is crucial to harnessing its full potential. 
This includes exploring cases where QAOA can offer a quantum advantage.
\textbf{Empirical studies} will also provide practical insights into the performance of QAOA. 
Simulator-based studies can shed light on how the algorithm performs in real-world scenarios. 
Further empirical investigations that complement theoretical studies can contribute to a more holistic understanding of the algorithm.
Finally, opportunities exist to \textbf{adapt or enhance classical algorithms} for quantum optimization. 
This could lead to significant advancements in quantum computing, including novel solutions that outperform existing quantum optimization algorithms.

Along these directions, some areas for further research include:

\begin{itemize}
    \item Machine Learning (ML) models have been employed to predict QAOA's performance and identify situations where QAOA is most likely to outperform classical algorithms~\cite{moussaQuantumNotQuantum2020, deshpandeCapturingSymmetriesQuantum2022}.
    Based on their findings, \citet{moussaQuantumNotQuantum2020} suggested exploring the potential of graph sparsification to enhance QAOA's performance.
    ML models like these provide valuable insights into the behavior of QAOA and offer guidance on leveraging its advantages, helping uncover opportunities for achieving quantum advantage.


    \item \citet{marwahaBoundsApproximatingMax2022} found that QAOA starts to outperform the threshold algorithm when $k>4$.
    However, this does not rule out the possibility that a different local tensor algorithm will match or outperform QAOA at larger $k$.
    Can QAOA outperform other local tensor algorithms at larger $k$?
 
    \item \citet{larkinEvaluationQAOABased2022} proposed a new metric for evaluating the runtime performance of QAOA and developed a method based on it to reduce the execution time of QAOA for solving MaxCut on 3-regular graphs.
    The effectiveness and scalability of this technique on other problems remains to be investigated. 
    Furthermore, studies on different performance metrics may provide more insights into how to improve QAOA's performance across different problems.

    \item Can the observed quantum speedup in specific problem domains be extended to more general cases? 
    For instance, the speedup shown by QAOA in the Maximum Independent Set (MIS) problem on 2D Rydberg atom arrays remains an open question for general cases~\cite{ebadiQuantumOptimizationMaximum2022}

        
    
\end{itemize}

\paragraph{Noise and error effects, mitigation techniques, and fault tolerance:} Understanding the impact of various noise types on QAOA's performance is essential. 
Developing noise mitigation strategies, error suppression schemes, and fault-tolerant techniques will improve the practicality of QAOA in real-world settings with decoherence and hardware limitations.
Specific strategies to \textbf{reduce noise} in QAOA and other Variational Quantum Algorithms (VQAs), such as optimizing SWAP networks, reducing the total number of CNOT gates, and leveraging equivalent circuit averaging, continue to be promising future directions.
Techniques such as symmetry verification have been proposed for \textbf{mitigating errors} in QAOA. 
This approach has shown effectiveness in improving output state fidelity, expected objective function value, and probability of sampling the optimal solution.

Some more specific potential areas for future research involve: 

\begin{itemize}
    \item According to \citet{camposTrainingSaturationLayerwise2021},  local coherent dephasing noise can remove training saturation in layer-wise learning, potentially aiding the optimization process. 
    However, it is unclear if there are other noise sources beyond the simple noise model that can play a similar role. Further investigation is needed to determine this.
    
    \item It was observed in numerical simulation that QAOA performance improves as noise correlation strength increases at fixed local error rates, suggesting a certain degree of noise resilience~\cite{kattemolleEffectsCorrelatedErrors2022}. 
    However, further studies are required to test the generalizability of this result.
       
    \item Optimizing SWAP networks can help reduce the negative impact of noise on QAOA and other VQAs~\cite{hashimOptimizedSWAPNetworks2022}. 
    More research should be done on improving SWAP network optimization techniques to enhance performance in practical settings.
     
    \item Is it possible to mitigate Noise-Induced Barren Plateaus (NIBP) with novel error mitigation strategies~\cite{wangNoiseinducedBarrenPlateaus2021}? 

    \item What is the prospect for quantum advantage of QAOA in the present of noise and errors?
    For example, a quantitative analysis of the error rates required to maintain the advantage of QAOA in noiseless scenarios would provide valuable insights.
    To further address this question, an empirical demonstration of quantum advantage of QAOA in real quantum devices is crucial.
    
\end{itemize}

\paragraph{Hardware-specific challenges and scalability:} 
Addressing hardware limitations such as connectivity, decoherence, and gate error rates is vital for scaling up QAOA to solve larger problems. 
\textbf{Leveraging specific hardware} to enhance QAOA performance has shown potential. 
However, different architectures possess their specific advantages and disadvantages that need thorough exploration.
Investigating QAOA's performance on different hardware platforms, including trapped ions, neutral atoms, superconducting qubits, photonic quantum computers, and the use of higher energy levels (qudits), will provide a deeper understanding of its capabilities and limitations.
For example, the utilization of higher energy levels beyond qubits presents an interesting prospect. 
In photonic quantum computers, qudits can be more efficiently mapped to the hardware for various problems.
Tailoring algorithms for photonic quantum computing also present many other advantages, such as potential improvements in terms of cluster state dimension compared to QAOA gate-based circuit algorithms.

Along these lines, several promising opportunities for further study consist of the following:

\begin{itemize}
  
    \item A systematic investigate the impact of compilation and routing qubits with swap networks on QAOA's performance in real hardware is needed. 
    This additional overhead can significantly impact the performance of the algorithm, especially when solving problems on graphs that differ from hardware connectivity~\cite{harriganQuantumApproximateOptimization2021}.
    
    \item When using ML models to predict the performance of QAOA~\cite{moussaQuantumNotQuantum2020}, adding some hardware-related features will assist in deciding whether or not to choose QAOA over classical algorithms on a specific quantum hardware.
    
    \item Recent observations showed that a decomposed implementation could significantly reduce the performance of QAOA\@.
Hence, it would be interesting to investigate the effect on QAOA by a direct hardware implementation of a combination of a controlled arbitrary phase and a SWAP gate~\cite{lacroixImprovingPerformanceDeep2020}.

    \item \citet{dlaskaQuantumOptimizationFourBody2022} concluded that we do not know how experimental investigations beyond numerical simulations for QAOA will perform using the innovative four-qubit Rydberg parity gate and parity architecture.

    \item According to \citet{proiettiNativeMeasurementbasedQuantum2022}, we should continue developing and tailoring algorithms for photonic quantum computing, as their MBQC QAOA for photonic quantum computers showed an up-to 30-fold improvement in terms of cluster state dimension when compared to QAOA gate-based circuit algorithms.
    
    \item Explore improvements in the surface code implementation, such as faster state distillation, to potentially produce quantum advantages on early fault-tolerant quantum processors for algorithms offering only quadratic speedups~\cite{sandersCompilationFaultTolerantQuantum2020,babbushFocusQuadraticSpeedups2021}.
    
\end{itemize}

\paragraph{Alternative optimization approaches and problem-specific algorithms:}
Exploring alternative optimization methods and tailoring algorithms to specific problem structures will undeniably enhance the performance and applicability of the QAOA across a diverse range of optimization problems.
It is worth investigating strategies such as FALQON, FQAOA, and others.
Furthermore, the concept of \textbf{shortcuts to adiabaticity} has played a crucial role in enhancing many ansatz designs.
As such, exploring this concept could lead to more efficient algorithms with faster convergence and reduced resource requirements.

In addition to these alternative optimization approaches, there is uncharted territory for potential improvements within the existing QAOA framework.
This includes adopting ``warm-starting'' approaches, which could provide superior capabilities, unique features, and enhanced practical use. 
These algorithmic enhancements could contribute significantly to the ongoing evolution of QAOA, broadening its scope and effectiveness.

\begin{itemize}

    \item New ansatz designs:
    research suggests exploring ma-QAOA to simplify parameter optimization~\cite{herrmanMultiangleQuantumApproximate2022,shiMultiAngleQAOADoes2022},
    investigating ab-QAOA for scalability and quantum advantage in combinatorial optimization problems~\cite{wurtzCounterdiabaticityQuantumApproximate2022}
    and leveraging QAOAnsatz for a broader set of problems~\cite{hadfieldQuantumApproximateOptimization2019}.

    \item \citet{herrmanRelatingMultiangleQuantum2022} investigated the connection between ma-QAOA and Continuous-Time Quantum Walks (CTQW) on dynamic graphs, showing their equivalence.
    Research could be done on investigating well-studied CTQW phenomena, such as hitting times, to improve our understanding of ma-QAOA and help find optimal parameters.
    
    \item Further research on WS-QAOA~\cite{eggerWarmstartingQuantumOptimization2021} to improve the initialization of QAOA parameters based on the solution to the relaxed QUBO problem, potentially enhancing performance at low depths, which is particularly important for implementation on NISQ devices.
    
    \item Research methods to deal with hard cases of combinatorial optimization problems where the energy landscape is rugged, such as the approach proposed by \citet{wangQuantumDropoutEfficient2022} using selective clause dropout.

    \item Applying QAOA to various Constraint Satisfaction Problems (CSPs) can deepen our understanding of the algorithm's effectiveness and applicability. 
    Evaluating its performance across these problems, especially those with unique properties is a valuable future research direction.

\end{itemize}

    \section{A Practical Guide to QAOA}%
    \label{sec:guide}
    The Quantum Approximate Optimization Algorithm is a versatile quantum algorithm that can be used to solve combinatorial optimization problems on quantum computers. 
This guide aims to provide researchers with detailed information on when and how to use QAOA effectively. 
It answers key questions such as which QAOA variant or ansatz to use for a specific problem, the benefits of using QAOA for different problem types, and how to tune and optimize the parameters of the QAOA ansatz. 
It also provides guidance to  apply QAOA to a real-world problem practically.

\subsection{Which QAOA Ansatz Variant Should I Use for My Problem?}

QAOA and its variants have been applied to a range of combinatorial optimization problems and various other problems relating to graph theory, constrained optimization, linear systems, and unstructured search. 
Some specific problems that QAOA and its variants have been successfully applied to include MaxCut, Traveling Salesman Problem, $k$-Vertex Cover, Discrete Portfolio Rebalancing, Sherrington-Kirkpatrick (SK) Hamiltonian, $k$-graph coloring, electric vehicle charging problems, and Max-$k$-Cut.

In light of this, when selecting the appropriate QAOA ansatz variant for a given problem, it is vital to consider the problem type, hardware constraints, and the trade-off between specificity and generality. 
Some QAOA variants seem particularly suitable for problems with hard constraints that always need to be satisfied and soft constraints that need to be minimized in their violations. 
They effectively solve optimization problems with rugged cost function landscapes and numerous local minima and have been demonstrated to sometimes outperform classical algorithms or provide some quantum advantage in such instances. 
In addition, QAOA's performance can benefit from leveraging problem symmetries, odd cycles, density, and other structural features. 
QAOA also seems to work well for hardware grid problems and problems where higher-depth versions are necessary for achieving satisfactory results.
Below we summarize the key features of some notable QAOA ansatz variants and which problems they might be good for.


\begin{description}
\item [Standard QAOA] This basic version works well for various combinatorial optimization problems like MaxCut, Minimum Vertex Cover (MVC), Constraint Satisfaction Problems (CSPs), and Maximum Independent Set (MIS). 
Use it as a starting point for problems with an unknown structure or simple instances.

\item [Multi-Angle QAOA (ma-QAOA)] This ansatz introduces new parameters into the circuit so that each element of the cost and mixer layers has its angle. 
It achieves better approximation ratios than vanilla QAOA and may require shallower circuits. It is suitable for combinatorial problems represented as graphs where a more complex parameter optimization is acceptable, as it may require shallower circuits but provides better or equal approximation ratios. 
The connection between ma-QAOA and Continuous-Time Quantum Walks (CTQW) on dynamic graphs might help find optimal parameters. 
Exploiting the natural symmetries of input graphs can reduce the number of ma-QAOA parameters by approximately 33\% with little to no impact on the objective function.

\item[QAOA+] This ansatz augments the traditional $p = 1$ QAOA with additional multi-parameter problem-independent layers of parameterized ZZ gates and a layer of mixer X gates. 
It is suitable for problems like MaxCut on random regular graphs and achieves higher approximation ratios than $p = 1$ QAOA while keeping the circuit depth low. 
The added circuit depth beyond the vanilla QAOA grows only in the number of qubits used as a set of $2N - 1$ parameters for $N$ qubits. 
QAOA+ outperforms the alternative multi-angle QAOA ansatz in most cases.

\item[Digitized Counterdiabatic QAOA (DC-QAOA)] This ansatz reduces computational complexity and circuit depth by utilizing counterdiabatic driving to speed up the optimization process. 
It is suitable for problems such as Ising models, classical optimization problems, and $k$-spin models. 
DC-QAOA seems to outperform the standard QAOA in all cases when applied to these problems.

\item[Adaptive Bias QAOA (ab-QAOA)] This ansatz incorporates adaptive bias fields into the mixer operators to accelerate convergence. 
The local fields (bias fields) are not optimized but updated according to a specific prescription. 
It is suitable for faster convergence and reduced computation time in combinatorial optimization problems and has a polynomially shorter computation time than vanilla QAOA\@. 
The improvement in computation time further increases with problem size.

\item[Adaptive Derivative-Assembled Problem-Tailored QAOA (ADAPT-QAOA)] This ansatz iteratively selects the QAOA mixer from a pre-defined pool of operators, maximizing the gradient of the commutator of the pool operator and the cost Hamiltonian over the ansatz of the previous step. 
It was reported to converge faster than standard QAOA while reducing the number of CNOT gates and optimization parameters by about 50\% each, particularly when entangling gates are included in the operator pool. 
The drawback of this method is that the selection of mixing operators requires additional measurements depending on the size of the operator pool.

\item[Recursive QAOA (RQAOA)] This non-local variant of QAOA iteratively reduces the size of the problem. It is designed to address the limitations caused by the $\mathbb Z_2$ symmetry of QAOA states and the geometric locality of the ansatz. 
It has shown promising results on NISQ devices and applies to combinatorial optimization problems like MaxCut.

\item[Quantum Alternating Operator Ansatzes (QAOAnsatz)] This ansatz alternates between a more general set of operators, allowing it to solve a broader range of problems, especially those with hard constraints defining feasible subspaces and soft constraints to minimize violations.
QAOAnsatz encompasses many variants such as XY mixer, Grover mixer, etc.~that can be tailored to different problems or constraints.

\item[Warm-Starting QAOA (WS-QAOA)] This approach initializes QAOA parameters based on the solution to the relaxed QUBO problem (continuous variables instead of binary ones). 
It provides initial states with the best approximation guarantee available classically in polynomial time and can be incorporated into the workflow of recursive QAOA. 
Suitable for recursive QAOA and provides initial states with the best approximation guarantee available classically in polynomial time.

\item[Feedback-based ALgorithm for Quantum OptimizatioN (FALQON)] This algorithm uses qubit measurements to assign values to quantum circuit variational parameters constructively, enabling approximate solutions without classical optimization. 
While it was designed for the nearly fault-tolerant era where deep circuits are feasible, it can also be used as a warm-start technique for NISQ devices, improving the initialization of standard QAOA\@.

\item[Fermionic QAOA (FQAOA)] This approach is suitable for solving combinatorial optimization problems with constraints, utilizing fermion particle number preservation to intrinsically impose these constraints and using fermion-based driver Hamiltonian for problem Hamiltonians with constraints.

\item[Spanning Tree QAOA (ST-QAOA)] In this ansatz, an approximate solution is constructed with $r$ rounds of gates explicitly tailored to the given problem instance using insights from the classical algorithm in the pre-computation routine. 
When $r = 1$, the ST-QAOA is guaranteed to match the performance of the classical solver. 
As the number of rounds increases, the ST-QAOA approaches the exact solution to the MaxCut problem. 
It achieves the same performance guarantee as the classical algorithm and can outperform vanilla QAOA at low depths.

\item[Ansatz Architecture Search (AAS)] This approach optimizes the QAOA ansatz and variational parameters by searching the discrete space of quantum circuit architectures near QAOA. 
It starts with a greedy search strategy and iteratively removes two-qubit gates from the best ansatz of the previous level, scored, and the best of them is selected as the output of this level. 
This method has shown significant improvement in the probability of finding low-energy states while using fewer two-qubit gates.

\item[Hardware-specific ansatz designs] Tailor your ansatz for better performance on specific platforms such as trapped ions, neutral atoms, superconducting qubits, and photonic quantum computers.
\end{description}

In the meantime, it is essential to note that while QAOA can be adapted and extended to various problems, it might not always be the most efficient or best-suited method for every problem.
For example, problems that do not have a well-defined graph representation, do not fit into the combinatorial optimization framework, or are not easily represented by Hamiltonians might prove more difficult to benefit from it. 
QAOA may also not be a good fit for certain problems due to performance limitations, locality constraints, and obstructions caused by specific properties of the problems~\cite{bravyiObstaclesVariationalQuantum2020,barakClassicalAlgorithmsQuantum2022,farhiQuantumApproximateOptimization2020}.
For example, it might not be suitable for problems that require a large depth or high entanglement due to limitations in quantum hardware and noise-related issues. 
Finally, the algorithm may also be less suitable when the graph topology differs significantly from the quantum hardware’s connectivity, as this may introduce substantial overhead in compiling and routing qubits.

Since the choice of QAOA variant deeply depends on the specific problem instance and available quantum hardware, the following guidelines can be helpful:

\begin{itemize}

\item MaxCut problems: Several QAOA variants have been suggested for MaxCut problems, including QAOA+, Warm-Starting QAOA (WS-QAOA), Recursive QAOA (RQAOA), Multi-Angle QAOA (ma-QAOA), Adaptive Bias QAOA (ab-QAOA), and Measurement Based Quantum Computing QAOA (MBQC QAOA). 
They all seem particularly well-suited for solving MaxCut problems, with multiple research papers mentioning their success in achieving a high approximation ratio.
QAOA+ improves over the traditional $p=1$ QAOA ansatz using additional layers of parameterized gates, providing higher approximation ratios. 
WS-QAOA leverages the best classical approximation guarantees, while RQAOA iteratively reduces the problem size, showing promise for NISQ devices. 
MBQC QAOA demonstrates improvements when applied to photonic quantum computers. 
The variant choice depends on the specific problem instance and available quantum hardware.

\item Combinatorial optimization problems with constraints: Adaptive Bias QAOA (ab-QAOA) and Fermionic QAOA (FQAOA) are suggested for these problems and are particularly applicable when the constraints are integral to the problem structure. 
ab-QAOA accelerates convergence by incorporating adaptive bias fields, while FQAOA utilizes fermion particle number preservation to impose constraints intrinsically. 
Grover Mixer QAOA (GM-QAOA) is suggested for CSPs, as it uses Grover-like selective phase shifting operators for efficient optimization and is not susceptible to Trotterization or Hamiltonian simulation errors.
Other approaches might also be suitable; the choice of optimal QAOA variant depends on the problem's structure and constraints.

\item Ising models: Digitized Counterdiabatic QAOA (DC-QAOA) and Ansatz Architecture Search (AAS) have been proposed as suitable variants for Ising models. 
DC-QAOA reduces computational complexity by leveraging counterdiabatic driving, while AAS modifies both the QAOA ansatz and parameter selection process to improve low-energy state identification.
DC-QAOA and AAS are not the only possible approaches for Ising models, but they are highlighted here for their unique approaches to reducing computational complexity and identifying low-energy states, respectively.

\item Graph-based problems: Recursive QAOA (RQAOA), Spanning Tree QAOA (ST-QAOA), and Graph Neural Networks (GNNs) initialization are recommended for graph-based problems. 
RQAOA iteratively reduces problem size and works well for graph-based problems, while ST-QAOA approaches the exact solution to the MaxCut problem as the number of rounds increases. 
GNNs-based initialization can speed up inference time across graphs.
Adaptive Derivative-Assembled Problem-Tailored (ADAPT-QAOA) is also a good choice for graph-based problems because it demonstrates faster convergence than the original QAOA due to shortcuts to adiabaticity.

\item Other problems:  
QAOA has also shown potential for solving problems such as the Sherrington-Kirkpatrick (SK) Hamiltonian problems, $k$-graph coloring and electric vehicle charging problems with global power contraints, Exact Cover problem, $k$-spin model, and ground-state energy calculations for small chemistry and material science problems.
Quantum Alternating Operator Ansatzes (QAOAnsatzes) has been proposed as a versatile extension of QAOA that works well for optimization problems with both hard and soft constraints.

\end{itemize}

\subsection{How to Optimize the QAOA Parameters?}

Initialization and optimization of QAOA parameters are crucial for achieving good performance. 
Different QAOA ansatz designs and hardware platforms will require different parameter optimization methods.
There is no universal solution for the selection of the parameter optimization method.
Nevertheless, it is possible to extract valuable insights and broad recommendations for choosing an appropriate method for QAOA applications.


\subsubsection*{Gradient-free methods}
Gradient-free methods are, in general, more suitable for problems with moderate circuit depths and limited computational resources.
Gradient-free methods often employed include BOBYQA, COBYLA, NEWUOA, Nelder-Mead, PRAXIS, and SBPLX.
BOBYQA within the APOSSM framework has shown the best performance for a fixed number of functional evaluations.

\noindent
\textbf{Strengths:}

\begin{itemize}
    \item Computationally efficient
    \item Requires fewer function evaluations when compared to gradient-based methods
    \item Some methods, such as SPSA and Powell's method, are less affected by noise than other methods
\end{itemize}

\noindent
\textbf{Weaknesses:}

\begin{itemize}
\item Performance can be challenged with an increasing number of layers
\item Some methods, like COBYLA, Nelder-Mead, and Conjugate-Gradient, are significantly affected by noise
\end{itemize}

\noindent
\textbf{When to use:} Gradient-free methods can be appropriate when computational efficiency is important and you are working with problems with modest circuit depths. 
Some gradient-free methods may be preferred when noise levels are low or moderate.

\subsubsection*{Gradient-based approaches}

Gradient-based approaches are more robust to variations in problems and noise but may require more computational resources. 
Often employed gradient-based approaches include Gradient Descent, Stochastic Gradient Descent (SGD), Model Gradient Descent (MGD), and BFGS.

\noindent
\textbf{Strengths:}

\begin{itemize}
\item Widely used for parameter optimization
\item Can handle problems with a smooth objective function
\item Can perform well even if the objective function is not smooth with respect to the error (policy-gradient-based reinforcement learning)
\item Some methods (e.g., AMSGrad and BFGS) perform better than gradient-free methods in the presence of shot-noise
\end{itemize}

\noindent
\textbf{Weaknesses:}

\begin{itemize}
\item Can be computationally expensive
\item Vulnerable to noise in NISQ devices
\item May require many measurements for each gradient component
\end{itemize}

\noindent
\textbf{When to use:} Gradient-based approaches may be appropriate when the objective function is smooth and when you require a method that can handle a wide range of problems. 
These methods may also be preferred when noise levels are low or when working with problems requiring a robust optimization process.

\subsubsection*{Machine Learning Approaches}
Machine Learning approaches can provide faster convergence and better optimization results but may require more training instances and computational resources.
Some ML approaches include Gaussian Process Regression (GPR), Linear Regression (LM), Regression Tree (RTREE), Support Vector Machine Regression (RSVM), Long Short-Term Memory (LSTM) neural networks, Graph Neural Networks (GNNs), and Reinforcement Learning (RL).

\noindent
\textbf{Strengths:}

\begin{itemize}
\item Can exploit correlations and patterns among parameters
\item Can accelerate QAOA optimization
\item Can generalize across different problem instances and graph sizes
\end{itemize}

\noindent
\textbf{Weaknesses:}

\begin{itemize}
\item Scalability can be an issue, particularly for more complex problems
\item May require many training instances for good performance
\end{itemize}

\noindent
\textbf{When to use:} Machine learning approaches can be appropriate when working with problems with correlations or patterns among parameters or when you require a method that can generalize across different problem instances and graph sizes. 
These approaches may also be suitable when working with large datasets or complex problems requiring advanced optimization techniques.

\subsubsection*{Other considerations}
Certain optimization problems can showcase specific attributes such as barren plateaus, transferability and reusability of parameters, or the existence of parameter symmetries. 
These characteristics can be harnessed to enhance the optimization procedure. 
For instance, adopting layerwise learning strategies can aid in bypassing barren plateaus, and exploiting parameter symmetries can simplify the optimization process.

Some general strategies and ideas include:

\begin{description}
    \item[Warm-starting QAOA] Initializing QAOA parameters based on the solution to the relaxed QUBO problem can provide a good starting point for optimization.
    It provides an advantage over standard QAOA at low depth, which is particularly important for NISQ devices.

   \item[FALQON's purely quantum optimization loop] By using FALQON's measurement-based feedback loop, one can assign values to quantum circuit variational parameters, improving the initialization of standard QAOA.
   It is primarily designed for fault-tolerant quantum devices that do not yet exist, limiting current applicability.

   \item[FQAOA] This method is designed to solve combinatorial optimization problems with constraints, utilizing fermion particle number preservation to impose these constraints intrinsically.

   \item[AAS] This technique searches the discrete space of quantum circuit architectures near QAOA to find a better ansatz.

    \item[Layerwise learning strategy] Layerwise learning strategies grow the circuit incrementally and update only subsets of parameters during optimization.
    This can help to avoid initializing on a plateau and reduce the probability of creeping onto a plateau during training. 
    This can improve the optimization process and help QAOA converge to better solutions.

    \item [Parameter Transferability and Reusability] Optimal QAOA parameters can be transferred and reused across different problem instances based on their local characteristics. 
    This can improve the quality of the solution and reduce the number of evaluations required to reach it.
    
    \item[Constraint-based optimization] Constraining QAOA circuit parameters to the range $(0, \pi)$ to exploit their symmetry can result in runtime acceleration of up to 5.5 times.

    \item[Parameter regression] Using parameter regression to optimize QAOA parameters can achieve an acceleration of more than 1.23 times.

    \item [Leveraging Parameter Symmetries] Parameter symmetries can simplify the optimization process and eliminate degeneracies in the parameter space. 
    Exploiting these symmetries can make the search for optimal QAOA parameters more efficient.

    \item[Exploit natural symmetries] Exploiting the natural symmetries of input graphs can reduce the number of parameters and improve QAOA performance by approximately 33\%.

    \item[Use Continuous-Time Quantum Walks] Relate ma-QAOA to CTQW on dynamic graphs to leverage well-studied CTQW phenomena for finding optimal parameters.

    \item[Heuristic strategies for parameter initialization] Optimal initialization of QAOA's parameters can be determined in $O(\text{poly}(p))$ time, while random initialization would necessitate $2^{O(p)}$ optimization runs to attain comparable performance.

    \item[Hardware-specific optimizations] Consider the unique properties and limitations of your chosen quantum hardware platform when optimizing QAOA parameters. 
	Depending on the hardware used, this may involve leveraging connectivity, qudits, global entangling operations, or higher energy levels.
\end{description}

\subsection{Practical Considerations for Implementing QAOA}

When applying QAOA to a problem, consider the following practical advice:

\begin{itemize}

\item Choose an ansatz that suits the problem type and hardware design while balancing specificity and generality.

\item Employ Ansatz Architecture Search (AAS) if you seek to optimize both the QAOA ansatz and variational parameters. 
This method uses a greedy search strategy, exploring the space of quantum circuit architectures to find a more suitable ansatz for your problem.

\item Assess the impact of different mixer designs on your QAOA algorithm and the trade-offs between improved performance and longer circuit depth.

\item Experiment with different parameter initialization and optimization methods, such as heuristic strategies, parameter fixing, and Graph Neural Networks (GNNs),  to find the best approach for your specific problem type and hardware platform.

\item Select an appropriate optimization algorithm for your problem, considering the trade-offs of gradient-free methods, gradient-based approaches, and machine learning techniques.

\item Explore the transferability and reusability of optimal parameters across different problem instances based on local characteristics of subgraphs.

\item Leverage parameter symmetries to simplify optimization, eliminate degeneracies in the parameter space, and improve QAOA performance. 
Investigate symmetries in objective functions, cost Hamiltonians, and QAOA parameters themselves.

\item To overcome barren plateaus in cost function landscapes, consider employing layerwise learning strategies, incremental circuit depth growth, or initializing closer to target parameters. 

\item When working with large-scale problems or complex optimization landscapes, be prepared to employ advanced optimization techniques, hybrid quantum-classical approaches, or specialized ansatz variations to overcome challenges and achieve desired results.

\item To overcome noise and error challenges, consider implementing error mitigation techniques such as gate count reduction, SWAP network optimization, symmetry verification, and leveraging hardware-specific features.

\end{itemize}

    \section*{Acknowledgements}
    We thank Dr.~Michele Grossi and Michał Stęchły for their invaluable discussions and insights that have greatly benefited this project.
We would like to express our gratitude to the Quantum Open Source Foundation (QOSF) for organizing the mentorship program that enabled the formation of our team.
We acknowledge the use of IBM Quantum services for this work. 
The views expressed are those of the authors, and do not reflect the official policy or position of IBM or the IBM Quantum team.
In this paper we used \verb|ibm_oslo|, \verb|ibm_lagos|, \verb|ibm_nairobi| and \verb|ibm_perth| which are some of the IBM Quantum Falcon r5.11H devices.
The authors disclosed receipt of financial support for the publication of this article from Quantum Neural Technologies S.A.

\paragraph{Author Contributions:}
Conceptualization, KB;
methodology, KB, AC, RL and AS;
software, major contributions: KB, DB, AC, RL, AS; minor contributions: CC, KP;
formal analysis and investigation, all authors contributed equally;
writing---original draft, and 
writing---review and editing, all authors contributed equally;
supervision, KB;
project administration, KB, AC and RL.

    \bibliography{main.bib}

    \appendix

\end{document}